\documentclass[10pt,aps,showpacs,nofootinbib,prd,aps,epsf,floats,
               amsmath,amssymb,amsfonts,axodraw,floatfix,graphicx,twocolumn]{revtex4-1}
\usepackage{amsmath, amssymb}
\usepackage{multirow}
\usepackage{paralist}
\newcommand{\mathsym}[1]{{}}

\usepackage{graphicx}
\usepackage{amsmath}
\usepackage{amssymb}
\usepackage{mathrsfs}
\newsavebox{\PSLASH}
 \sbox{\PSLASH}{$p$\hspace{-1.8mm}/}
 
\renewcommand{\theequation}{\thesection.\arabic{equation}}
\newcounter{saveeqn}
\newcommand{\add}{\addtocounter{equation}{1}}
\newcommand{\alphaeqn}{\setcounter{saveeqn}{\value{equation}}%
\setcounter{equation}{0}%
\renewcommand{\theequation}{\mbox{\thesection.\arabic{saveeqn}{\alpha{equation}}}}}
\newcommand{\reseteqn}{\setcounter{equation}{\value{saveeqn}}%
\renewcommand{\theequation}{\thesection.\arabic{equation}}}

 \newsavebox{\notrightarrow}
 \sbox{\notrightarrow}{$\to$\hspace{-4mm}/}
 
 \newsavebox{\PARTIALSLASH}
 \sbox{\PARTIALSLASH}{$\partial$\hspace{-1.6mm}/}
 
 \newsavebox{\ASLASH}
 \sbox{\ASLASH}{$A$\hspace{-2.1mm}/}
 
 \newsavebox{\KSLASH}
 \sbox{\KSLASH}{$k$\hspace{-1.8mm}/}
 
 \newsavebox{\LSLASH}
 \sbox{\LSLASH}{$\ell$\hspace{-1.8mm}/}
 
 \newsavebox{\QSLASH}
 \sbox{\QSLASH}{$q$\hspace{-1.8mm}/}
 
 \newsavebox{\DSLASH}
 \sbox{\DSLASH}{$D$\hspace{-2.2mm}/}
 
 \newsavebox{\DbfSLASH}
 \sbox{\DbfSLASH}{${\mathbf D}$\hspace{-2.8mm}/}
 
 \newsavebox{\DELVECRIGHT}
 \sbox{\DELVECRIGHT}{$\stackrel{\rightarrow}{\partial}$}
 
 \newcommand{\blue}{\IfColor{\textCadetBlue}{}}
\newcommand{\black}{\IfColor{\textBlack}{}}
\newcommand{\red}{\IfColor{\textRed}{}}
\newcommand{\green}{\IfColor{\textOliveGreen}{}}
\newcommand{\lil}{\IfColor{\textRedViolet}{}}

\begin{document}
\title{Inverse magneto-rotational catalysis and the phase diagram of a rotating hot and magnetized quark matter}
\author{N. Sadooghi$^{a}$}\email{Corresponding author: sadooghi@physics.sharif.ir}
\author{S. M. A. Tabatabaee Mehr$^b$}\email{tabatabaee@ipm.ir}
\author{F. Taghinavaz$^b$}\email{ftaghinavaz@ipm.ir}
\affiliation{${\ }^a$Department of Physics, Sharif University of Technology,
P.O. Box 11155-9161, Tehran, Iran}
\affiliation{${\ }^{b}$School of Particles and Accelerators, Institute for Research in
Fundamental Sciences (IPM), P.O. Box 19395-5531, Tehran, Iran}
\begin{abstract}
	We study the properties of a hot and magnetized quark matter in a rotating cylinder in the presence of a constant magnetic field. To do this, we solve the corresponding Dirac equation using the Ritus eigenfunction method. This leads to the energy dispersion relation, Ritus eigenfunctions, and the quantization relation for magnetized fermions. To avoid causality-violating effects, we impose a certain global boundary condition, and study its effect, in particular, on the energy eigenmodes and the quantization relations of fermions. Using the fermion propagator arising from this method, we then solve the gap equation at zero and nonzero temperatures. At zero temperature, the dynamical mass $\bar{m}$ does not depend on the angular frequency, as expected. We thus study its dependence on the distance $r$ relative to the axis of rotation and the magnetic field $B$, and explore the corresponding finite size effect for various couplings $G$. We then consider the finite temperature case. The dependence of $\bar{m}$ on the temperature $T$, magnetic field $B$, angular frequency $\Omega$, and distance $r$ for various $G$ is studied. We show that $\bar{m}$ decreases, in general, with $B$ and $\Omega$. This is the ''inverse magneto-rotational catalysis (IMRC)'' or the ''rotational magnetic inhibition'', previously discussed in the literature. To explore the evidence of this effect in the phase diagrams of our model, we examine the phase portraits of the critical temperature $T_c$ as well as the critical angular frequency $\Omega_c$ with respect to $G, B,\Omega$, and $r$ as well as $G, B, T$, and $r$, respectively. We show that $T_{c}$ and $\Omega_c$ decrease, in particular, with $B$. This is interpreted as clear evidence for IMRC. 
\end{abstract}
\maketitle
\section{Introduction}\label{sec1}
\setcounter{equation}{0}
One of the main goals of modern experiments of heavy-ion collision (HIC) at the Relativistic Heavy Ion Collider (RHIC) and Large Hadron Collider (LHC) is to study the quark matter under extreme conditions. These include extremely high temperatures ($>10^{12}$ K \cite{rajagopal2018}), very large densities (up to $5\rho_{0}$ with $\rho_0=2.7\times 10^{14}$ gr/cm$^{3}$ \cite{lacey2002}), extremely large electromagnetic fields ($10^{18}-10^{20}$ G \cite{warringa2007}), and, in particular, large angular frequencies ($10^{22}$ Hz \cite{becattini2016}). The aim is to imitate the circumstances of the early Universe, which is believed to be made of a hot plasma of free quarks and gluons. It is known that the plasma of quarks and gluons undergoes certain Quantum Chromodynamics (QCD) phase transitions upon cooling, and this leads to hadronization. These transitions include a deconfinement/confinement and, in particular, a chiral phase transition. Theoretically, the QCD phase transitions can be studied using various effective QCD-like models, e.g. the Nambu-Jona--Lasinio (NJL) model \cite{klevansky1992} and its extensions. Numerically, it is the merit of lattice QCD simulations at zero density \cite{bazavov2013}, that show, \textit{inter alia}, that these transitions occur at the same critical temperature, and they are nothing other than a smooth crossover. As concerns the QCD matter under high density/high baryon chemical potential, it is shown, via model building, that it undergoes a certain spontaneous color symmetry breaking, that leads to the formation of diquarks in a color superconductive medium \cite{fayazbakhsh2011}. Another important feature of noncentral HICs is the generation of very strong magnetic fields, which has many exciting effects on the Quark matter created in these collisions \cite{warringa2007, skokov2009, huang2015}. These effects, including the (inverse) magnetic catalysis (see \cite{shovkovy2015} and the references therein)
and the chiral magnetic effect \cite{fukushima2008}, are the subject of intensive studies in recent years. In particular, the impact of a constant magnetic field on the QCD phase diagram is studied intensively in the literature \cite{fayazbakhsh2011, cao2021}. The main focus here is on the catalytic effect of constant magnetic fields. This enhances the formation of chiral condensates and thus leads, in comparison to the field-free case, to an increase of the critical temperature of the chiral phase transition $T_c$. There are, however, pieces of evidence from lattice QCD simulations that in the absence of baryonic chemical potential, $T_c$ decreases with increasing the strength of the magnetic field \cite{bali2012-1, bali2012-2, delia2013,bruckmann2013}. This effect, which is previously dubbed "inverse magnetic catalysis" \cite{rebhan2011}, is shown to be present in dense quark matter \cite{fayazbakhsh2011}, or once the anomalous magnetic moment of the quark matter is nonzero \cite{fayazbakhsh2014}, or an axial vector interaction is present
\cite{huang2014}, or when the scalar coupling constant of effective models depends on the magnetic field \cite{farias2014, ferrer2014}, or for nonlocal chiral quark models, \cite{scoccola2017}. The true reason for the inverse magnetic catalysis is still under debate (see \cite{cao2021} and the references therein).
\par
Apart from extreme temperatures, densities, and external electromagnetic fields, the plasma of quarks and gluons created at RHIC and LHC possesses extremely large vorticity. This is the purpose of the present paper to focus on the interplay between rotation, magnetic field, and temperature on the chiral symmetry breaking (see below for more explanation). 
A simple estimate of the nonrelativistic vorticity $\boldsymbol{\Omega}=\frac{1}{2}\boldsymbol{\nabla}\times\boldsymbol{v}$ is made in  \cite{becattini2016}. Assuming that the difference between the $z$ component of the collective velocity in a HIC close to the target and projectile spectators is  about 0.1 (in the units of the speed of light), and that the transverse size of the system is about $5$ fm, the vorticity $\Omega$ turns out to be of the order 0.02 fm$^{-1}\sim 10^{22}$ Hz \cite{becattini2016}. Many interesting transport phenomena are related to a rotating quark matter, whose macroscopic description is mainly made by relativistic hydrodynamics. Some of them are the chiral vortical effect and wave, in analogy to chiral magnetic effect and wave (see \cite{kharzeev2015} and references therein). Similar to the case of magnetic fields, there are several attempts to study the phase structure of QCD under rotation. In \cite{liao2016}, the effect of rotation on the formation of two different condensates in a hot and dense QCD matter, the chiral condensate and the color superconductivity, are studied, and the $T$-$\Omega$ phase portrait is presented. It is found that a generic rotational suppression effect occurs, in particular, on the scalar pairing states. This effect is supposed to be caused by a rotational polarization effect induced by the global rotation. In order to check whether pairing states with nonzero angular momentum are favorable, the effect of rotation on the chiral phase transition in an NJL model with a vector interaction is studied in \cite{huang2018}.  It is shown that whereas the phase structure in the $T$-$\mu$ plane is sensitive to the coupling strength in the vector channel, the phase structure in $T$-$\Omega$ plane is not. 
The aforementioned suppression of the chiral condensate is originally found in \cite{fukushima2015}. Here, the Dirac equation of a single flavor fermionic system is solved in the presence of rotation and magnetic field, and the corresponding energy dispersion relation is found. The latter indicates a close analogy between the rotation and the chemical potential because the energy spectrum is shifted similarly by a term proportional to the angular frequency $\Omega$ of the fermionic system.
In \cite{fukushima2015}, after solving the Dirac equation in the presence of rotation and magnetic field, and after determining the energy dispersion relation, the authors introduce the temperature and magnetic field in a system without boundary conditions. The zero temperature case is then derived by taking the limit $T\to 0$. In this way, the dynamical mass exhibits $\Omega$ dependence, and decreases with increasing $\Omega$. At a certain critical $\Omega$ the dynamical mass vanishes, and the chiral symmetry is restored. 
The fact that the chiral condensate is suppressed in the presence of finite rotation is interpreted as the inverse magnetic catalysis, a phenomenon which occurs, in general, in low energy effective models at finite densities \cite{fayazbakhsh2011}.  
It is referred to as "rotational magnetic inhibition".\footnote{In this paper, we use the term "inverse magneto-rotational effect".} In \cite{ebihara2017}, it is, however, shown that in an explicit computation at zero temperature, the dynamical mass does not depend on $\Omega$. A fact that is also confirmed in the present paper. In the absence of magnetic fields, the authors in \cite{ebihara2017} also introduce a global boundary condition to avoid causality-violating problems. This is also systematically done in a series of papers by Chernodub et al. \cite{chernodub2016-1, chernodub2016-2, chernodub2016-3} in the absence and presence of magnetic fields. Here, another MIT boundary condition is imposed on the fermions on the surface of the cylinder, and its effect of the phase diagram of a QCD-like model in the presence of rotation is studied. The spectral and MIT boundary conditions are originally introduced in \cite{kdrothe1980} and \cite{chodos1974, lutken1984}. Various effects of these boundary conditions on the thermal expectation values of the fermion condensate, neutrino charge, and stress-energy tensor are studied intensively in \cite{ambrus2016}. Other recent studies of the effect of rotation on the confinement/deconfinement phase transition and mesonic condensation are studied in \cite{chernodub2020, fukushima2021, braguta2021} and \cite{zhang2020, cao2019}. 
\par
In the present paper, we continue studying the interplay between rotation and magnetic field at zero and finite temperatures using a global boundary condition, and gain additional insights into IMRC. To do this, we use a one flavor NJL model, and solve numerically the corresponding gap equation for different fixed parameters $T,eB,\Omega$, and $r$. The aim is, in particular, to find pieces of evidence for this effect in the phase diagrams of our model. The organization of the paper is as follows: In Sec. \ref{sec2}, we solve the  Dirac equation within a cylinder using the Ritus eigenfunction method \cite{ritus1972}. In Sec. \ref{sec2a}, the solution is presented for a system with no boundary condition, and in  Sec. \ref{sec2b}, it is given for a system with a global boundary condition. In Secs. \ref{sec2a3} and \ref{sec2b3}, the quantization of fermionic fields in a system without and with boundary conditions is demonstrated. It is then used in Sec. \ref{sec2b4} to derive the fermion propagator of fermions in a bounded, rotating, and magnetized system. 
In this context, the Ritus eigenfunction formalism is introduced as a methodical novelty in the present paper, though the same notations as previously introduced and utilized in \cite{fayazbakhsh2011, fayazbakhsh-ritus, sadooghi2016, tabatabaee2020} are used. 
\par
In Sec. \ref{sec3}, the numerical solutions of the gap equation at zero and nonzero temperatures are presented (see Secs. \ref{sec3a} and \ref{sec3b}). At zero temperature, we mainly focus on the $r$ and $eB$ dependence of the dynamical mass for different values of NJL couplings. We show, in particular, that the $eB$ dependence of the dynamical mass at some fixed distance relative to the rotation axis and for a relatively large coupling exhibits certain oscillations. These are due to successive filling of the Landau levels. We then focus on the $T,eB,\Omega$ and $r$ dependence of the dynamical mass at finite temperature.  We show, in particular, that the $eB$ dependence of $\bar{m}$  decreases with increasing $eB$. Moreover, $\bar{m}$ decreases with increasing $\Omega$. These are clear pieces of evidence of IMRC. We study the $G,eB,\Omega$, and $r$ dependence of the critical temperature $T_c$, and show that it decreases with $eB$ and $\Omega$. We finally examine the $G,eB, T$, and $r$ dependence of the critical angular frequency $\Omega_c$, and show that it decreases with $eB$ and $T$. The latter results can be viewed as a new piece of evidence of the IMRC. We devote Sec. \ref{sec4} to a number of concluding remarks.       
\section{Ritus Eigenfunction formalism and rotating fermions in a constant magnetic field}\label{sec2}
\setcounter{equation}{0}
In this section, we use the Ritus eigenfunction method \cite{ritus1972} to solve the Dirac equation of a charged and massive fermion in the presence of a constant magnetic in a system that uniformly rotates with a constant angular velocity $\Omega$ about a fixed axis. Being interested on the boundary effects, we set the system under certain global boundary condition, and explore its consequences for the solution of the corresponding Dirac equation. We assume that the magnetic field is aligned in the $z$-direction, and that all spatial regions of the system have the same angular velocity about the same axis (rigid rotation). This system is thus cylindrical symmetric around this axis, and is naturally described by the cylindrical coordinate system $x^{\mu}=(t,x,y,z)=(t,r\cos\varphi,r\sin\varphi,z)$. The corresponding line element reads \cite{chernodub2016-1}
\begin{eqnarray}\label{N1}
ds^2&=&g_{\mu\nu}dx^{\mu}dx^{\nu}=\left(1-r^{2}\Omega^{2}\right)dt^2-dx^2\nonumber\\
&&+2\Omega ydtdx-dy^2-2\Omega xdtdy-dz^{2}.
\end{eqnarray}
This is equivalent to the metric
\begin{eqnarray}\label{N2}
\hspace{-0.5cm}g_{\mu\nu}=\left(
\begin{array}{cccc}
1-\Omega^2(x^2+y^2)&+\Omega y&-\Omega x&0\\
+\Omega y&-1&0&0\\
-\Omega x&0&-1&0\\
0&0&0&-1
\end{array}
\right).
\end{eqnarray}
Adopting the conventional notations in the curved space, we use the vierbein $e^{\mu}_{~a}$ to connect the general coordinate with the Cartesian coordinate in the local rest frame (tangent space), $x^{\mu}=e^{\mu}_{~a}x^{a}$. Here, the Greek indices $\mu=t,x,y,z$ refer to the general coordinate in the rotating frame, while the Latin indices $a=0,1,2,3$ to the Cartesian coordinate in the local rest frame. We choose the nonvanishing components of $e^{\mu}_{~a}$ as \cite{chernodub2016-1, fukushima2015}
\begin{eqnarray}\label{N3}
e^{\mu}_{~a}:&\quad& e^{t}_{~0}=e^{x}_{~1}=e^{y}_{~2}=e^{z}_{~3}=1,\nonumber\\
&& e^{x}_{~0}=+y\Omega,\quad e^{y}_{~0}=-x\Omega.
\end{eqnarray}
They lead together with $g_{\mu\nu}$ from \eqref{N2} to the metric $\eta_{ab}=g_{\mu\nu}e^{\mu}_{~a}e^{\nu}_{~b}=\mbox{diag}\left(1,-1,-1,-1\right)$.
\par
In a curved spacetime, the Dirac equation of a charged massive fermion in a constant background magnetic field is given by
\begin{eqnarray}\label{N4}
\bigg[i\gamma^{\mu}\left(D_{\mu}^{(q)}+\Gamma_{\mu}\right)-m_{q}\bigg]\psi(x)=0,
\end{eqnarray}
with $D_{\mu}^{(q)}\equiv \partial_{\mu}-iqeA_{\mu}$. Here, $m_{q}$ is the mass of the fermion with charge $eq, e>0$. The gauge field $A_{\mu}$ in the rotating frame is defined by $A_{\mu}=e^{a}_{~\mu}A_{a}$. Here, $e^{a}_{~\mu}$s satisfy $e^{a}_{~\mu}e^{\mu}_{~b}=\delta^{a}_{~b}$, and are given by
\begin{eqnarray*}
e^{a}_{~\mu}:&\quad& e^{0}_{~t}=e^{1}_{~x}=e^{2}_{~y}=e^{3}_{~z}=1, \nonumber\\
&& e^{t}_{~1}=-y\Omega,\quad e^{t}_{~2}=+x\Omega.
\end{eqnarray*}
Choosing $A_{a}=\left(0,-\boldsymbol{A}\right)=\left(0,By/2,-Bx/2,0\right)$, we arrive at a magnetic field aligned in the $z$-direction $\boldsymbol{B}=B\boldsymbol{\hat{z}}$ with $B>0$. In \eqref{N4}, the affine connection $\Gamma_{\mu}$ is defined in terms of the spin connection $\omega_{\mu ab}$ and vierbeins $e^{\mu}_{~a}$ as
\begin{eqnarray}\label{N5}
\Gamma_{\mu}&\equiv&-\frac{i}{4}\omega_{\mu ab}\sigma^{ab},
\end{eqnarray}
with
\begin{eqnarray}\label{N6}
\omega_{\mu ab}\equiv g_{\alpha\beta}e^{\alpha}_{~a}\left(\partial_{\mu}e^{\beta}_{~b}+\Gamma^{\beta}_{\mu\nu}e^{\nu}_{~b}\right),
\end{eqnarray}
and $\sigma^{ab}\equiv\frac{i}{2}[\gamma^{a},\gamma^{b}]$. In \eqref{N6}, the Christoffel connection $\Gamma^{\beta}_{\mu\nu}\equiv \frac{1}{2}g^{\beta\sigma}\left(\partial_{\mu}g_{\sigma\nu}+\partial_{\nu}g_{\mu\sigma}-\partial_{\sigma}g_{\mu\nu}\right)$. As it turns out, for the metric \eqref{N2}, the nonvanishing components of $\Gamma^{\beta}_{\mu\nu}$ are given by
\begin{eqnarray}\label{N7}
\Gamma_{tt}^{x}=-\Omega^{2}x,&\quad&\Gamma_{tt}^{y}=-\Omega^{2}y,\nonumber\\
\Gamma_{tx}^{y}=\Gamma_{xt}^{y}=\Omega,&\quad& \Gamma_{ty}^{x}=\Gamma_{yt}^{x}=-\Omega.
\end{eqnarray}
The affine connection $\Gamma_{\mu}$ is then given by
\begin{eqnarray}\label{N8}
\Gamma_{t}=-\frac{i}{2}\Omega\sigma^{12},\quad\Gamma_{x}=\Gamma_{y}=\Gamma_{z}=0.
\end{eqnarray}
Moreover, the $\gamma$-matrices in \eqref{N4} are defined by $\gamma^{\mu}=e^{\mu}_{~a}\gamma^{a}$. For $e^{\mu}_{~a}$ given in \eqref{N3}, they read \cite{chernodub2016-1}
\begin{eqnarray}\label{N9}
\begin{array}{rclcrcl}
\gamma^{t}&=&\gamma^{0},&\quad&\gamma^{x}&=&y\Omega\gamma^{0}+\gamma^{1},\\
\gamma^{y}&=&-x\Omega \gamma^{0}+\gamma^{2},&\quad& \gamma^{z}&=&\gamma^{3}.
\end{array}
\end{eqnarray}
Plugging $\Gamma_{\mu}$ from \eqref{N8} and $\gamma^{\mu}$ from \eqref{N9} into \eqref{N4}, the explicit form of the Dirac equation of a rotating fermionic system in  a constant magnetic field reads
\begin{eqnarray}\label{N10}
\left(\gamma\cdot \Pi^{(q)}-m_{q}\right)\psi^{(q)}=0,
\end{eqnarray}
where
\begin{eqnarray}\label{N11}
\gamma\cdot\Pi^{(q)}&\equiv&
i\gamma^{0}\left(\partial_{t}-i\Omega \hat{J}_{z}\right)+i\gamma^{1}\left(\partial_x+iqeBy/2\right)
\nonumber\\
&&+i\gamma^{2}\left(\partial_y-iqeBx/2\right)+i\gamma^{3}\partial_{z},
\end{eqnarray}
and $\hat{J}_{z}\equiv\hat{L}_{z}+\Sigma_{z}/2$ with $\hat{L}_{z}\equiv-i\left(x\partial_{y}-y\partial_{x}\right)$, the total angular momentum in the $z$-direction, and $\Sigma_{z}\equiv \mathbb{I}_{2\times 2}\otimes \sigma^{3}$. Here, we used the Weyl representation of the $\gamma$-matrices
\begin{eqnarray}\label{N12}
\gamma^{0}=\left(\begin{array}{cc}
0&1\\
1&0
\end{array}
\right),\quad\boldsymbol{\gamma}=\left(
\begin{array}{cc}
0&\boldsymbol{\sigma}\\
-\boldsymbol{\sigma}&0
\end{array}
\right),
\end{eqnarray}
with $\boldsymbol{\sigma}=\left(\sigma^{1},\sigma^{2},\sigma^{3}\right)$ are the Pauli matrices, and $[\sigma^{i},\sigma^{j}]=2i\epsilon^{ijk}\sigma^{k}$ to get $\sigma^{12}=\frac{i}{2}[\gamma^{1},\gamma^{2}]=\Sigma_{z}$.   Moreover, $\mathbb{I}_{2\times 2}\equiv \text{diag}(1,1)$.
\par
Similar to the description presented in \cite{tabatabaee2020}, in the Ritus eigenfunction method, we start solving \eqref{N10} by making use of the Ansatz $\psi_{+}^{(q)}=\mathbb{E}_{\lambda,\ell,+}^{(q)}u\left(\tilde{p}_{\ell,+}\right)$ for the positive frequency solution and $\psi_{-}^{(q)}=\mathbb{E}_{\lambda,\ell,-}^{(q)}v\left(\tilde{p}_{\ell,-}\right)$ for the negative frequency solution. Here, $\mathbb{E}_{\lambda,\ell,\kappa}^{(q)}$ with $\kappa=\pm 1$ satisfies the Ritus eigenfunction relation
\begin{eqnarray}\label{N13}
\left(\gamma\cdot \Pi^{(q)}\right)\mathbb{E}_{\lambda,\ell,\kappa}^{(q)}=\kappa \mathbb{E}_{\lambda,\ell,\kappa}^{(q)}\left(\gamma\cdot\tilde{p}^{(q)}_{\lambda,\ell,\kappa}\right),
\end{eqnarray}
where $\Pi^{(q)}$ is defined in \eqref{N11}. The aim is to determine the Ritus function $\mathbb{E}_{\lambda,\ell,\kappa}^{(q)}$ and the Ritus momentum $\tilde{p}_{\lambda,\ell,\kappa}^{(q)}$ in terms of $\lambda$. The latter plays the role of Landau levels in a rotating system (see below). Using the Weyl basis \eqref{N12} for the $\gamma$-matrices, the operator $\gamma\cdot \Pi^{(q)}$ turns out to be
\begin{eqnarray}\label{N14}
\gamma\cdot \Pi^{(q)}=\left(
\begin{array}{cc}
0&\Pi^{(q)}_{R}\\
\Pi^{(q)}_{L}&0
\end{array}
\right),
\end{eqnarray}
with
\begin{widetext}
\begin{eqnarray}\label{N15}
\Pi^{(q)}_{R}&=&\left(
\begin{array}{ccc}
i\partial_t+\Omega\left(\hat{L}_{z}+1/2\right)+i\partial_z&&+i\left(\partial_{x}+iqeBy/2\right)+\left(\partial_{y}-iqeBx/2\right)\\
+i\left(\partial_{x}+iqeBy/2\right)-\left(\partial_{y}-iqeBx/2\right)&&i\partial_t+\Omega\left(\hat{L}_{z}-1/2\right)-i\partial_z
\end{array}
\right),\nonumber\\
\Pi^{(q)}_{L}&=&\left(
\begin{array}{ccc}
i\partial_t+\Omega\left(\hat{L}_{z}+1/2\right)-i\partial_z&~~&-i\left(\partial_{x}+iqeBy/2\right)-\left(\partial_{y}-iqeBx/2\right)\\
-i\left(\partial_{x}+iqeBy/2\right)+\left(\partial_{y}-iqeBx/2\right)&~~&i\partial_t+\Omega\left(\hat{L}_{z}-1/2\right)+i\partial_z
\end{array}
\right).
\end{eqnarray}
In a cylinder coordinate system $(r,\varphi,z)$ with $\left(x=r\cos\varphi, y=r\sin\varphi, z\right)$, \eqref{N15} is equivalently given by
\begin{eqnarray}\label{N16}
\Pi^{(q)}_{R}&=&\left(
\begin{array}{ccc}
i\partial_{t}+\Omega\left(-i\partial_{\varphi}+\frac{1}{2}\right)+i\partial_z&~~&+ie^{-i\varphi}\left(\partial_{r}-\frac{i}{r}\partial_{\varphi}-\frac{qeB}{2}r\right)\\
+ie^{+i\varphi}\left(\partial_{r}+\frac{i}{r}\partial_{\varphi}+\frac{qeB}{2}r\right)&~~&
i\partial_{t}+\Omega\left(-i\partial_{\varphi}-\frac{1}{2}\right)-i\partial_{z}
\end{array}
\right),\nonumber\\
\Pi^{(q)}_{L}&=&\left(
\begin{array}{ccc}
i\partial_{t}+\Omega\left(-i\partial_{\varphi}+\frac{1}{2}\right)-i\partial_z&~~&-ie^{-i\varphi}\left(\partial_{r}-\frac{i}{r}\partial_{\varphi}-\frac{qeB}{2}r\right)\\
-ie^{+i\varphi}\left(\partial_{r}+\frac{i}{r}\partial_{\varphi}+\frac{qeB}{2}r\right)&~~&
i\partial_{t}+\Omega\left(-i\partial_{\varphi}-\frac{1}{2}\right)+i\partial_{z}
\end{array}
\right).
\end{eqnarray}
\end{widetext}
To arrive at \eqref{N16}, we used
\begin{eqnarray}\label{N17}
i\partial_{x}\pm\partial_{y}=ie^{\mp i\varphi}\left(\partial_{r}\mp \frac{i}{r}\partial_{\varphi}\right),
\end{eqnarray}
and replaced $\hat{L}_{z}$ with $\hat{L}_{z}=-i\partial_{\varphi}$. Plugging $\psi_{+}^{(q)}=\mathbb{E}_{\lambda,\ell,+}^{(q)}u\left(\tilde{p}_{\ell,+}\right)$ and $\psi_{-}^{(q)}=\mathbb{E}_{\lambda,\ell,-}^{(q)}v\left(\tilde{p}_{\ell,-}\right)$ into \eqref{N10}, and using \eqref{N13}, we arrive at
\begin{eqnarray}\label{N18}
\left(\gamma\cdot \tilde{p}_{\lambda,\ell,+}^{(q)}-m_{q}\right)u\left(\tilde{p}_{\ell,+}\right)&=&0,\nonumber\\
\left(\gamma\cdot \tilde{p}_{\lambda,\ell,-}^{(q)}+m_{q}\right)v\left(\tilde{p}_{\ell,-}\right)&=&0.
\end{eqnarray}
The solutions are the standard Dirac spinors of free electrons with $p^{\mu}=(p_0,\boldsymbol{p})$ replaced with $\tilde{p}_{\lambda,\ell,\kappa}^{(q)}$, where $\kappa=+1$ ($\kappa=-1$) denotes the positive (negative) frequency solution of the Dirac equation. 
\par
In what follows, we first determine $\mathbb{E}_{\lambda,\ell, \kappa}^{(q)}$ and  $\tilde{p}_{\lambda,\ell,\kappa}^{(q)}$ in a system with no boundary condition. We then consider a certain global boundary condition, and determine $\mathbb{E}_{\lambda,\ell, \kappa}^{(q)}$ and  $\tilde{p}_{\lambda,\ell,\kappa}^{(q)}$. In both cases, we present the quantization relations for fermionic field operators $\bar{\psi}^{(q)}$ and $\psi^{(q)}$.
\subsection{Rotating magnetized fermions in a system with no boundary condition}\label{sec2a}
\subsubsection{Determination of $\mathbb{E}_{\lambda,\ell,\kappa}^{(q)}$}\label{sec2a1}
To determine $\mathbb{E}_{\lambda,\ell,\kappa}^{(q)}$ in this case, we use, similar to the nonrotating case \cite{tabatabaee2020}, the Ansatz
\begin{eqnarray}\label{N19}
\mathbb{E}_{\lambda,\ell,\kappa}^{(q)}=e^{-i\kappa\left(E_{\lambda,\ell,\kappa} t-p_{z}z\right)}\mathbb{P}_{\lambda,\ell}^{(q)},
\end{eqnarray}
with the projector defined by
\begin{eqnarray}\label{N20}
\mathbb{P}_{\lambda,\ell}^{(q)}\equiv P_{+}f_{\lambda,\ell,s_q}^{+}+P_{-}f_{\lambda,\ell,s_q}^{-},
\end{eqnarray}
$s_{q}\equiv \text{sgn}\left(qeB\right)$ and the spin projector
\begin{eqnarray}\label{N21}
P_{\pm}\equiv \frac{1\pm i\gamma^{1}\gamma^{2}}{2}.
\end{eqnarray}
In \eqref{N19}, $E_{\lambda,\ell,\kappa}$ and $p_{z}$ are the zeroth and fourth components of the Ritus momentum $\tilde{p}_{\lambda,\ell,\kappa}^{(q)}$. Because of the specific structure of the $\gamma$-matrices in the Weyl representation, $\mathbb{E}_{\lambda,\ell,\kappa}^{(q)}$ reduces to a block diagonal matrix in the form
\begin{eqnarray}\label{N22}
\mathbb{E}_{\lambda,\ell,\kappa}^{(q)}=\left(
\begin{array}{cc}
\mathscr{E}_{\lambda,\ell,\kappa}^{(q)}&0\\
0&\mathscr{E}_{\lambda,\ell,\kappa}^{(q)}
\end{array}
\right),
\end{eqnarray}
with
\begin{eqnarray}\label{N23}
\hspace{-1cm}\mathscr{E}_{\lambda,\ell,\kappa}^{(q)}=e^{-i\kappa\left(E_{\lambda,\ell,\kappa}^{(q)}t-p_z z\right)}\left(
\begin{array}{cc}
f_{\lambda,\ell,s_q}^{+}&0\\
0&f_{\lambda,\ell,s_q}^{-}
\end{array}
\right).
\end{eqnarray}
Plugging this Ansatz into
\begin{eqnarray}\label{N24}
\hat{J}_{z}\mathbb{E}_{\lambda,\ell,\kappa}^{(q)}=\left(\ell+\frac{1}{2}\right)\mathbb{E}_{\kappa,\lambda}^{(q)},
\end{eqnarray}
with $\hat{J}_{z}=\hat{L}_{z}+\Sigma_{z}/2$, we arrive at
\begin{eqnarray}\label{N25}
\hat{L}_{z}f_{\lambda,\ell,s_q}^{+}&=&\ell f_{\lambda,\ell,s_q}^{+},\nonumber\\
\hat{L}_{z}f_{\lambda,\ell,s_q}^{-}&=&\left(\ell+1\right) f_{\lambda,\ell,s_q}^{-}.
\end{eqnarray}
Plugging $\hat{L}_{z}=-i\partial_{\varphi}$ into \eqref{N25}, we arrive immediately at
\begin{eqnarray}\label{N26}
f_{\lambda,\ell,s_q}^{+}&=&e^{i\ell\varphi}\chi_{\lambda,\ell,s_q}^{+},\nonumber\\
f_{\lambda,\ell,s_q}^{-}&=&e^{i\left(\ell+1\right)\varphi}\chi_{\lambda,\ell,s_q}^{-},
\end{eqnarray}
with unknown functions $\chi_{\lambda,\ell,s_q}^{\pm}$. To determine these functions, we consider first the quadratic equation
\begin{eqnarray}\label{N27}
\left(\gamma\cdot \Pi^{(q)}\right)^2\mathbb{E}_{\lambda,\ell,\kappa}^{(q)}= \mathbb{E}_{\lambda,\ell,\kappa}^{(q)}\left(\gamma\cdot\tilde{p}^{(q)}_{\lambda,\ell,\kappa}\right)^2.
\end{eqnarray}
Plugging $\gamma\cdot \Pi^{(q)}$ from \eqref{N14} and $\mathbb{E}_{\lambda,\ell,\kappa}^{(q)}$ from \eqref{N22} into \eqref{N27}, and using $\tilde{p}_{\lambda,\ell,\kappa}^{(q)2}=m_{q}^{2}$ as well as $\Pi_{L}^{(q)}\Pi_{R}^{(q)}=\Pi_{R}^{(q)}\Pi_{L}^{(q)}$, we arrive at
\begin{eqnarray}\label{N28}
\Pi_{L}^{(q)}\Pi_{R}^{(q)}\mathscr{E}_{\lambda,\ell,\kappa}^{(q)}=m_{q}^{2}\mathscr{E}_{\lambda,\ell,\kappa}^{(q)},
\end{eqnarray}
with  $\Pi_{L}^{(q)}\Pi_{R}^{(q)}$ given by
\begin{eqnarray}\label{N29}
\Pi_{L}^{(q)}\Pi_{R}^{(q)}=\left(\begin{array}{cc}
\mathscr{O}_{+}&0\\
0&\mathscr{O}_{-}
\end{array}
\right).
\end{eqnarray}
Here,
\begin{eqnarray}\label{N30}
\mathscr{O}_{\pm}&\equiv&\left(i\partial_{t}-\Omega\left(i\partial_{\varphi}\mp 1/2\right)\right)^{2}+\partial_{r}^{2}+\frac{1}{r}\partial_{r}+\frac{\partial_{\varphi}^{2}}{r^{2}}\nonumber\\
&&-qeB\left(i\partial_{\varphi}\mp 1\right)-\left(\frac{qeB}{2}\right)^{2}r^{2}+\partial_{z}^{2}.
\end{eqnarray}
The differential equation for $\chi_{\lambda,\ell,s_q}^{\pm}$ arise by plugging $\mathscr{E}_{\lambda,\ell,\kappa}^{(q)}$ from \eqref{N23} with $f_{\lambda,\ell,s_q}^{\pm}$ from \eqref{N26} into \eqref{N28}. We thus arrive at
\begin{eqnarray}\label{N31}
\bigg[x\partial_{x}^{2}+\partial_{x}+\lambda-\frac{\ell^{2}}{4x}+\frac{s_{q}(\ell+1)}{2}-\frac{x}{4}\bigg]\chi_{\lambda,\ell,s_q}^{+}=0,\nonumber\\
\bigg[x\partial_{x}^{2}+\partial_{x}+\lambda-\frac{(\ell+1)^{2}}{4x}+\frac{s_{q}\ell}{2}-\frac{x}{4}\bigg]\chi_{\lambda,\ell,s_q}^{-}=0,\nonumber\\
\end{eqnarray}
where $x\equiv \frac{|qeB|r^{2}}{2}$ and
\begin{eqnarray}\label{N32}
\lambda\equiv \frac{\left(E_{\lambda,\ell,\kappa}^{(q)}+\kappa\Omega j\right)^{2}-p_z^2-m_q^2}{2|qeB|},
\end{eqnarray}
with $j\equiv \ell+1/2$. Hence, according to \eqref{N32}, the energy dispersion relation for a rotating and magnetized fermionic system reads
\begin{eqnarray}\label{N33}
\hspace{-1cm}E_{\lambda,\ell,\kappa}^{(q)}=-\kappa \Omega j\pm\sqrt{2\lambda|qeB|+p_{z}^{2}+m_{q}^{2}}.
\end{eqnarray}
To solve the differential equations \eqref{N31}, we use the Ansatz
\begin{eqnarray}\label{N34}
\chi_{\lambda,\ell,s_q}^{+}&=&e^{-x/2}x^{|\ell|/2}g_{\lambda,\ell,s_q}^{+},\nonumber\\
\chi_{\lambda,\ell,s_q}^{-}&=&e^{-x/2}x^{|\ell+1|/2}g_{\lambda,\ell,s_q}^{-},
\end{eqnarray}
which leads to
\begin{eqnarray}\label{N35}
\hspace{-1cm}&&\big[x\partial_{x}^{2}+\left(|\ell|+1-x\right)\partial_{x}+\mathscr{N}_{\lambda,s_q}^{+}\big]g_{\lambda,\ell,s_q}^{+}=0,\nonumber\\
\hspace{-1cm}&&\big[x\partial_{x}^{2}+\left(|\ell+1|+1-x\right)\partial_{x}+\mathscr{N}_{\lambda,s_q}^{-}\big]g_{\lambda,\ell,s_q}^{-}=0,
\end{eqnarray}
upon plugging \eqref{N34} into \eqref{N31}. Here,
\begin{eqnarray}\label{N36}
\mathscr{N}_{\lambda,s_q}^{+}&\equiv&\lambda+\frac{s_{q}\left(\ell+1\right)-|\ell|-1}{2},\nonumber\\
\mathscr{N}_{\lambda,s_q}^{-}&\equiv&\lambda+\frac{s_{q}\ell-|\ell+1|-1}{2}.
\end{eqnarray}
Comparing the differential equations \eqref{N35} with Kummer's differential equation
$$
\left(z\partial_{z}^{2}+\left(b-z\right)\partial_z-a\right)g(z)=0,
$$
whose solution
$$
g(z)=A~{}_{1}F_{1}\left(a;b;z\right)+B~ U\left(a;b;z\right),
$$
is a linear combination of a hypergeometric function of the first and second kind ${}_{1}F_{1}\left(a;b;x\right)$ and $U\left(a;b;x\right)$, and requiring that $g^{\pm}_{\lambda,\ell,s_q}$ are regular at $x\to 0$,\footnote{This is equivalent to $r\to 0$.} we arrive at
\begin{eqnarray}\label{N37}
g_{\lambda,\ell,s_q}^{+}&=&\mathscr{A}^{+}~{}_{1}F_{1}\left(-\mathscr{N}_{\lambda,s_q}^{+};|\ell|+1;x\right),\nonumber\\
g_{\lambda,\ell,s_q}^{-}&=&\mathscr{A}^{-}~{}_{1}F_{1}\left(-\mathscr{N}_{\lambda,s_q}^{-};|\ell+1|+1;x\right).
\end{eqnarray}
Here, $\mathscr{A}^{\pm}$ are appropriate normalization factors, which are determined by using the orthonormality relation
\begin{eqnarray}\label{N38}
\lefteqn{\hspace{-0.8cm}
\int d^{4}r~\bar{\mathbb{E}}_{\lambda^{\prime},\ell^{\prime},\kappa}^{(q)}(r)
\mathbb{E}_{\lambda,\ell,\kappa}^{(q)}(r)}\nonumber\\
&&\hspace{-0.5cm}=\delta\left(E_{\lambda^{\prime},\ell^{\prime},\kappa}^{(q)}-E_{\lambda,\ell,\kappa}^{(q)}\right)\delta\left(k_{z}-k_{z}^{\prime}\right)
\delta_{\lambda,\lambda^{\prime}}\delta_{\ell,\ell^{\prime}}.
\end{eqnarray}
In cylinder coordinate system, we have $d^{4}r=dt r dr d\varphi dz$. Moreover, $\bar{\mathbb{E}}_{\lambda^{\prime},\ell^{\prime},\kappa}^{(q)}(r)=\gamma^{0}\mathbb{E}_{\lambda,\ell,\kappa}^{(q)\dagger}\gamma^{0}$. Here, $\mathbb{E}_{\lambda,\ell,\kappa}^{(q)}$ is given in \eqref{N22} with $\mathscr{E}_{\lambda,\ell,\kappa}^{(q)}$
from \eqref{N23}, and $f_{\lambda,\ell,s_q}^{\pm}$ read
\begin{eqnarray}\label{N39}
f_{\lambda,\ell,s_q}^{+}&=&\mathscr{A}^{+}e^{i\ell\varphi}e^{-x/2}x^{|\ell|/2}~{}_{1}F_{1}\left(-\mathscr{N}_{\lambda,s_q}^{+};|\ell|+1;x\right),\nonumber\\
f_{\lambda,\ell,s_q}^{-}&=&\mathscr{A}^{-}e^{i
\left(\ell+1\right)\varphi}e^{-x/2}x^{|\ell+1|/2}\nonumber\\
&&\times{}_{1}F_{1}\left(-\mathscr{N}_{\lambda,s_q}^{-};|\ell+1|+1;x\right).
\end{eqnarray}
These solutions are general and valid for both cases of a rotating fermionic system without and with a boundary condition. Let us now assume, that the rotating magnetized fermions are in a system with no spatial boundary condition. As it turns out, in this case, the parameter $\lambda$ in \eqref{N32} is a positive integer, i.e. $\lambda\in\mathbb{N}_{0}$. Thus $\lambda$ plays the role of Landau levels similar to the case of nonrotating fermions in a magnetic field. On the other hand, if the first argument $-\mathscr{N}_{\lambda,s_q}^{\pm}$ in ${}_{1}F_{1}$ appearing in \eqref{N39} is a nonpositive integer, the hypergeometric function can be replaced by the associated Laguerre polynomials,
 \begin{eqnarray*}
{}_{1}F_{1}\left(-n;m+1;z\right)=\frac{m! n!}{(m+n)!}L_{n}^{m}(z).
\end{eqnarray*}
The solutions \eqref{N39} thus read
\begin{eqnarray}\label{N40}
f_{\lambda,\ell,s_q}^{+}(x)&=&\frac{\mathscr{A}^{+}\mathscr{N}_{\lambda,s_q}^{+}!|\ell|!}{\left(\mathscr{N}_{\lambda,s_q}^{+}+|\ell|\right)!}e^{i\ell\varphi}e^{-x/2}x^{|\ell|/2}L_{\mathscr{N}_{\lambda,s_q}^{+}}^{|\ell|}\left(x\right),\nonumber\\
f_{\lambda,\ell,s_q}^{-}(x)&=&\frac{\mathscr{A}^{-}\mathscr{N}_{\lambda,s_q}^{-}!|\ell+1|!}{\left(\mathscr{N}_{\lambda,s_q}^{-}+|\ell+1|\right)!}e^{i
\left(\ell+1\right)\varphi}e^{-x/2}x^{|\ell+1|/2}\nonumber\\
&&\times L_{\mathscr{N}_{\lambda,s_q}^{-}}^{|\ell+1|}\left(x\right).
\end{eqnarray}
Using then \eqref{N38} for $r\in[0,\infty[$ and the orthonormality relations of the Laguerre polynomial
\begin{eqnarray*}
\int_{0}^{\infty}dz z^{\alpha} e^{-z}L_{n}^{\alpha}(z)L_{m}^{\alpha}(z)=\frac{(n+\alpha)!}{n!}~\delta_{m,n},
\end{eqnarray*}
for $\mbox{Re}(\alpha)>-1$, $\mathscr{A}^{\pm}$ are determined. We finally arrive at
\begin{eqnarray}\label{N41}
f_{\lambda,\ell,s_q}^{+}&=&\left(\frac{|qeB|}{2\pi}\frac{\mathscr{N}_{\lambda,s_q}^{+}!}{\left(\mathscr{N}_{\lambda,s_q}^{+}+|\ell|\right)!}\right)^{1/2}e^{i\ell\varphi}e^{-x/2}x^{|\ell|/2}\nonumber\\
&&\times L_{\mathscr{N}_{\lambda,s_q}^{+}}^{|\ell|}(x),\nonumber\\
f_{\lambda,\ell,s_q}^{-}&=&
\left(\frac{|qeB|}{2\pi}\frac{\mathscr{N}_{\lambda,s_q}^{-}!}{\left(\mathscr{N}_{\lambda,s_q}^{-}+|\ell+1|\right)!}\right)^{1/2}
e^{i
\left(\ell+1\right)\varphi}e^{-x/2}\nonumber\\
&&\times x^{|\ell+1|/2}L_{\mathscr{N}_{\lambda,s_q}^{-}}^{|\ell+1|}\left(x\right).
\end{eqnarray}
Here,  $P_{\pm}^{\dagger}=\gamma_{0}P_{\pm}\gamma_{0}$, $P_{\pm}^{2}=P_{\pm}$, and $P_{\pm}P_{\mp}=0$ are also used. In Table \ref{table1}, $\mathscr{N}_{\lambda,s_q}^{\pm}$ for $s_q=+1$ and $s_q=-1$, corresponding to $q>0$ and $q<0$, are listed. Let us notice that the Laguerre polynomials $L_{\mathscr{N}_{\lambda,s_q}^{\pm}}^{\cdots}$ appearing in \eqref{N41} are defined only for $\mathscr{N}_{\lambda,s_q}^{\pm}\geq 0$. This constraints the choice for $\ell$ for positively and negatively charged particles with positive and negative spins, $s=+1$ and $s=-1$, respectively.\footnote{Let us remind that the positive $+$ and negative $-$ upper indices on $\mathscr{N}^{\pm}_{s_q}$ denote the up ($s=+1$) and down ($s=-1$) spin orientations.} In Table \ref{table2}, the allowed values of $\ell$ for different choices of $\lambda$ are demonstrated. According to this table, the lowest energy level (LEL) with $\lambda=0$ is only occupied either with positively charged particles with $s=+1$ and $\ell\geq 0$ or with negatively charged particles with $s=-1$ and $\ell\leq -1$. As concerns the higher energy levels with $\lambda \geq 1$, they can be occupied with positively and negatively charged particles with both spin orientations $s=+1$ (spin up) and $s=-1$ (spin down). For positively charged particles the allowed values for $\ell$ are $\ell=-\lambda,-\lambda+1,\cdots,-2,-1,0,1,2,\cdots$ and for negatively charged particles $\ell=\cdots,-2,-1,0,1,2,\cdots,\lambda-2,\lambda-1$.
\begin{center}
\begin{table}[ht]
    \centering
\caption{The values for $\mathscr{N}_{\lambda,s_q}^{\pm}$ appearing in the first argument of the hypergeometric functions ${}_{1}F_{1}(a;b;x)$ in \eqref{N39}. Assuming $eB>0$, $s_{q}=+1$ and $s_{q}=-1$ correspond to $q>0$  and $q<0$, respectively.}\label{table1}    \vspace{1ex}
    \vspace{1ex}
        \begin{tabular}{c|rclcrcl}
            \hline\hline
&$\ell$&$\leq$&-1&\qquad\qquad&$\ell$&$\geq$&$0$\\
\hline
  $q>0,s=+1$&$\mathscr{N}_{\lambda,+}^{+}$&$=$&$\lambda+\ell$&\qquad\qquad&$\mathscr{N}_{\lambda,+}^{+}$&=&$\lambda$ \\
  $q>0,s=-1$&$\mathscr{N}_{\lambda,+}^{-}$&$=$&$\lambda+\ell$&\qquad\qquad&$\mathscr{N}_{\lambda,+}^{-}$&=&$\lambda-1$ \\
  $q<0,s=+1$&$\mathscr{N}_{\lambda,-}^{+}$&$=$&$\lambda-1$&\qquad\qquad&$\mathscr{N}_{\lambda,-}^{+}$&=&$\lambda-\ell-1$ \\
  $q<0,s=-1$&$\mathscr{N}_{\lambda,-}^{-}$&$=$&$\lambda$&\qquad\qquad&$\mathscr{N}_{\lambda,-}^{-}$&=&$\lambda-\ell-1$ \\
            \hline\hline
        \end{tabular}
\end{table}
\end{center}
\begin{table*}[ht]
    \centering
    \caption{The allowed values of $\ell$ for the LEL, $\lambda=0$, and higher energy levels, $\lambda\geq 1$, and for different combinations of particles' charge $Q$ and spin $s$. In an infinitely extended rotating fermionic system in a constant magnetic field, we have $\lambda\in \mathbb{N}_{0}$.}
    \label{table2}
    \vspace{1ex}
        \begin{tabular}{c|rclcl}
            \hline\hline
            \multirow{2}{*}{$q>0,s=+1$}
            &$\lambda$&$=$&$0$&\qquad\qquad&$\ell=0,1,2,\cdots$\\
            &$\lambda$&$\geq$&$1$&\qquad\qquad&$\ell=-\lambda,-\lambda+1,\cdots,-2,-1,0,1,2,\cdots$ \\
            \hline
            \multirow{2}{*}{$q>0,s=-1$}
            &$\lambda$&$=$&$0$&\qquad\qquad&---\\
            &$\lambda$&$\geq$&$1$&\qquad\qquad&$\ell=-\lambda,-\lambda+1,\cdots,-2,-1,0,1,2,\cdots$ \\
            \hline\hline
            \multirow{2}{*}{$q<0,s=+1$}
            &$\lambda$&$=$&$0$&\qquad\qquad&---\\
            &$\lambda$&$\geq$&$1$&\qquad\qquad&$\ell=\cdots,-2,-1,0,1,2,\cdots,\lambda-2,\lambda-1$ \\
            \hline
            \multirow{2}{*}{$q<0,s=-1$}
            &$\lambda$&$=$&$0$&\qquad\qquad&$\ell=\cdots,-2,-1$\\
            &$\lambda$&$\geq$&$1$&\qquad\qquad&$\ell=\cdots,-2,-1,0,1,2,\cdots,\lambda-2,\lambda-1$ \\
            \hline\hline
        \end{tabular}
\end{table*}
Plugging $f_{\lambda,\ell,s_q}^{\pm}$ from \eqref{N40} into \eqref{N23} and the resulting expression into \eqref{N22}, 
the Ritus function $\mathbb{E}_{\lambda,\ell,\kappa}^{(q)}$ for an infinitely large fermionic system in a constant magnetic field is determined. 
Let us notice, however, that since the Ansatz \eqref{N20} for $\mathbb{P}_{\lambda,\ell}^{(q)}$ does not take the boundaries for $\ell$ demonstrated
 in Table \ref{table2} into account, it has to be accordingly modified. This is done in Sec. \ref{sec2a3}, where the final expression for the quantization of the 
 fermionic fields in a multiflavor system under rotation and constant magnetic field with no boundary condition is presented.
\subsubsection{Determination of $\tilde{p}_{\lambda,\ell,\kappa}^{(q)}$}\label{sec2a2}
To determine $\tilde{p}_{\lambda,\ell,\kappa}^{(q)}$, let us first consider the Ritus relation \eqref{N13} with $\gamma\cdot \Pi^{(q)}$ from \eqref{N14}, and $\Pi_{R/L}^{(q)}$ from \eqref{N16}. Plugging $\mathbb{E}_{\lambda,\ell,\kappa}^{(q)}$ from \eqref{N22} into the left hand side of \eqref{N13}, we obtain
\begin{eqnarray}\label{N42}
\hspace{-1cm}\left(\gamma\cdot \Pi^{(q)}\right)\mathbb{E}_{\lambda,\ell,\kappa}^{(q)}=
\left(\begin{array}{cc}
0&\Pi_{R}^{(q)}\mathscr{E}_{\lambda,\ell,\kappa}^{(q)}\\
\Pi_{L}^{(q)}\mathscr{E}_{\lambda,\ell,\kappa}^{(q)}&0
\end{array}
\right),
\end{eqnarray}
with $\mathscr{E}_{\lambda,\ell,\kappa}^{(q)}$ given in terms of $f_{\lambda,\ell,s_q}^{\pm}$ [see \eqref{N23}]. For the sake of generality, we use $f_{\lambda,\ell,s_q}^{\pm}$, from \eqref{N39} in terms of the hypergeometric function ${}_{1}F_{1}(a;b;z)$. To determine the resulting differential equations $-ie^{\pm i\varphi}\left(\partial_{r}\pm \frac{i}{r}\partial_{\varphi}\pm\frac{qeB}{2}r\right)$ which appear on the right hand side (r.h.s.) of \eqref{N42}, we use following relations
\begin{eqnarray}\label{N43}
&&
b{}_{1}F_{1}\left(a;b;z\right)-b{}_{1}F_{1}\left(a-1;b;z\right)-z{}_{1}F_{1}\left(a;b+1;z\right)=0,\nonumber\\
&&
{}_{1}F_{1}\left(a;b;z\right)-\frac{(b+z)}{b}{}_{1}F_{1}\left(a;b+1;z\right)-\frac{\left(a-b-1\right)}{b(b+1)}\nonumber\\
&&\qquad \times{}_{1}F_{1}\left(a;b+2;z\right)=0, \nonumber\\
&&\left(a-b+1\right){}_{1}F_{1}\left(a;b;z\right)-a~{}_{1}F_{1}\left(a+1;b;z\right)\nonumber\\
&&\qquad-\left(1-b\right){}_{1}F_{1}\left(a;b-1;z\right)=0,
\end{eqnarray}
and arrive after some work at
\begin{eqnarray}\label{N44}
\lefteqn{\hspace{-2cm}-ie^{\pm i\varphi}\left(\partial_{r}\pm \frac{i}{r}\partial_{\varphi}\pm\frac{qeB}{2}r\right)f_{\lambda,\ell,s_q}^{\pm}
}\nonumber\\
&&=\pm is_{\ell}\sqrt{2\lambda|qeB|}f_{\lambda,\ell,s_q}^{\mp},
\end{eqnarray}
with $s_{\ell}\equiv \mbox{sgn}(\ell)=1$ for $\ell\geq 0$ and $=-1$ for $\ell\leq -1$. Plugging these results into the r.h.s. of \eqref{N42}, we obtain
\begin{eqnarray}\label{N45}
\Pi_{R}^{(q)}\mathscr{E}_{\lambda,\ell,\kappa}^{(q)}=\kappa\mathscr{E}_{\lambda,\ell,\kappa}^{(q)}\Xi_{R}^{(q)},\qquad
\Pi_{L}^{(q)}\mathscr{E}_{\lambda,\ell,\kappa}^{(q)}=\kappa\mathscr{E}_{\lambda,\ell,\kappa}^{(q)}\Xi_{L}^{(q)},\nonumber\\
\end{eqnarray}
where
\begin{eqnarray}\label{N46}
\Xi_{R}^{(q)}&\equiv& \left(
\begin{array}{cc}
\epsilon_{\lambda}^{(q)}-p_{z}&+i\kappa s_\ell\sqrt{2\lambda|qeB|}\\
+i\kappa s_\ell\sqrt{2\lambda|qeB|}&\epsilon_{\lambda}^{(q)}+p_{z}
\end{array}
\right),\nonumber\\
\Xi_{L}^{(q)}&\equiv&\left(
\begin{array}{cc}
\epsilon_{\lambda}^{(q)}+p_{z}&-i\kappa s_\ell\sqrt{2\lambda|qeB|}\\
+i\kappa s_\ell\sqrt{2\lambda|qeB|}&\epsilon_{\lambda}^{(q)}-p_{z}
\end{array}
\right),\nonumber\\
\end{eqnarray}
with $\epsilon_{\lambda}^{(q)}\equiv E_{\lambda,\ell,\kappa}^{(q)}+\kappa\Omega j$.
These lead eventually to
\begin{eqnarray}\label{N47}
\left(\gamma\cdot \Pi^{(q)}\right)\mathbb{E}_{\lambda,\ell,\kappa}^{(q)}=\kappa\mathbb{E}_{\lambda,\ell,\kappa}^{(q)}\left(
\begin{array}{cc}
0&\Xi_{R}^{(q)}\\
\Xi_{L}^{(q)}&0
\end{array}
\right),
\end{eqnarray}
with $\Xi_{R/L}^{(q)}$ from \eqref{N46}. Comparing, at this stage, \eqref{N47} with the r.h.s. of \eqref{N13}, we arrive immediately at the Ritus momentum in a rotating fermionic system
\begin{eqnarray}\label{N48}
\tilde{p}_{\lambda,\ell,\kappa}^{(q)\mu}=\left(\epsilon_{\lambda}^{(q)}, 0, \kappa s_\ell\sqrt{2\lambda|qeB|},p_{z}\right),
\end{eqnarray}
where $\epsilon_{\lambda}^{(q)}=E_{\lambda,\ell,\kappa}^{(q)}+\kappa\Omega j$ with $j\equiv \ell+\frac{1}{2}$. Let us notice that since $\tilde{p}_{\lambda,\ell,\kappa}^{(q)2}=m_{q}^{2}$,  \eqref{N48} leads to
\begin{eqnarray}\label{N49}
\epsilon_{\lambda}^{(q)}=\pm\left(m_{q}^{2}+2\lambda|qeB|+p_{z}^{2}\right)^{1/2}.
\end{eqnarray}
Thus, $\epsilon_{\lambda}^{(q)}$ depends only on $\lambda$,\footnote{Here, $\kappa^{2}=1$ and $\mbox{sgn}^{2}(\ell)=1$ are used.} which, for a rotating fermionic system with no boundary condition, is a positive integer, i.e. $\lambda\in \mathbb{N}_{0}$. Hence, the expression under the squared root in \eqref{N49} turns out to be always positive. Let us also notice that according to our construction, $E_{\lambda,\ell,\kappa}^{(q)}$ is always positive, while $\epsilon_{\lambda}$ is allowed to be positive and negative.
\subsubsection{Quantization of fermionic fields in an infinitely extended rotating system}\label{sec2a3}
Combining the above results, the quantization relations for a magnetized fermion in a rotating system without boundary read
\begin{eqnarray}\label{N50}
\psi_{\alpha}^{(q)}(x)&=&\sum_{\ell,\lambda,s}\int \frac{dp_z}{2\pi}\frac{1}{\sqrt{2|\epsilon_{\lambda}^{(q)}}|}\left\{
e^{-i\left(E_{\lambda,\ell,+}^{(q)}t-p_{z}z\right)}a_{p_{z}}^{\lambda,\ell,s}\right.\nonumber\\
&&\left.\times \big[\widetilde{\mathbb{P}}_{\lambda,\ell}^{(q)}\left(x\right)\big]_{\alpha\rho}u_{s,\rho}\left(\tilde{p}_{\ell,+}\right)\Theta\left(E_{\lambda,\ell,+}^{(q)}\right)\right.\nonumber\\
&&\left.+
e^{+i\left(E_{\lambda,\ell,-}^{(q)}t-p_{z}z\right)}b_{p_{z}}^{\lambda,\ell,s\dagger}\big[\widetilde{\mathbb{P}}_{\lambda,\ell}^{(q)}\left(x\right)\big]_{\alpha\rho}^{\dagger}\right.\nonumber\\
&&\left.\times v_{s,\rho}\left(\tilde{p}_{\ell,-}\right)\Theta\left(E_{\lambda,\ell,-}^{(q)}\right)\right\}. \nonumber\\
\bar{\psi}_{\alpha}^{(q)}(x)&=&\sum_{\ell,\lambda,s}\int \frac{dp_z}{2\pi}\frac{1}{\sqrt{2|\epsilon_{\lambda}^{(q)}}|}\left\{
e^{+i\left(E_{\lambda,\ell,+}^{(q)}t-p_{z}z\right)}a_{p_{z}}^{\lambda,\ell,s\dagger}\right.\nonumber\\
&&\left.\times \bar{u}_{s,\rho}\left(\tilde{p}_{\ell,+}\right) \big[\widetilde{\mathbb{P}}_{\lambda,\ell}^{(q)}\left(x\right)\big]_{\rho\alpha}^{\dagger}\Theta\left(E_{\lambda,\ell,+}^{(q)}\right)\right.\nonumber\\
&&\left.+
e^{-i\left(E_{\lambda,\ell,-}^{(q)}t-p_{z}z\right)}b_{p_{z}}^{\lambda,\ell,s}
\bar{v}_{s,\rho}\left(\tilde{p}_{\ell,-}\right)\right. \nonumber\\
&&\left.\times
\big[\widetilde{\mathbb{P}}_{\lambda,\ell}^{(q)}\left(x\right)\big]_{\rho\alpha}
\Theta\left(E_{\lambda,\ell,-}^{(q)}\right)\right\}.
\end{eqnarray}
Here, $a_{p_{z}}^{\lambda,\ell,s\dagger}$ and $a_{p_{z}}^{\lambda,\ell,s}$, as well as
$b_{p_{z}}^{\lambda,\ell,s\dagger}$ and $b_{p_{z}}^{\lambda,\ell,s}$ are the
creation and annihilation operators of particles and antiparticles. They satisfy the commutation relations
\begin{eqnarray}\label{N51}
\hspace{-0.9cm}
\{a_{p_{z}}^{\lambda,\ell,s},a_{p_{z}^{\prime}}^{\lambda^{\prime},\ell^{\prime},s^{\prime}\dagger}\}&=&2\pi\delta\left({p_{z}}-p_{z}^{\prime}\right)\delta_{\lambda,\lambda^{\prime}}\delta_{\ell,\ell^{\prime}}\delta_{s,s^{\prime}},\nonumber\\
\hspace{-0.9cm}
\{b_{p_{z}}^{\lambda,\ell,s},b_{p_{z}^{\prime}}^{\lambda^{\prime},\ell^{\prime},s^{\prime}\dagger}\}&=&2\pi\delta\left({p_{z}}-p_{z}^{\prime}\right)\delta_{\lambda,\lambda^{\prime}}\delta_{\ell,\ell^{\prime}}\delta_{s,s^{\prime}}.
\end{eqnarray}
In \eqref{N50}, $\widetilde{\mathbb{P}}_{\lambda,\ell}^{(q)}$, the modified version of $\mathbb{P}_{\lambda,\ell}^{(q)}$ from \eqref{N20}, reads
\begin{eqnarray}\label{N52}
\hspace{-0.5cm}\widetilde{\mathbb{P}}_{\lambda,\ell}^{(q)}=\left(\mathscr{P}_{+}^{(q)}f_{\lambda,\ell,s_q}^{+s_q}+\Pi_{\lambda}\mathscr{P}_{-}^{(q)}f_{\lambda,\ell,s_q}^{-s_{q}}\right)\Gamma_{\lambda,\ell,q},
\end{eqnarray}
with
\begin{eqnarray}\label{N53}
\mathscr{P}^{(q)}_{\pm}\equiv \frac{1\pm is_{q}\gamma_1\gamma_{2}}{2},
\end{eqnarray}
which leads to $\mathscr{P}_{\pm}^{(+)}=\mathscr{P}_{\pm}$ and $\mathscr{P}_{\pm}^{(-)}=\mathscr{P}_{\mp}$, and
\begin{eqnarray}\label{N54}
\Pi_{\lambda}&\equiv& 1-\delta_{\lambda,0},\nonumber\\
\Gamma_{\lambda,\ell,q}&\equiv&\Theta\left(q\right)\Theta\left(\ell+\lambda\right)+\Theta\left(-q\right)\Theta\left(-\ell+\lambda-1\right). \nonumber\\
\end{eqnarray}
Here, $\Pi_{\lambda}$ considers the degeneracy of Landau levels. The fact that positively charged particles with negative spins, and negatively charged particles with positive spins do not occupy the LEL (see Table \ref{table2}) is considered by $\Gamma_{\lambda,\ell,q}$.
The above definitions are written in the same language as previously presented in \cite{tabatabaee2020} for the solutions of the Dirac equation of nonrotating fermions in a constant magnetic field. 
\par
In \eqref{N52}, $f_{\lambda,\ell,s_q}^{\pm}$ are given in \eqref{N41}. Moreover, in \eqref{N50} and \eqref{N54}, the Heaviside function $\Theta(z)=+1$  and $=0$ for $z\geq 0$ and  $z<0$, respectively.
As aforementioned, the modification of $\mathbb{P}_{\lambda,\ell}^{(q)}$ according to Table \ref{table2} is necessary because in this way, the allowed values of $\ell$ for positively and negatively charged fermions with up or down spins are considered directly in the solutions of the Dirac equation as well as the quantization of the Dirac fields for a  system without boundary conditions. In what follows, we introduce the global boundary condition for the fermionic field to avoid the system having a velocity that exceeds the speed of light \cite{fukushima2017}.
\subsection{Rotating magnetized fermions in a system with a global boundary condition}\label{sec2b}
\subsubsection{Imposing a global boundary condition}\label{sec2b1}
\begin{figure*}
\includegraphics[width=8cm,height=6cm]{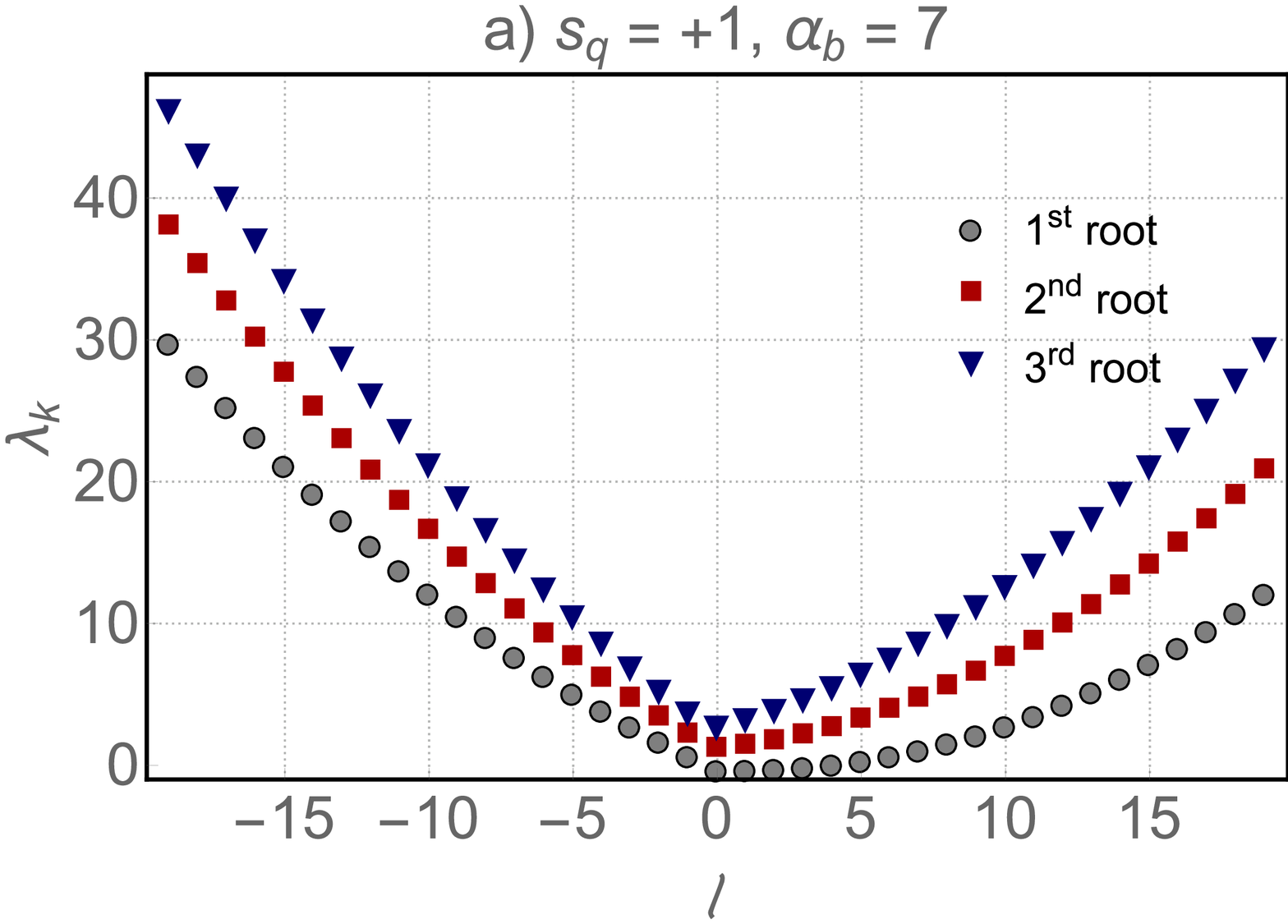}
\includegraphics[width=8cm,height=6cm]{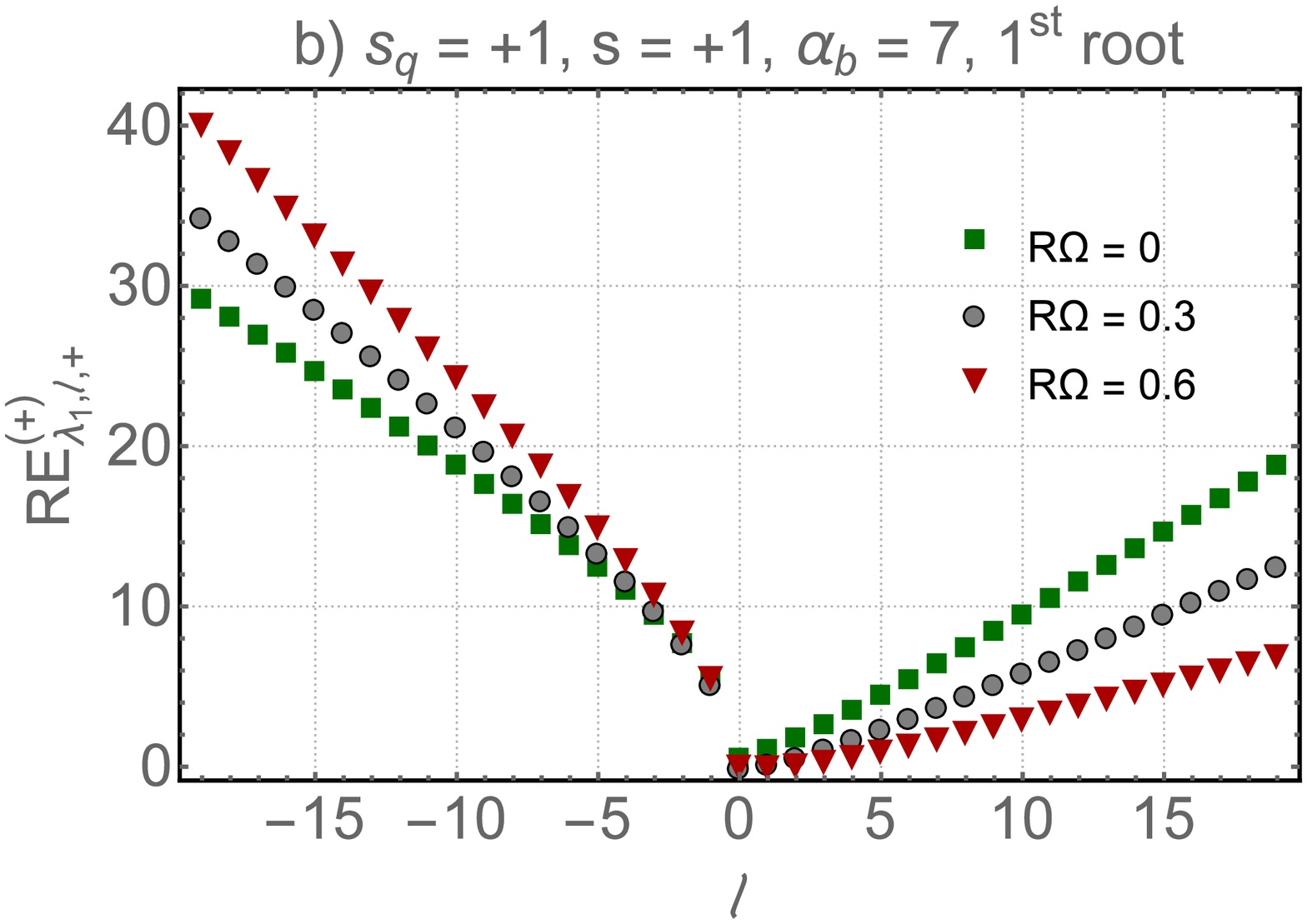}
\caption{color online. (a) The $\ell$ dependence of the first, second, and third roots $\lambda_k, k=1,2,3$ of the hypergeometric functions appearing in \eqref{N61} ($\ell\geq 0$) and \eqref{N62} ($\ell\leq -1$). (b) The $\ell$ dependence of the energies $RE_{\lambda_1,\ell,+}^{(+)}$ of the first root $\lambda_1$ for different $R\Omega=0, 0.3,0.6$. The energy $E_{\lambda_1,\ell,+}^{(+)}$ corresponds to a positively charged particle ($s_q=+1, \kappa=+1$).}\label{fig1}
\end{figure*}
\begin{figure*}
	\includegraphics[width=8cm,height=6cm]{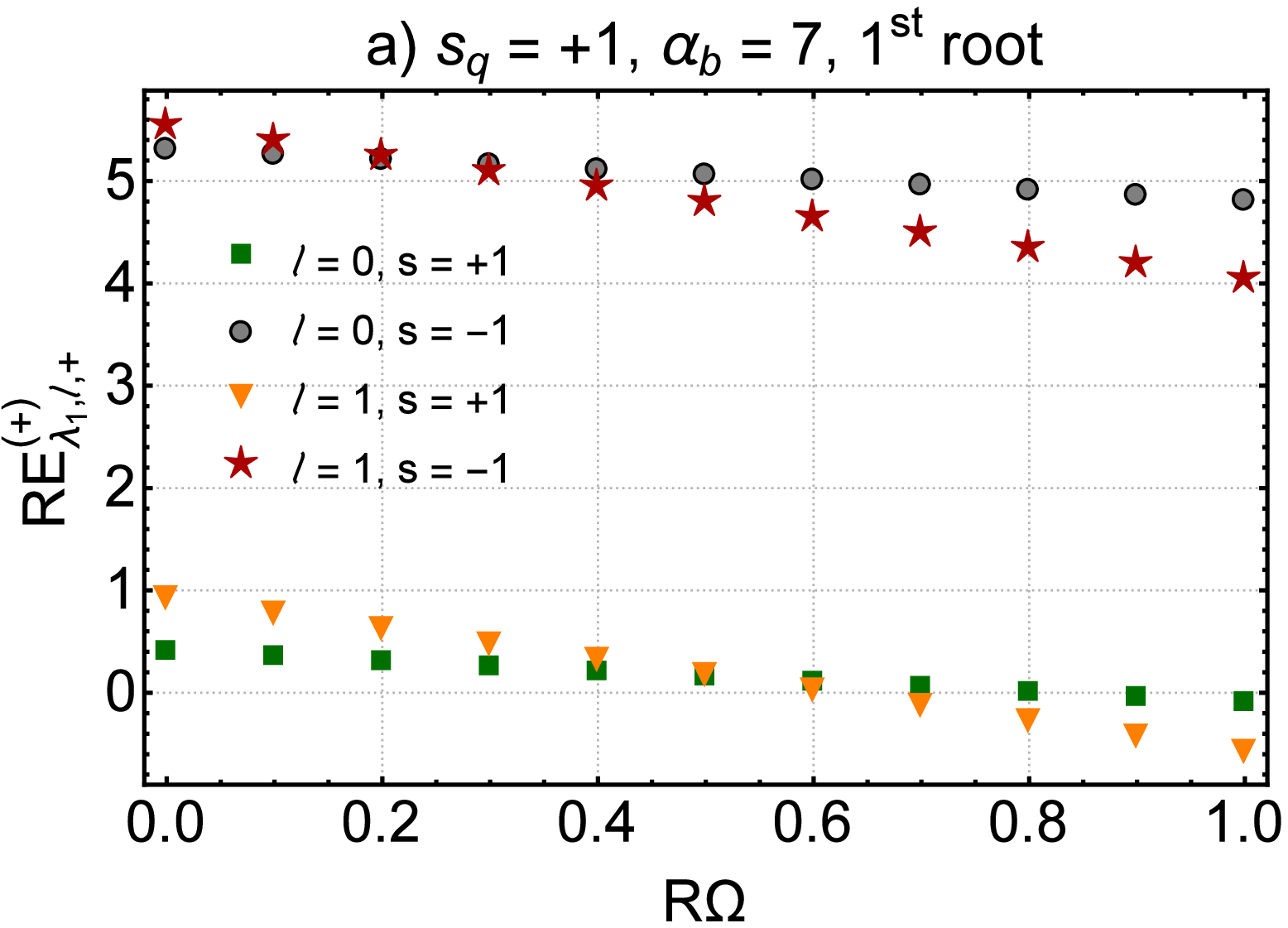}
	\includegraphics[width=8cm,height=6cm]{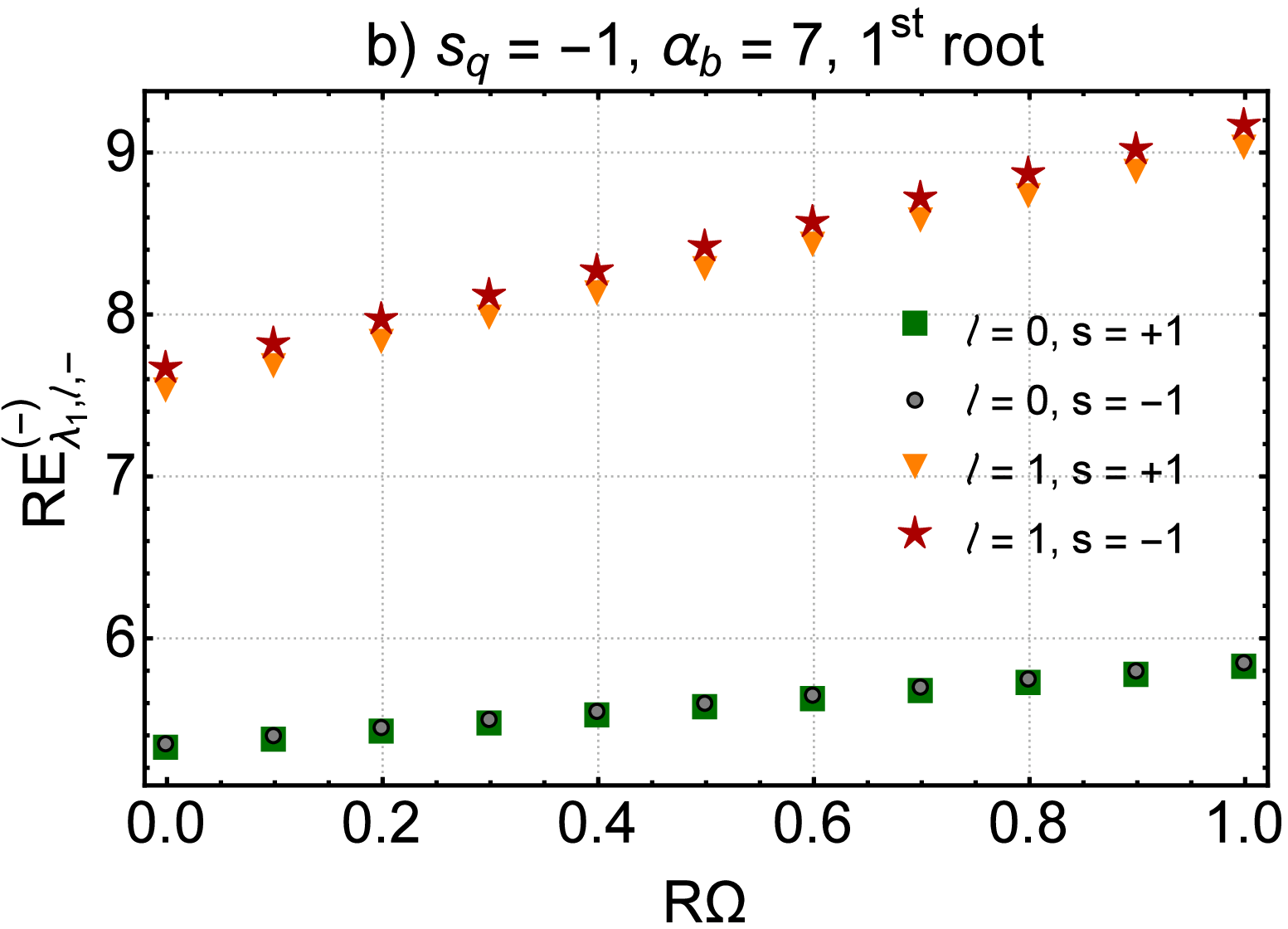}
	\caption{color online. The $R\Omega$ dependence of $E_{\lambda_1,\ell,+}^{(+)}$ (panel a) and $E_{\lambda_1,\ell,-}^{(-)}$ (panel b). }\label{fig2}
\end{figure*}
The solutions of the Dirac equation for magnetized fermions in a nonrotating system with the global boundary condition are already presented in \cite{fukushima2017}. In what follows, we use the solutions from Sec. \ref{sec2a} for a rotating quark matter with no boundary, and impose the same global boundary condition at $r=R$ as in \cite{fukushima2017},
\begin{eqnarray}\label{N55}
I\equiv \int_{-\infty}^{+\infty}dz\int_{0}^{2\pi} d\varphi \bar{\psi}^{(q)}\gamma^{r}\psi^{(q)}\bigg|_{r=R}=0,
\end{eqnarray}
with $R$  the cylinder radius  and $\gamma^{r}=\gamma^{1}\cos\varphi+\gamma^{2}\sin\varphi$. In contrast to \cite{fukushima2017}, the solutions for $\psi^{(q)}$ and $\bar{\psi}^{(q)}$ are derived using the Ritus eigenfunction method (see Sec. \ref{sec2a}).
Plugging  $\psi^{(q)}=\mathbb{E}_{\lambda,\ell,+}^{(q)}u(\tilde{p}_{\ell,+})$ and
$\bar{\psi}^{(q)}=\bar{u}(\tilde{p}_{\ell,+})\mathbb{E}_{\lambda,\ell,+}^{(q)}$
for positive frequency solution ($\kappa=+1$) and $\psi^{(q)}=\mathbb{E}_{\lambda,\ell,-}^{(q)}v(\tilde{p}_{\ell,-})$ as well as  $\bar{\psi}^{(q)}=\bar{v}(\tilde{p}_{\ell,-})\mathbb{E}_{\lambda,\ell,-}^{(q)}$ for negative frequency solution ($\kappa=-1$), we arrive first at
\begin{eqnarray}\label{N56}
\text{I}_{+}
&\equiv&\int_{-\infty}^{+\infty}dz\int_{0}^{2\pi} d\varphi \bar{u}\left(\tilde{p}^{\prime}_{\ell^{\prime},+}\right)\nonumber\\
&&\times\mathbb{E}_{\lambda^{\prime},\ell^{\prime},+}^{(q)}\gamma^{r}\mathbb{E}_{\lambda,\ell,+}^{(q)}u\left(\tilde{p}_{\ell,+}\right)\bigg|_{r=R}=0, \nonumber\\
\text{I}_{-}&\equiv&\int_{-\infty}^{+\infty}dz\int_{0}^{2\pi} d\varphi \bar{v}\left(\tilde{p}^{\prime}_{\ell^{\prime},-}\right)\nonumber\\
&&\times\mathbb{E}_{\lambda^{\prime},\ell^{\prime},-}^{(q)}\gamma^{r}\mathbb{E}_{\lambda,\ell,-}^{(q)}v\left(\tilde{p}_{\ell,-}\right)\bigg|_{r=R}=0.
\end{eqnarray}
Using then $\mathbb{E}_{\lambda,\ell,\kappa}^{(q)}$ from \eqref{N19} with $\mathbb{P}_{\lambda,\ell}^{(q)}$ from \eqref{N20}, and $f_{\lambda,\ell,s_{q}}^{\pm}$ from \eqref{N40} as well as $\gamma^{r}P_{\pm}=P_{\mp}\gamma^{r}$, $I_{\pm}$ become proportional to
\begin{eqnarray}\label{N57}
0=\text{I}_{+}&\propto& \bar{u}\left(\tilde{p}_{\lambda,\ell,+}^{(q)}\right)\mathscr{H}_{\lambda^{\prime},\lambda,\ell}u\left(\tilde{p}_{\lambda^{\prime},\ell,+}^{(q)}\right),\nonumber\\
0=\text{I}_{-}&\propto& \bar{v}\left(\tilde{p}_{\lambda,\ell,+}^{(q)}\right)\mathscr{H}_{\lambda^{\prime},\lambda,\ell}v\left(\tilde{p}_{\lambda^{\prime},\ell,+}^{(q)}\right),
\end{eqnarray}
where $\alpha_b\equiv x(r={R})=|qeB|R^{2}/2$, and
\begin{eqnarray}\label{N58}
\mathscr{H}_{\lambda^{\prime},\lambda,\ell}\equiv
\left(
\begin{array}{cccc}
0&0&0&+h_{\lambda^{\prime},\lambda,\ell}^{(1)}\\
0&0&+h_{\lambda^{\prime},\lambda,\ell}^{(2)}&0\\
0&-h_{\lambda^{\prime},\lambda,\ell}^{(1)}&0&0\\
-h_{\lambda^{\prime},\lambda,\ell}^{(2)}&0&0&0
\end{array}
\right),\nonumber\\
\end{eqnarray}
with
\begin{eqnarray}\label{N59}
h_{\lambda^{\prime},\lambda,\ell}^{(1)}&\equiv& {}_{1}F_{1}\left(-\mathscr{N}_{\lambda^{\prime},s_q}^{+};|\ell|+1;\alpha_{b}\right)\nonumber\\
&&\times{}_{1}F_{1}\left(-\mathscr{N}_{\lambda,s_q}^{-};|\ell+1|+1;\alpha_{b}\right),
\end{eqnarray}
and $h_{\lambda^{\prime},\lambda,\ell}^{(2)}\equiv h_{\lambda,\lambda^{\prime},\ell}^{(1)}$. In order to fulfill the boundary condition $I=0$ with $I$ from \eqref{N55}, we have to find the solution of
\begin{eqnarray}\label{N60}
{}_{1}F_{1}\left(-\mathscr{N}_{\lambda,s_q}^{+};|\ell|+1;\alpha_{b}\right)&=&0,\nonumber\\
{}_{1}F_{1}\left(-\mathscr{N}_{\lambda,s_q}^{-};|\ell+1|+1;\alpha_{b}\right)&=&0.
\end{eqnarray}
But, before doing this let us notice that the hypergeometric function ${}_{1}F_{1}(a;b;z)$, being defined as
\begin{eqnarray*}
{}_{1}F_{1}(a;b;z)=\sum_{k=0}^{\infty}\frac{\left(a\right)_{k}}{\left(b\right)_{k}}\frac{z^{k}}{k!},
\end{eqnarray*}
yields only a polynomial with a finite number of terms, when $a<0$, and either $b>0$ or $b<a$ \cite{mathworld}. Here, $(x)_{n}\equiv\Gamma(x+n)/\Gamma(x)$ is the Pochhammer symbol. In Table \ref{table3}, we summarize the intervals for $-\mathscr{N}_{\lambda,s_q}^{+}<0$ and $|\ell|+1>0$ or $|\ell|+1<-\mathscr{N}_{\lambda,s_q}^{+}$  corresponding to
${}_{1}F_{1}\left(-\mathscr{N}_{\lambda,s_q}^{+};|\ell|+1;\alpha_{b}\right)$ as well as
$-\mathscr{N}_{\lambda,s_q}^{-}<0$ and $|\ell+1|+1>0$ or $|\ell+1|+1<-\mathscr{N}_{\lambda,s_q}^{-}$ corresponding to
${}_{1}F_{1}\left(-\mathscr{N}_{\lambda,s_q}^{-};|\ell+1|+1;\alpha_{b}\right)$.
\begin{table*}[ht]
    \centering
    \caption{ The allowed intervals of $\ell$ for which the hypergeometric functions ${}_{1}F_{1}\left(-\mathscr{N}_{\lambda,s_q}^{+};|\ell|+1;\alpha_{b}\right)$ and ${}_{1}F_{1}\left(-\mathscr{N}_{\lambda,s_q}^{-};|\ell+1|+1\right)$ yield polynomials with a finite number of terms. Here, $\lambda>0$ is assumed.}
    \label{table3}
    \vspace{1ex}
        \begin{tabular}{c|rclcl}
            \hline\hline
            \multirow{2}{*}{${}_{1}F_{1}\left(-\mathscr{N}_{\lambda,+}^{+};|\ell|+1;\alpha_{b}\right)$}
            &$\lambda$&$=$&$0$&\qquad\qquad&$\ell=0,1,2,\cdots$\\
            &$\lambda$&$\geq$&$1$&\qquad\qquad&$\ell=-\lambda,-\lambda+1,\cdots,-2,-1,0,1,2,\cdots$ \\
            \hline
            \multirow{2}{*}{${}_{1}F_{1}\left(-\mathscr{N}_{\lambda,+}^{-};|\ell+1|+1;\alpha_{b}\right)$}
            &$\lambda$&$=$&$0$&\qquad\qquad&---\\
            &$\lambda$&$\geq$&$1$&\qquad\qquad&$\ell=-\lambda,-\lambda+1,\cdots,-2,-1,0,1,2,\cdots$ \\
            \hline\hline
            \multirow{2}{*}{${}_{1}F_{1}\left(-\mathscr{N}_{\lambda,-}^{+};|\ell|+1;\alpha_{b}\right)$}
            &$\lambda$&$=$&$0$&\qquad\qquad&---\\
            &$\lambda$&$\geq$&$1$&\qquad\qquad&$\ell=\cdots,-2,-1,0,1,2,\cdots,\lambda-2,\lambda-1$ \\
            \hline
            \multirow{2}{*}{${}_{1}F_{1}\left(-\mathscr{N}_{\lambda,-}^{-};|\ell+1|+1;\alpha_{b}\right)$}
            &$\lambda$&$=$&$0$&\qquad\qquad&$\ell=\cdots,-2,-1$\\
            &$\lambda$&$\geq$&$1$&\qquad\qquad&$\ell=\cdots,-2,-1,0,1,2,\cdots,\lambda-2,\lambda-1$ \\
            \hline\hline
        \end{tabular}
\end{table*}
\par
As concerns the solutions of the hypergeometric functions appearing in \eqref{N60}, we first consider $s_q=+1$, and choose $\alpha_{b}=7$,\footnote{The numerical results presented in Sec. \ref{sec3} correspond to $q>0$ leading to $s_q=1$.}. Then, setting
\begin{eqnarray}\label{N61}
\begin{array}{rclcccc}
{}_{1}F_{1}\left(-\mathscr{N}_{\lambda,+}^{+};|\ell|+1;\alpha_{b}\right)&=&0,&~~&\mbox{for}&~~&\ell\geq 0,
\end{array}\nonumber\\
\end{eqnarray}
and
\begin{eqnarray}\label{N62}
\begin{array}{rclcccc}
{}_{1}F_{1}\left(-\mathscr{N}_{\lambda,+}^{-};|\ell+1|+1;\alpha_{b}\right)&=&0,&~~&\mbox{for}&~~&\ell\leq -1,
\end{array}\nonumber\\
\end{eqnarray}
we determine numerically the roots of these two functions.
In Fig. \ref{fig1}(a), the results of the first, second, and third roots of the hypergeometric functions in \eqref{N61} and \eqref{N62} are plotted. The roots are not symmetrically distributed around $j=0$, i.e., there is a certain asymmetry with respect to $j\to -j$ or equivalently $\ell \to -\ell-1$, which is also observed in \cite{fukushima2017}. Here, it is argued that this is because of broken $\mathcal{C}$ and $\mathcal{CP}$ symmetry in the case of nonvanishing magnetic field. Let us denote these roots with $\lambda_{k}, k=1,2,\cdots$. In contrast to the previous case of no boundary condition, $\lambda_{k}\in \mathbb{R}$ (i.e. they are not necessarily integers). Plugging $\lambda_{k}$ into the energy dispersion equation [see also \eqref{N40}],
\begin{eqnarray}\label{N63}
\hspace{-0.4cm}E_{\lambda_{k},\ell,\kappa}^{(q)}=-\kappa \Omega j+\sqrt{\frac{4\lambda_{k}\alpha_b}{R^2}+p_{z}^{2}+m_{q}^{2}},
\end{eqnarray}
where $j=\ell+1/2$ and $\alpha_b=|qeB|R^2/2$, yields the value of each energy level. At this stage, it is important to check whether $E_{\lambda_{k},\ell,\kappa}^{(q)}$ for given values of $\kappa, \Omega, m_q,p_z,R,\alpha_b$ and for $\ell\in(-\infty,+\infty)$ remains positive. To do this, let us consider $\kappa=+1, s_q=+1 (\text{or equivalently}~q>0), m_{q}=0, p_{z}=0$, and $\alpha_{b}=7$. Plugging $\lambda_{1}$ ($k=1$) from Fig. \ref{fig1}(a) into \eqref{N63}, and assuming that the transverse size of the QGP in the early stage of the collision to be $R=6$ fm \cite{becattini2016}, we arrive at $E_{\lambda_{1},\ell,+}^{(+)}$ as a function of $\ell$ and $\Omega$. In Fig. \ref{fig1}(b), $RE_{\lambda_{1},\ell,+}^{(+)}$ is plotted for $R\Omega=0, 0.3, 0.6$ and $-19\leq\ell\leq19$. Since according to \cite{becattini2016}, $\Omega=0.02\text{~fm}^{-1}\sim 10^{22}\text{~s}^{-1}$, $R\Omega=0.3$ and $R\Omega=0.6$ with $R=6$ fm correspond to $2.5\times 10^{22}\text{~s}^{-1}$ and $5\times 10^{22}\text{~s}^{-1}$, respectively. The energy  $E_{\lambda_{1},\ell,+}^{(+)} $ turns out to be positive in the whole interval of $\ell$.
\par
In Fig. \ref{fig2}, we have shown that the LEL is affected by the choice of the angular frequency $\Omega$. In Fig. \ref{fig2}(a), the $R\Omega$ dependence of $RE_{\lambda_1,\ell,+}^{(+)}$ is demonstrated for $\ell=0$ and $\ell=1$. The data corresponding to up (down) spin $s=+1$ ($s=-1$) arise by plugging $m_{q}=p_z=0, R=6~\text{fm},\kappa=+$ ($\kappa=-$) into $\eqref{N63}$ with $\lambda_1$ determined from \eqref{N61} for $s=+1$ as well as \eqref{N62} for $s=-1$ with $s_q=+1$ (or equivalently a positively charged particle) and $\ell=0$ and $\ell=1$.\footnote{Let us remind that the superscripts $\pm$ in $\mathscr{N}^{+}_{\lambda,s_q}$ and $\mathscr{N}^{-}_{\lambda,s_q}$ appearing in \eqref{N61} and \eqref{N62} correspond to fermions with spin up ($+$) and down ($-$).} In Fig. \ref{fig2}(b), the same is done for $\kappa=-1$ and $s_q=-1$~(or equivalently a negatively charged antiparticle).  As it turns out, for positively charged particles in the regime $R\Omega<0.6$, the energy level corresponding to $\ell=0$ and $s=+1$ is lower than $\ell=0, s=+1$ (green squares) and $\ell=1, s=+1$  (yellow triangles). For $R\Omega= 0.6$, however, the energy level for $\ell=+1$ becomes lower than that corresponding to $\ell=0$, and for $R\Omega>0.7$ negative $E_{\lambda_1,\ell,+}^{(+)}$ appear, which are unacceptable. The same effect
is observed in Fig. \ref{fig2}, for $\ell=0, s=-1$ (gray circles) and $\ell=1, s=-1$ (red stars). The same plot shows that for $s_q=+1$ (or equivalently $q>1$), in general, the energy levels for $s=+1$ is lower than the energy levels for $s=-1$.
\par
As concerns the results for a negatively charged antiparticle in Fig. \ref{fig2}(b), it turns out that, in contrast to the positively charged particle, the energies corresponding to $\ell=0$ and spin orientations $s=+1$ and $s=-1$ are lower than the energies corresponding to $\ell=1$ with $s=\pm 1$. We thus conclude that in general, the spin degeneracy in the LEL for magnetized and rotating Dirac fermions with a global boundary condition is to be determined numerically.
\subsubsection{Normalization of the wave functions with the global boundary condition}\label{sec2b2}
In Sec. \ref{sec2a}, we used the Ritus eigenfunction method, and derived the solutions to the Dirac equation in a rotating system of fermions in a constant background magnetic field. When the system is infinitely extended, i.e. when no boundary conditions are imposed, the Ritus function $\mathbb{E}_{\lambda,\ell,\kappa}^{(\kappa)}$ is given by \eqref{N19}-\eqref{N20}, with $f^{\pm}_{\lambda,\ell,s_q}$ given in \eqref{N41}. Here, the normalization factors $\mathscr{A}^{\pm}$ from \eqref{N40} are determined by using the orthonormality of the Laguerre polynomials. In what follows, we determine $\mathscr{A}^{\pm}$ for a fermionic system under the global boundary condition \eqref{N55}.
\par
To do this, we follow the method introduced in \cite{fukushima2017}. Here, $f^{\pm}_{\lambda_{k},\ell,\kappa}$ from \eqref{N39} are given by
\begin{eqnarray}\label{N64}
f_{\lambda_k,\ell,s_q}^{+}&=&\mathscr{C}_{k,\ell,s_q}^{+}e^{i\ell\varphi}\Phi_{\lambda_k,\ell,s_q}^{+},\nonumber\\
f_{\lambda_k,\ell,s_q}^{-}&=&\mathscr{C}^{-}e^{i(\ell+1)\varphi}\Phi_{\lambda_k,\ell,s_q}^{-},
\end{eqnarray}
where two functions $\Phi_{\lambda_k,\ell,s_q}^{\pm}$ are defined as\footnote{Let us remind that in this case ($R\to \infty$) the hypergeometric functions appearing in \eqref{N65} are to be replaced with the Laguerre function $L_{n}^{m}$, as described in Sec. \ref{sec2a}.}
\begin{eqnarray}\label{N65}
\Phi_{\lambda_{k},\ell,s_q}^{+}&\equiv&\frac{1}{|\ell|!}\left(\frac{|qeB|}{2\pi}\frac{\left(\mathscr{N}_{\lambda_k,s_q}^{+}+|\ell|\right)!}{\mathscr{N}_{\lambda_k,s_q}^{+}!}\right)^{1/2}e^{-x/2}\nonumber\\
&&\times x^{|\ell|/2}{}_{1}F_{1}\left(-\mathscr{N}_{\lambda_k,s_q}^{+};|\ell|+1;x\right),\nonumber\\
\Phi_{\lambda_{k},\ell,s_q}^{-}&\equiv&\frac{1}{|\ell+1|!}\left(\frac{|qeB|}{2\pi}\frac{\left(\mathscr{N}_{\lambda_k,s_q}^{-}+|\ell+1|\right)!}{\mathscr{N}_{\lambda_k,s_q}^{-}!}\right)^{1/2}\nonumber\\
&&\times e^{-x/2}x^{|\ell+1|/2}{}_{1}F_{1}\left(-\mathscr{N}_{\lambda_k,s_q}^{-};|\ell+1|+1;x\right). \nonumber\\
\end{eqnarray}
The normalization factors  $\mathscr{C}_{k,\ell,s_q}^{\pm}$ in a cylinder with a finite radius $R$ are defined so that by taking the limit $R\to \infty$, the results from \eqref{N41} of an unbounded rotating system are reproduced \cite{fukushima2017},
 \begin{eqnarray}\label{N66}
\mathscr{C}_{k,\ell,s_q}^{+}&=&\left(\frac{|qeB|}{2\pi\int\limits_{0}^{\alpha_b}dx\left(\Phi_{\lambda_k,\ell,s_q}^{+}(x)\right)^{2}}\right)^{1/2},\nonumber\\
\mathscr{C}_{k,\ell,s_q}^{-}&=&\left(\frac{|qeB|}{2\pi\int\limits_{0}^{\alpha_b}dx\left(\Phi_{\lambda_k,\ell,s_q}^{-}(x)\right)^{2}}\right)^{1/2}.
\end{eqnarray}
For $\lambda_{k}$s that satisfy two conditions in \eqref{N61} and \eqref{N62}, it turns out that $\mathscr{C}_{k,\ell,s_q}^{+}=\mathscr{C}_{k,\ell,s_q}^{-}\equiv \mathscr{C}_{k,\ell,s_q}$. This can be shown numerically. This is the same result as previously reported in \cite{fukushima2017} for a nonrotating quark matter in the presence of a constant magnetic field. 
\subsubsection{Quantization of fermionic fields in a system with global boundary conditions}\label{sec2b3}
In Sec. \ref{sec2a3}, we presented the quantization relations of fermions in a rotating system without boundary condition [see \eqref{N50}-\eqref{N54}]. Imposing a global boundary condition does not change this quantization too much. The exact quantization relations for fermionic fields $\psi$ and $\bar{\psi}$ are given by
\begin{eqnarray}\label{N67}
\psi_{\alpha}^{(q)}(x)&=&\sum_{k,\ell,s}\int \frac{dp_z}{2\pi}\frac{\mathscr{C}_{k,\ell,s_q}}{\sqrt{2|\epsilon_{\lambda_{k}}^{(q)}}|}\left\{
e^{-i\left(E_{\lambda_{k},\ell,+}^{(q)}t-p_{z}z\right)}a_{p_{z}}^{\lambda_{k},\ell,s}\right.\nonumber\\
&&\left.\times \big[\widetilde{\mathbb{P}}_{\lambda_{k},\ell}^{(q)}\left(x\right)\big]_{\alpha\rho}u_{s,\rho}\left(\tilde{p}_{\ell,+}\right)\Theta\left(E_{\lambda_{k},\ell,+}^{(q)}\right)\right.\nonumber\\
&&\left.+
e^{+i\left(E_{\lambda_{k},\ell,-}^{(q)}t-p_{z}z\right)}b_{p_{z}}^{\lambda_{k},\ell,s\dagger}\big[\widetilde{\mathbb{P}}_{\lambda_{k},\ell}^{(q)}\left(x\right)\big]_{\alpha\rho}^{\dagger}\right.\nonumber\\
&&\left.\times v_{s,\rho}\left(\tilde{p}_{\ell,-}\right)\Theta\left(E_{\lambda_{k},\ell,-}^{(q)}\right)\right\}. \nonumber\\
\bar{\psi}_{\alpha}^{(q)}(x)&=&\sum_{k,\ell,s}\int \frac{dp_z}{2\pi}\frac{\mathscr{C}_{k,\ell,s_q}}{\sqrt{2|\epsilon_{\lambda_{k}}^{(q)}}|}\left\{
e^{+i\left(E_{\lambda_{k},\ell,+}^{(q)}t-p_{z}z\right)}a_{p_{z}}^{\lambda_{k},\ell,s\dagger}\right.\nonumber\\
&&\left.\times \bar{u}_{s,\rho}\left(\tilde{p}_{\ell,+}\right) \big[\widetilde{\mathbb{P}}_{\lambda_{k},\ell}^{(q)}\left(x\right)\big]_{\rho\alpha}^{\dagger}\Theta\left(E_{\lambda_{k},\ell,+}^{(q)}\right)\right.\nonumber\\
&&\left.+
e^{-i\left(E_{\lambda_{k},\ell,-}^{(q)}t-p_{z}z\right)}b_{p_{z}}^{\lambda_{k},\ell,s}
\bar{v}_{s,\rho}\left(\tilde{p}_{\ell,-}\right)\right. \nonumber\\
&&\left.\times
\big[\widetilde{\mathbb{P}}_{\lambda_{k},\ell}^{(q)}\left(x\right)\big]_{\rho\alpha}
\Theta\left(E_{\lambda_{k},\ell,-}^{(q)}\right)\right\}.
\end{eqnarray}
Here, $a_{p_{z}}^{\lambda_{k},\ell,s\dagger}$ and $a_{p_{z}}^{\lambda_{k},\ell,s}$ as well as
$b_{p_{z}}^{\lambda_{k},\ell,s\dagger}$ and $b_{p_{z}}^{\lambda_{k},\ell,s}$ are the
creation and annihilation operators of particles and antiparticles, and satisfy the commutation relations
\begin{eqnarray}\label{N68}
\hspace{-0.9cm}
\{a_{p_{z}}^{\lambda_{k},\ell,s},a_{p_{z}^{\prime}}^{\lambda_{k}^{\prime},\ell^{\prime},s^{\prime}\dagger}\}&=&2\pi\delta\left({p_{z}}-
p_{z}^{\prime}\right)\delta_{\lambda_{k},\lambda_{k}^{\prime}}\delta_{\ell,\ell^{\prime}}\delta_{s,s^{\prime}},\nonumber\\
\hspace{-0.9cm}
\{b_{p_{z}}^{\lambda_{k},\ell,s},b_{p_{z}^{\prime}}^{\lambda_{k}^{\prime},\ell^{\prime},s^{\prime}\dagger}\}&=&2\pi\delta\left({p_{z}}-
p_{z}^{\prime}\right)\delta_{\lambda_{k},\lambda_{k}^{\prime}}\delta_{\ell,\ell^{\prime}}\delta_{s,s^{\prime}}.
\end{eqnarray}
In \eqref{N50}, $\widetilde{\mathbb{P}}_{\lambda_{k},\ell}^{(q)}$, the modified version of $\mathbb{P}_{\lambda_{k},\ell}^{(q)}$ from \eqref{N20}, reads
\begin{eqnarray}\label{N69}
\hspace{-0.5cm}\widetilde{\mathbb{P}}_{\lambda_{k},\ell}^{(q)}=\left(\mathscr{P}_{+}^{(q)}f_{\lambda_{k},\ell,s_q}^{+s_q}+\Pi_{\lambda_{k}}\mathscr{P}_{-}^{(q)}f_{\lambda_{k},\ell,s_q}^{-s_{q}}\right)\Gamma_{\lambda_{k},\ell,q},\nonumber\\
\end{eqnarray}
with $\mathscr{P}^{(q)}_{\pm}$ given in \eqref{N53} with $f_{\lambda_{k},\ell,s_q}^{\pm}$ from \eqref{N64}. Here, $\epsilon_{\lambda_k}^{(q)}=E_{\lambda_k,\ell,\kappa}^{(q)}+\kappa\Omega j$ with $j=\ell+1/2$, as in the previous section.
Hence, according to these results, \eqref{N50}-\eqref{N53} are still valid, except that in \eqref{N52}, two factors $\Pi_{\lambda}$ and $\Gamma_{\lambda_{k},\ell,q}$ are to determined numerically [see our descriptions in \ref{sec2b2}].\footnote{In what follows, we use instead of $\widetilde{\mathbb{P}}_{\lambda_k,\ell}^{(q)}$ from \eqref{N69}, $\mathbb{P}_{\lambda_k,\ell}^{(q)}$ from \eqref{N20}, keeping in mind that the restrictions for $\ell$ are automatically dictated by the properties of the hypergeometric functions appearing in $f_{\lambda_k,\ell,s_q}^{\pm}$  from \eqref{N64}.} Moreover, the summation over $k$ and $\ell$ shall be performed according to the description in this section. Let us also remind that $\Pi_{\lambda}$ was introduced to consider the degeneracy of the energy levels, and $\Gamma_{\lambda,\ell,q}$ to consider the lower and upper bounds of $\lambda$ for positive and negative charges according to Table \ref{table2}. We also notice that index $k$ in the above expressions counts the number of the roots of the hypergeometric functions \eqref{N61} and \eqref{N62}, $\lambda_{k}, k=1,2,\cdots$.
\subsubsection{The fermion propagator in a magnetized rotating fermionic system with boundary condition}\label{sec2b4}
The main purpose of this paper is to compute the chiral condensate at zero and finite temperature $T$ and zero chemical potential, and to study the effect of rotation on its $T$ dependence for a fixed magnetic field. To do this, we use in Sec. \ref{sec3}, the mass gap relation
\begin{eqnarray}\label{N70}
\bar{m}_{q}=G\lim_{r\to r^{\prime}} \mbox{Tr}\left(S^{(q)}\left(r,r^{\prime}\right)\right),
\end{eqnarray}
where $G$ is a dimensionful coupling constant, and $S\left(x,x^{\prime}\right)$ is the fermion propagator of magnetized Dirac fermions in a rotating system with a global boundary condition. In what follows, we show that the fermion propagator is given by
\begin{widetext}
\begin{eqnarray}\label{N71}
\hspace{-1cm}
S_{\alpha\beta}^{(q)}(r,r^{\prime})=i\sum_{k,\ell}\int\frac{dp_{0}dp_z}{\left(2\pi\right)^{2}}\mathscr{C}_{k,\ell,s_q}^{2} e^{-ip_0\left(t-t^{\prime}\right)+ip_z\left(z-z^{\prime}\right)}[\mathbb{P}_{\lambda_{k},\ell}^{(q)}(x)]_{\alpha\rho}\left(\frac{\gamma\cdot \tilde{p}_{\lambda_{k},\ell,+}^{(q)}+m_{q}}{\left(p_{0}+\Omega j\right)^{2}-\epsilon_{\lambda_{k}}^{(q)2}}\right)_{\rho\sigma}[\mathbb{P}_{\lambda_{k,\ell}}^{(q)}(x^{\prime})]^{\dagger}_{\sigma\beta},
\end{eqnarray}
\end{widetext}
with $\epsilon_{\lambda_{k}}^{(q)2}=m_{q}^{2}+2\lambda_{k}|qeB|+p_{z}^{2}$ from \eqref{N49}. The functions $\mathbb{P}_{\lambda_{k,\ell}}^{(q)}$ are given in \eqref{N20} with $f^{\pm}_{\lambda_{k,\ell,s_q}}$ from \eqref{N64}. To show this, let us start with the definition of the fermion propagator
\begin{eqnarray}\label{N72}
S_{\alpha\beta}^{(q)}=\theta(t-t')\langle \psi_{\alpha}(x)\bar{\psi}_{\beta}(x')\rangle-\theta(t-t')\langle\bar{\psi}_{\beta}(x')\psi_{\alpha}(x)\rangle.\nonumber\\
\end{eqnarray}
Using the quantization relation \eqref{N67}, we arrive first at
\begin{widetext}
\begin{eqnarray}\label{N73}
\langle \psi_{\alpha}(x)\bar{\psi}_{\beta}(x')\rangle&=&\sum_{k,\ell,s}\int\frac{dp_z}{2\pi}\frac{\mathscr{C}_{k,\ell,s_q}^{2}}{2|\epsilon_{\lambda_k}^{(q)}|}
\left\{e^{-iE_{\lambda_k,+}^{(q)}(t-t')+ip_{z}(z-z')}
[\mathbb{P}_{\lambda_k,\ell}^{(q)}(x)]_{\alpha\rho}u_{s,\rho}\left(\tilde{p}_{\ell,+}\right)\bar{u}_{\sigma,s}\left(\tilde{p}_{\ell,+}\right)[\mathbb{P}_{\lambda,k}^{(q)}]^{\dagger}_{\sigma\beta}\right\},\nonumber\\
\langle \bar{\psi}_{\beta}(x')\psi_{\alpha}(x)\rangle&=&\sum_{k,\ell,s}\int\frac{dp_z}{2\pi}\frac{\mathscr{C}_{k,\ell,s_q}^{2}}{2|\epsilon_{\lambda_k}^{(q)}|}
\left\{e^{+iE_{\lambda_k,-}^{(q)}(t-t')-ip_{z}(z-z')}
[\mathbb{P}_{\lambda_k,\ell}^{(q)}(x)]_{\alpha\rho} u_{s,\rho}\left(\tilde{p}_{\ell,+}\right)\bar{v}_{\rho,s}\left(\tilde{p}_{\ell,-}\right)[\mathbb{P}_{\lambda,k}^{(q)}]^{\dagger}_{\sigma\beta}\right\}.
\end{eqnarray}
Plugging then these expressions into \eqref{N72}, and using $E_{\lambda_k,\ell,\kappa}^{(q)}=\epsilon_{\lambda_k}^{(q)}-\kappa\Omega j$,
\begin{eqnarray}\label{N74}
\theta\left(\pm z\right)=\lim\limits_{\varepsilon\to 0^{+}}\mp\int\frac{dp_0}{2\pi}\frac{e^{izt}}{p_0\mp i\varepsilon},
\end{eqnarray}
as well as \eqref{N18}, we obtain
\begin{eqnarray}\label{N75}
S_{\alpha\beta}^{(q)}\left(x,x'\right)&=&-i\sum\limits_{k,\ell}\int\frac{dp_0 dp_z}{(2\pi)^{2}}\frac{\mathscr{C}_{k,\ell,s_q}^{2}e^{i\Omega j\left(t-t'\right)}}{2\epsilon_{\lambda_{k}}^{(q)}}[\mathbb{P}_{\lambda_k,\ell}^{(q)}(x)]_{\alpha\rho}\bigg\{\frac{\gamma\cdot\tilde{p}_{\lambda_k,\ell,+}^{(q)}+m_q}{p_0-i\varepsilon}e^{i\left(p_0-\epsilon_{\lambda_k}^{(q)}\right)(t-t')+ip_z\left(z-z'\right)}\nonumber\\
&& +\frac{\gamma\cdot\tilde{p}_{\lambda_k,\ell,-}^{(q)}-m_q}{p_0+i\varepsilon}e^{i\left(p_0+\epsilon_{\lambda_k}^{(q)}\right)(t-t')-ip_z\left(z-z'\right)}\bigg\}_{\rho\sigma}
[\mathbb{P}_{\lambda_k,\ell}^{(q)}(x')]_{\sigma\beta}.
\end{eqnarray}
\end{widetext}
Performing a shift of variables $p_0\to  -p_0+\epsilon_{\lambda_k}^{(q)}$ and $p_{0}\to -p_{0}-\epsilon_{\lambda_k}^{(q)}$, and eventually $p_{0}\to p_{0}+\Omega j$, we arrive at \eqref{N71}, as claimed. In the next section, \eqref{N71} is used to determine the chiral condensate at zero and finite temperature.
\section{Inverse magneto-rotational Catalysis at zero and finite temperature; Numerical results}\label{sec3}
\setcounter{equation}{0}
One of the aims of this paper is to elaborate on the interplay between the rotation and the presence of a constant magnetic field, in particular, on the formation of bound states. It is known that external magnetic fields enhance chiral symmetry breaking. This is the well-established magnetic catalysis. There are a number of attempts exploring the effect of the rigid rotation of a system of quark matter on magnetic catalysis. In this section, after reviewing the results for zero temperature by shedding light on some new aspects, which are not discussed before in the literature, we introduce the temperature $T$, and explore the $T,eB,\Omega$, and $r$ dependence of the dynamical mass. We then present numerical results for the $G,eB,\Omega$, and $r$ dependence of the critical temperature $T_c$, and $G,eB,T$, and $r$ dependence of certain critical frequency $\Omega_c$.      
\subsection{Zero temperature}\label{sec3a}
\setcounter{equation}{0}
In this section, after presenting the relations which are used to study the effect of rotation on the magnetic catalysis in a fermionic system with boundary at zero temperature, we explore, in particular, the $r$ dependence of the dynamical mass $\bar{m}$. Here, it is shown that the angular frequency plays no role in the behavior of $\bar{m}$.  
\par
First, we focus on the mass gap relation \eqref{N70}. Plugging the propagator $S^{(q)}$ from \eqref{N71} into \eqref{N70}, and performing the trace over $\gamma$-matrices, we arrive at
\begin{eqnarray}\label{E1}
\frac{\bar{m}_{q}}{G}=\frac{i\bar{m}_{q}}{2\pi^{2}}\sum\limits_{k,\ell}\mathscr{C}_{k,\ell,s_q}^{2}\Phi^{2}_{\lambda_{k},\ell,s_q}\int\frac{ dp_0 dp_{z}}{\big[\left(p_0+\Omega j\right)^{2}-\epsilon_{\lambda_{k}}^{(q)2}\big]}, \nonumber\\
\end{eqnarray}
with
$$
\Phi_{\lambda_{k,\ell,s_q}}^{2}\equiv \Phi^{+2}_{\lambda_{k},\ell,s_q}+\Phi^{-2}_{\lambda_{k},\ell,s_q}.
$$
Performing then a shift of variable $p_{0}\to p_{0}-\Omega j$, the integration over $p_0$ can be immediately carried out. The resulting expression reads
\begin{eqnarray}\label{E2}
\frac{\bar{m}_{q}}{G}=\frac{i\bar{m}_{q}}{2\pi^{2}}\sum\limits_{k,\ell}\mathscr{C}_{k,\ell,s_q}^{2}\Phi^{2}_{\lambda_{k},\ell,s_q}\int dp_{z}\frac{1}{\epsilon_{\lambda_{k}}^{(q)}}.
\end{eqnarray}
As it turns out, the angular frequency $\Omega$ is canceled from the computation, and has indeed no effect on the mass $m_q$, arising from \eqref{E2}. This is in contrast to the results presented in \cite{fukushima2015}, where the zero temperature case is considered as a limit of the finite temperature case. In this case, the $p_{0}$ integration appearing in \eqref{E1} is replaced with a sum over Matsubara frequencies, and the $\Omega j$ dependence thus appears in a Heaviside $\theta$-function, arising from 
$$
\lim\limits_{T\to 0}T\ln\left(1+e^{-\frac{x}{T}}\right)=-x\theta(-x). 
$$  
The integration over $p_z$ is carried out by introducing the ultraviolet smooth cutoff \cite{fukushima2015,fukushima2017,shovkovy2011}
\begin{eqnarray}\label{E3}
\hspace{-0.5cm}f\left(p;\Lambda,\delta\Lambda\right)=\frac{\sinh\left(\Lambda/\delta\Lambda\right)}{[\cosh\left(p/\delta\Lambda\right)+\cosh\left(\Lambda/\delta\Lambda\right)]},
\end{eqnarray}
with $p=\sqrt{2\lambda_{k,\ell,s_q}|qeB|+p_{z}^{2}}$. For the limit $\delta\Lambda/\Lambda\to 0$, the function $f(p;\Lambda,\delta\Lambda)$ approaches the Heaviside $\Theta$-function
\begin{eqnarray}\label{E4}
\hspace{-1cm}\lim\limits_{\delta\Lambda/\Lambda\to 0}f\left(p;\Lambda,\delta\Lambda\right)\to \Theta\left(\Lambda^2-2\lambda_{k,\ell,s_q}|qeB|-p_z^2\right).\nonumber\\
\end{eqnarray}
Plugging the $\Theta$-function into the remaining $p_{z}$ integral in \eqref{E1}, and integrating $p_z$ from $-\left(\Lambda^2-2\lambda_{k,\ell,s_q}|qeB|\right)^{1/2}$ to $+\left(\Lambda^2-2\lambda_{k,\ell,s_q}|qeB|\right)^{1/2}$, dictated by the $\Theta$-function, we arrive at
\begin{eqnarray}\label{E5}
\frac{\bar{m}_q}{G}&=&\frac{\bar{m}_{q}}{\pi}\sum\limits_{k,\ell}\mathscr{C}_{k,\ell,s_q}^{2}\Phi^{2}_{\lambda_{k},\ell,s_q}
\nonumber\\
&&\times \mbox{tanh}^{-1}\left(\frac{\sqrt{\Lambda^{2}-2\lambda_{k,\ell,s_q}|qeB|}}{\Lambda^{2}+m_{q}^{2}}\right)\nonumber\\
&&\times\Theta\left(\Lambda^{2}-2\lambda_{k,\ell,s_q}|qeB|\right).
\end{eqnarray}
Assuming $\bar{m}_{q}\neq 0$, the nontrivial solutions to \eqref{E5} can be determined numerically by fixing $q, \Lambda,R, eB$ and $G$.
In what follows we choose
\begin{eqnarray}\label{E6}
q=+1,\qquad\Lambda=1~\mbox{GeV},\qquad R=6~\mbox{fm}.
\end{eqnarray}
Instead of $eB$, it is more appropriate to work with the dimensionless quantity $\alpha_b= eB R^2/2$ which is introduced in the previous section. To generate our data, we use $\alpha_b=1,\cdots,10$ that correspond to $eB/m_{\pi}^{2}$ given in Table \ref{tab4} for $R=6$ fm.\footnote{Here, we use $1$ fm$^{-1}\sim m_{\pi}$ in MeV, where $m_{\pi}\sim 200$ MeV is the pion mass.}
\begin{figure}[h]
\includegraphics[width=8cm,height=6cm]{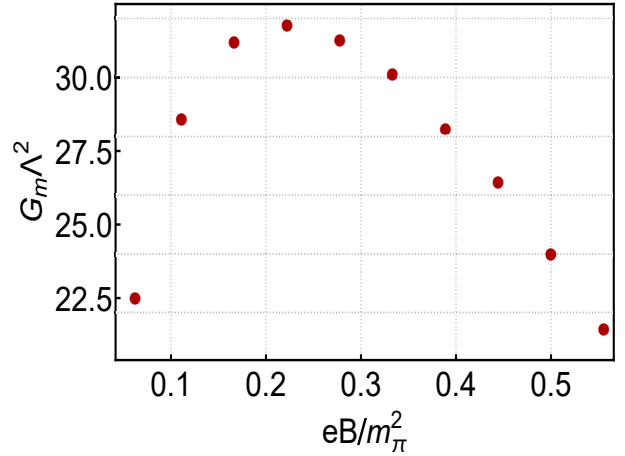}
\caption{color online. The minimum value of the coupling constant for which the mass gap possesses nonvanishing solution in the interval $x\in[1,\alpha_{b}]$, with $x=eBr^2/2$ and $\alpha_{b}=eBR^2/2$. }\label{fig3x}
\end{figure}

\begin{figure}
\includegraphics[width=8cm,height=6cm]{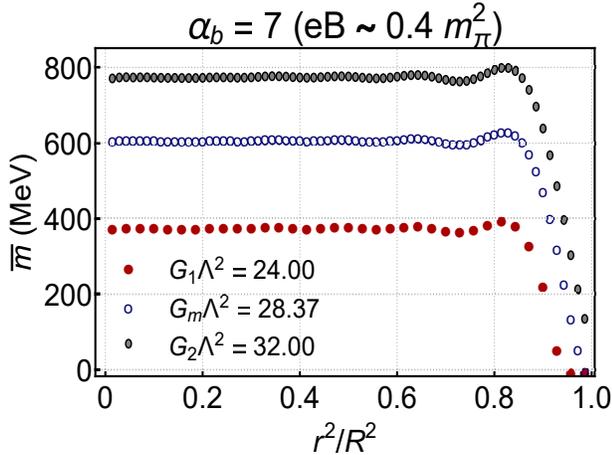}
\caption{color online. The $r^{2}/R^{2}$ dependence of the mass gap $\bar{m}$ is plotted for $\alpha_{b}=7$ ($eB\sim 0.4 m_{\pi}^{2}$) and three different values of $G$. As it turns out, for a fixed $eB$, increasing $G$ enhances the chiral symmetry breaking.}\label{fig4}
\end{figure}
\begin{figure*}
\includegraphics[width=8cm,height=6cm]{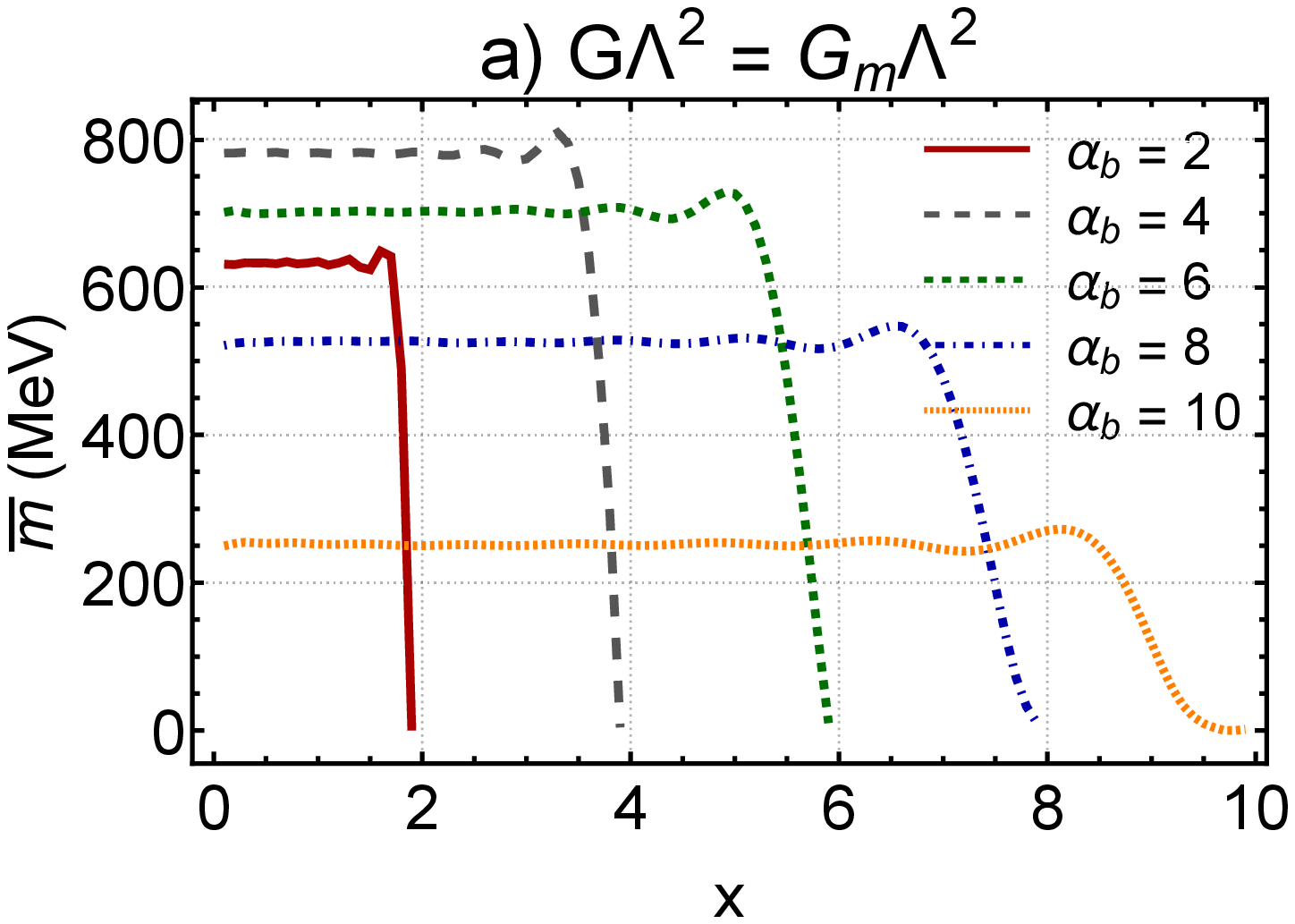}
\includegraphics[width=8cm,height=6cm]{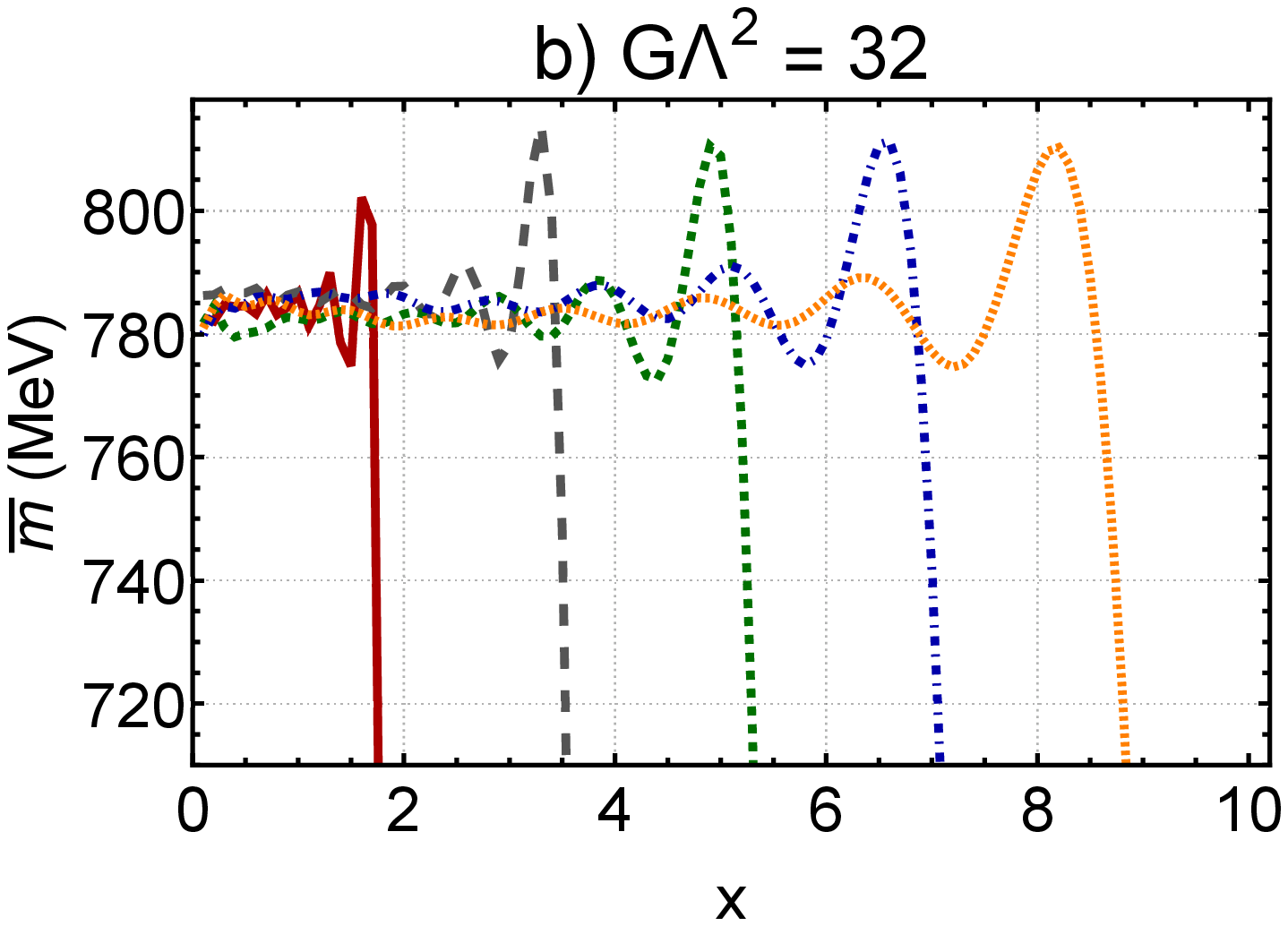}
\caption{color online. The $x$ dependence of $\bar{m}$ is plotted for $\alpha_{b}=2,4,6,8,10$ and $G\Lambda^2$ equal to $G_{m}\Lambda^2$ from Table \ref{tab5} (panel a) and a constant $G\Lambda^2=32$ (panel b). }\label{fig5}
\end{figure*}
\begin{figure*}
\includegraphics[width=8cm,height=6cm]{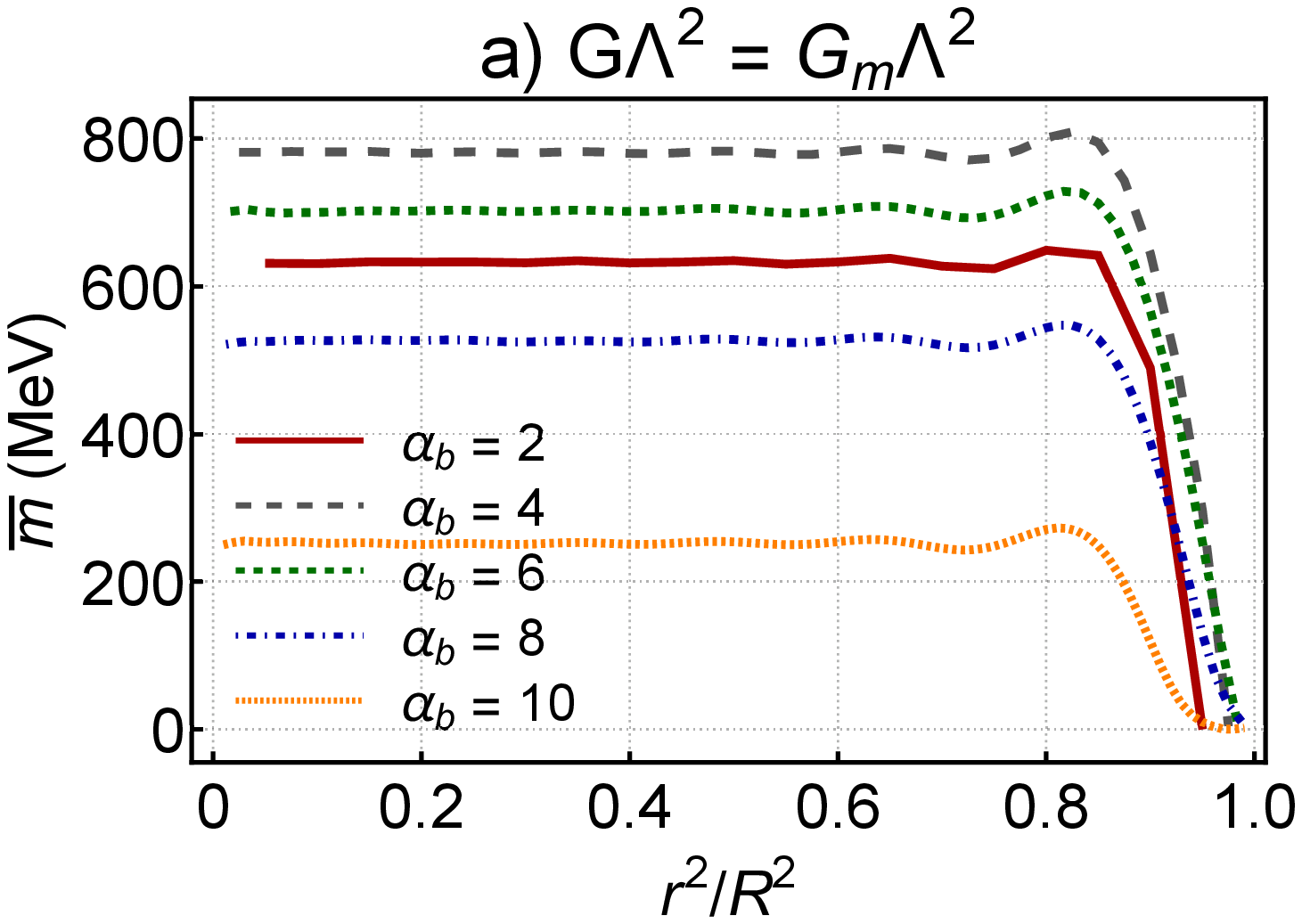}
\includegraphics[width=8cm,height=6cm]{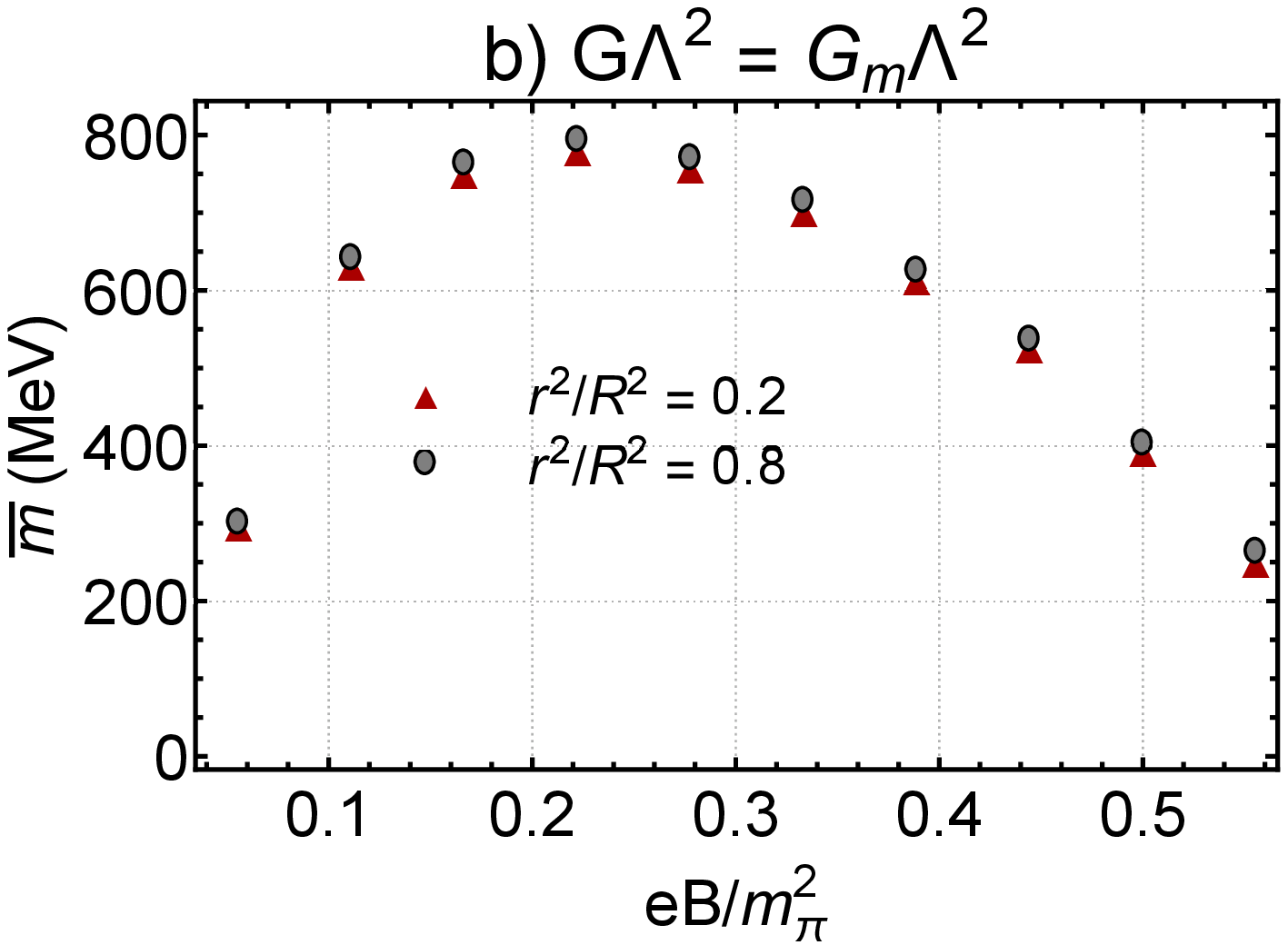}
\caption{color online. (a) The $r^2/R^2$ dependence of $\bar{m}$ is plotted for $\alpha_b=2,4,6,8,10$, and $G\Lambda^2=G_m\Lambda^2$. The coupling are chosen so that $\bar{m}$ remains almost constant in the range $r^2<0.8 R^2$. (b) Using same couplings $G_m\Lambda^2$ from Table \ref{tab4}, the $eB/m_{\pi}^2$ dependence of $\bar{m}$ is plotted for fixed $r^2=0.2R^2$ and $r^2=0.8 R^2$. The behavior reflects the dependence of $G_m\Lambda^2$ on $eB/m_{\pi}^2$ (see Fig. \ref{fig3x}).    }\label{fig6}
\end{figure*}
\begin{figure*}
\includegraphics[width=8cm,height=6cm]{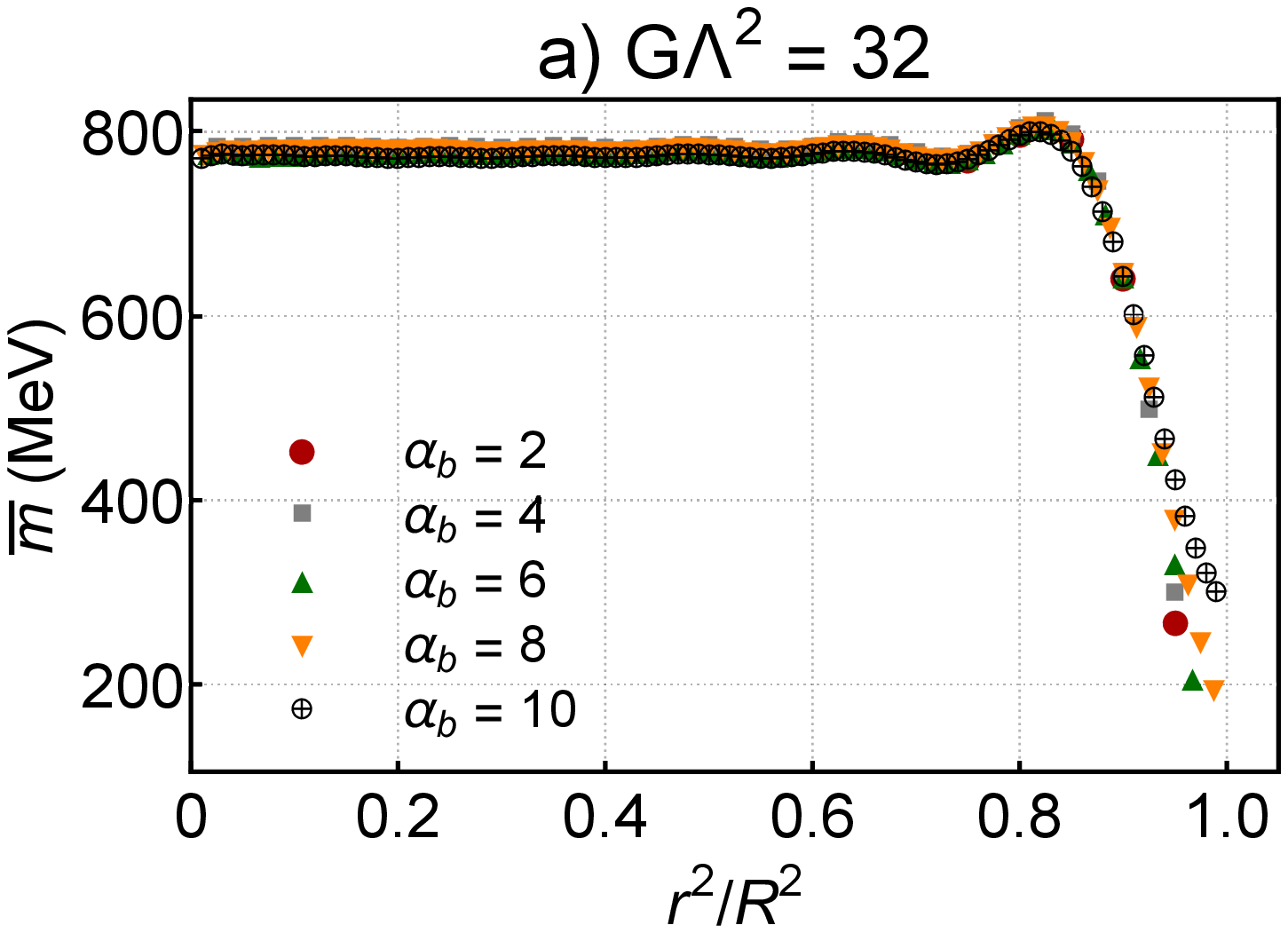}
\includegraphics[width=8cm,height=6cm]{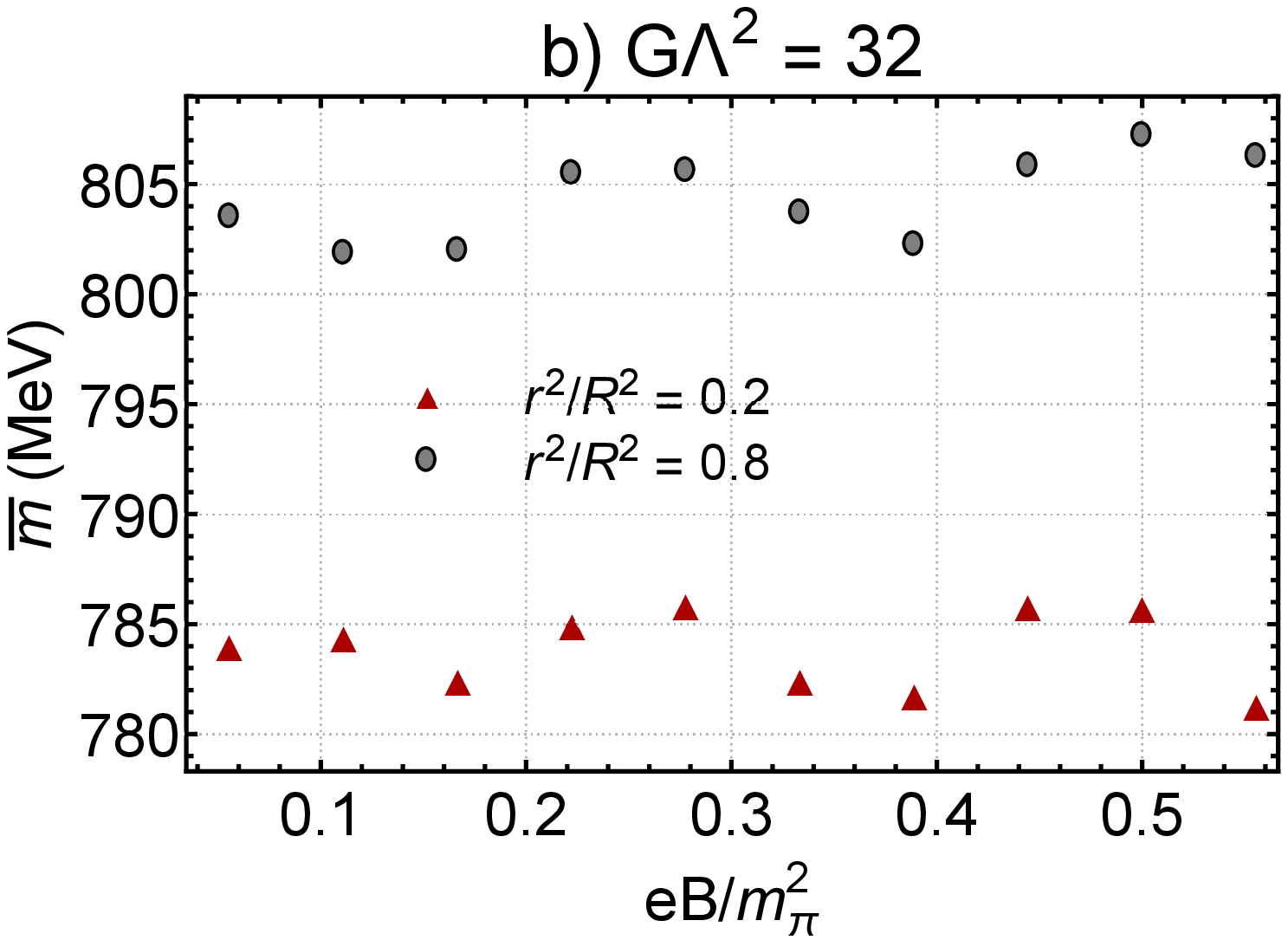}
\caption{color online. (a) The $r^2/R^2$ dependence of $\bar{m}$ is plotted for $\alpha_b=2,4,6,8,10$, and $G\Lambda^2=32$. In contrast to the results from Fig. \ref{fig6}(a), weak magnetic fields do not affect the dependence of $\bar{m}$ on $r$. Only small perturbations occur, which are explicitly shown in panel (b), where the $eB/m_{\pi}^2$ dependence of $\bar{m}$ is plotted for fixed $r^2=0.2R^2$ and $r^2=0.8 R^2$. The oscillations are due to successive filling of the Landau levels (de Haas-van Alfven effect). }\label{fig7}
\end{figure*}
\begin{table}[hbt]
\begin{tabular}{ccccccc}
\hline
$\alpha_{b}$&$\quad$&$eB/m_{\pi}^{2}$&$\qquad$&$\alpha_{b}$&$\quad$&$eB/m_{\pi}^{2}$\\
\hline\hline
$1$&&$0.05$&$\qquad\quad$&$6$&&$0.33$\\
$2$&&$0.10$&$\qquad\quad$&$7$&&$0.39$\\
$3$&&$0.17$&$\qquad\quad$&$8$&&$0.44$\\
$4$&&$0.22$&$\qquad\quad$&$9$&&$0.50$\\
$5$&&$0.28$&$\qquad\quad$&$10$&&$0.55$\\
\hline
\end{tabular}
\caption{The values of $eB/m_{\pi}^{2}$ for $R=6$ fm and given $\alpha_{b}$s. }\label{tab4}
\end{table}
As concerns G, let us notice that the mass gap $\bar{m}_{q}$ arising in \eqref{N70} is related to the chiral condensate which is created as a result of a spontaneous chiral symmetry breaking in a QCD-like model. An appropriate example is the Lagrangian density of one flavor NJL model
\begin{eqnarray}\label{E7}
\mathscr{L}=\bar{\psi}\left(\gamma\cdot \Pi-m\right)\psi+\frac{G}{2}[\left(\bar{\psi}\psi\right)^{2}+\left(\bar{\psi}i\gamma^{5}\psi\right)^{2}].\nonumber\\
\end{eqnarray}
Here, $\Pi\equiv \Pi^{(q=1)}$ is given in \eqref{N11}, $m\equiv m_{q=1}$ is the current mass of a particle with $q=1$. For the sake of simplicity, we assume the fermion to be massless ($m=0$). The solution of the mass gap $\bar{m}$ is related to the value of chiral condensate $\langle\bar{\psi}\psi\rangle$ through
\begin{eqnarray}\label{E8}
\bar{m}=-G\langle \bar{\psi}\psi\rangle.
\end{eqnarray}
For vanishing magnetic fields and in a nonrotating fermionic system the condensate is built only when $G\Lambda^2$ is large enough \cite{klevansky1992,miransky1995}. However, it is known that external magnetic fields enhance the condensation so that even moderate values of $G\Lambda^2$ would be enough for the formation of the condensate \cite{miransky1995}. It is thus interesting to determine the minimum value of $G\Lambda^2$ for which the gap equation \eqref{E5} possesses a nontrivial solution in a rotating system for a given $\alpha_b$ and for a relatively large interval $x\in[0,\alpha_{b}]$ with $x=eB r^2/2$. Denoting these kinds of $G$'s by $G_m$, we plotted them as a function of $eB/m_{\pi}^{2}$ in Fig. \ref{fig3x}. Their values are listed in Table \ref{tab5}. As it turns out, for small values of $\alpha_b\leq 4$ ($eB\leq 0.22 m_{\pi}^{2}$) the NJL coupling $G_m$ increases with increasing $\alpha_{b}$. This means that the magnetic field is not yet strong enough to hold the constituent mass nonvanishing in the interval $x\leq 10$ fm. However, once $\alpha_{b}$ increases, it becomes strong enough and enhances the production of the dynamical mass even when the coupling is not very large. This is why $G_{m}$ decreases with increasing $\alpha_{b}\geq 4$.
 \begin{table}[hbt]
\begin{tabular}{ccccccccccc}
\hline
$\alpha_{b}$&&$eB/m_{\pi}^{2}$&&$G_m$&$\qquad\quad$
&$\alpha_{b}$&&$eB/m_{\pi}^{2}$&&$G_{m}$\\
\hline\hline
$1$&&$0.05$&&$22.60$&$\qquad\quad$&$6$&&$0.33$&&$30.22$\\
$2$&&$0.10$&&$28.71$&$\qquad\quad$&$7$&&$0.39$&&$28.37$\\
$3$&&$0.17$&&$31.30$&$\qquad\quad$&$8$&&$0.44$&&$26.55$\\
$4$&&$0.22$&&$31.89$&$\qquad\quad$&$9$&&$0.50$&&$24.11$\\
$5$&&$0.28$&&$31.37$&$\qquad\quad$&$10$&&$0.55$&&$21.57$\\
\hline
\end{tabular}
\caption{The $eB/m_{\pi}^{2}$ dependence of $G_{m}$ as the minimum value of G for which a nonvanishing constituent mass $\bar{m}$ arises in the interval $x\in[1,\alpha_b]$.}\label{tab5}
\end{table}
\par
In Fig. \ref{fig4}, the mass gap $\bar{m}$ is plotted as a function of $x/\alpha_{b}=r^{2}/R^{2}$ for $\alpha_{b}=7$ ($eB \sim 0.4 m_{\pi}^{2}$), and three different $G\Lambda^{2}$, $G_{1}\Lambda^{2}=24, G_{2}\Lambda^{2}=32$, and $G_{m}\Lambda^{2}=28.37$. It is demonstrated how larger values of $G$ enhances the chiral symmetry breaking. The qualitative dependence of the constituent mass $\bar{m}$ on the position relative to $R$ does not change dramatically by increasing $G$.
\par
In Fig. \ref{fig5}, the $x$ dependence of $\bar{m}$ is plotted for $\alpha_{b}=2,4,6,8,10$. In Fig. \ref{fig5}(a), we used the corresponding $G_{m}\Lambda^2$ to each $\alpha_b$ (see Table \ref{tab4}), while in Fig. \ref{fig5}(b), $G\Lambda^{2}=32$ is used. The color code in Fig. \ref{fig5}(b) is the same as in Fig. \ref{fig5}(a). As it is shown, the quantitative dependence of $\bar{m}$ does not change by increasing $\alpha_{b}$, but the position where $\bar{m}$ starts to decrease depends on $\alpha_{b}$, because, according to its definition, the maximum value of $x=eB r^{2}/2$ is equal to $\alpha_{b}=eBR^2/2$. In Fig. \ref{fig5}(b), we consider only the interval $\bar{m}\in [700,820]$ in the vertical axis for fixed $G\Lambda^{2}=32$ and the same values of $\alpha_{b}$ as demonstrated in Fig. \ref{fig5}(a). It is shown that independent of $\alpha_{b}$, $\bar{m}$ exhibits small oscillations as a function of $x$. The amplitudes of the oscillations become large in the vicinity of the boundary. At boundary $R$ ($x=\alpha_{b}$), $\bar{m}$ decreases rapidly.
\par
Being a function of $eB$, the parameter $x$ is not a natural quantity to demonstrate the dependence of $\bar{m}$ on the position $r$ relative to the boundary $R$. This is why, we plotted $\bar{m}$ as a function of $r^2/R^2$ in Fig. \ref{fig6}(a) for $\alpha_{b}=2,4,6,8,10$. Here, $G\Lambda^2$ is chosen to be $G_{m}\Lambda^2$ which are different for different $\alpha_{b}$ (see Table \ref{tab4}). As expected from the previous results in Fig. \ref{fig5}(a), independent of $\alpha_b$, $\bar{m}$ remains relatively constant for a large interval $r^2 \in [0,0.8R^2]$ before it starts decreasing at the boundary $r\sim R$. However, for a fixed $r/R$, it has a non-monotonic dependence on $\alpha_b$. First, it increases and then decreases with $\alpha_{b}$. To scrutinize this dependence, we plotted in Fig. \ref{fig6}(b) the $eB$ dependence of $\bar{m}$ for fixed $r^2/R^2=0.2,0.8$ and for $G\Lambda^2=G_{m}\Lambda^2$. As it is demonstrated here, it increases first as a function of $eB$ and then decreases with increasing $eB$. This specific behavior is mainly related to the $\alpha_{b}$-dependence of the NJL coupling $G_m$, demonstrated in Fig. \ref{fig3x}.
\par
The $r^2/R^2$ dependence of $\bar{m}$ for $\alpha_{b}=2,4,6,8,10$ and a fixed $G\Lambda^2=32$ is plotted in Fig. \ref{fig7}. As it is shown, $\bar{m}$ remains almost constant for $r^{2}<0.8 R^2$, and rapidly decreases for $r\to R$. The values of $\bar{m}$ corresponding to $\alpha_{b}$ are slightly different. In order to see the difference between $\bar{m}$'s in the interval $r^2 \in [0,0.8R^2]$, the $eB$ dependence of $\bar{m}$ is plotted for $r^2=0.2R^2, 0.8R^2$ and relatively large $G\Lambda^{2}=32$ in Fig. \ref{fig7}(b). In the interval $\bar{m}\in[780,810]$, the constituent mass oscillates with $eB$. This is in contrast to the behavior of $\bar{m}$ demonstrated in Fig. \ref{fig6}. The positions of the maxima and minima appearing in Fig. \ref{fig7} do not change by increasing $r^{2}$ from $r^{2}=0.2 R^2$ to $r^2=0.8 R^2$. The oscillations are related to the de Haas-Alfven effect, and are because of the successive filling of Landau levels.  
\subsection{Finite temperature}\label{sec3b} 
In this section, we generalize our previous results to the case of finite temperature. We demonstrate the $T,R\Omega,eB$ and $r^{2}/R^{2}$ dependence of $\bar{m}$ for a fixed set of parameters $\{\alpha_{b}, x,R\Omega,G\Lambda^2,T\}$. We also determine the phase diagram $T_{c}$ ($\Omega_{c}$) versus $R\Omega, eB$ ($T,eB$) and $r^2/R^2$ for a fixed set of parameters
$\{\alpha_{b}, x,R\Omega,G\Lambda^2\}$ ($\{\alpha_{b}, x,T,G\Lambda^2\}$). We demonstrate, in particular, the IMRC, in which a finite rotation neutralizes the magnetic catalysis induced by a constant magnetic field.  As a consequence $\bar{,m}$ decreases with increasing $eB$ for relatively large $R\Omega$ and small coupling $G\Lambda^2$. Moreover, in exploring the phase diagram of $T_{c}$ versus $R\Omega$, this effect is reflected in reducing $T_{c}$ as a function of $eB$ for large value of $R\Omega$. The same effect is also demonstrated in the phase diagram $R\Omega_{c}$ versus $eB$ and $T$. The dependence of $T_c$ and $\Omega_c$ on the coupling $G\Lambda^2$ and $r^2/R^2$ is also explored. 
\par
To introduce the temperature, let us consider \eqref{E1}, and use
\begin{eqnarray}\label{E9}
p_0\to i\omega_{n}=i\pi T(2n+1),\qquad \int\frac{dp_0}{2\pi}\to iT\sum_{n=-\infty}^{+\infty}, \nonumber\\
\end{eqnarray}
where $\omega_{n}$ is the corresponding Matsubara frequencies for fermions. We arrive first at
\begin{eqnarray}\label{E10}
\frac{\bar{m}_{q}}{G}=\frac{\bar{m}_{q}}{2\pi^{2}}\sum\limits_{k,\ell,n}\mathscr{C}_{k,\ell,s_q}^{2}\Phi^{2}_{\lambda_{k},\ell,s_q}\int\frac{dp_{z}}{\big[\left(p_0-i\Omega j\right)^{2}+\epsilon_{\lambda_{k}}^{(q)2}\big]}, \nonumber\\
\end{eqnarray}
with $\epsilon_{\lambda_{k}}^{(q)}$ from \eqref{N49}. Using then
\begin{eqnarray}\label{E11}
T\sum_{n=-\infty}^{+\infty}\frac{1}{\left(\omega_{n}-i\mu\right)^{2}+\epsilon^2}=\frac{1-f\left(\epsilon+\mu\right)-f(\epsilon-\mu)}{2\epsilon},\nonumber\\
\end{eqnarray}
where $f(\epsilon\pm\mu)\equiv\left(e^{\left(\epsilon\pm\mu\right)/T}+1\right)^{-1}$
is the Fermi-Dirac distribution function, the gap equation \eqref{E10} is separated into a  $T$ independent and a $T$ dependent part,
\begin{eqnarray}\label{E12}
\lefteqn{
\frac{\bar{m}_q}{G}=\frac{\bar{m}_{q}}{\pi}\sum\limits_{k,\ell}\mathscr{C}_{k,\ell,s_q}^{2}\Phi^{2}_{\lambda_{k},\ell,s_q}
}
\nonumber\\
&&\times \mbox{tanh}^{-1}\left(\frac{\sqrt{\Lambda^{2}-2\lambda_{k,\ell,s_q}|qeB|}}{\Lambda^{2}+m_{q}^{2}}\right)\nonumber\\
&&\times\Theta\left(\Lambda^{2}-2\lambda_{k,\ell,s_q}|qeB|\right)\nonumber\\
&&-\frac{\bar{m}_{q}}{\pi}\sum\limits_{k,\ell,n}\mathscr{C}_{k,\ell,s_q}^{2}\Phi^{2}_{\lambda_{k},\ell,s_q}\nonumber\\
&&\times \int_{0}^{\infty} dp_{z}\frac{f(\epsilon_{\lambda_{k}}^{(q)}+\Omega j)+f(\epsilon_{\lambda_{k}}^{(q)}-\Omega j)}{\epsilon_{\lambda_{k}}^{(q)}}.
\end{eqnarray}
The $T$ independent part of the gap equation is regularized in the same manner as in \eqref{E2}. This yields \eqref{E5}, which appears again in the first term on the r.h.s.  of \eqref{E12}. Here, $\Lambda$ is the corresponding cutoff, as appears also in \eqref{E5}. Concerning the $T$ dependent part of the gap equation, the distribution functions prevent the corresponding integrals to be divergent. In what follows, we present first the numerical results for the gap equation \eqref{E12} for fixed parameters $\{q,\Lambda,R\}$ from \eqref{E6}. Then, focusing on the critical temperature as well as angular frequency, we study, in particular, their $eB$ dependence. 
\subsubsection{The constituent mass as a function of $T, eB,  R\Omega$,  and $r^2/R^2$}\label{sec3b1} 
\begin{figure*}
	\includegraphics[width=8cm,height=6cm]{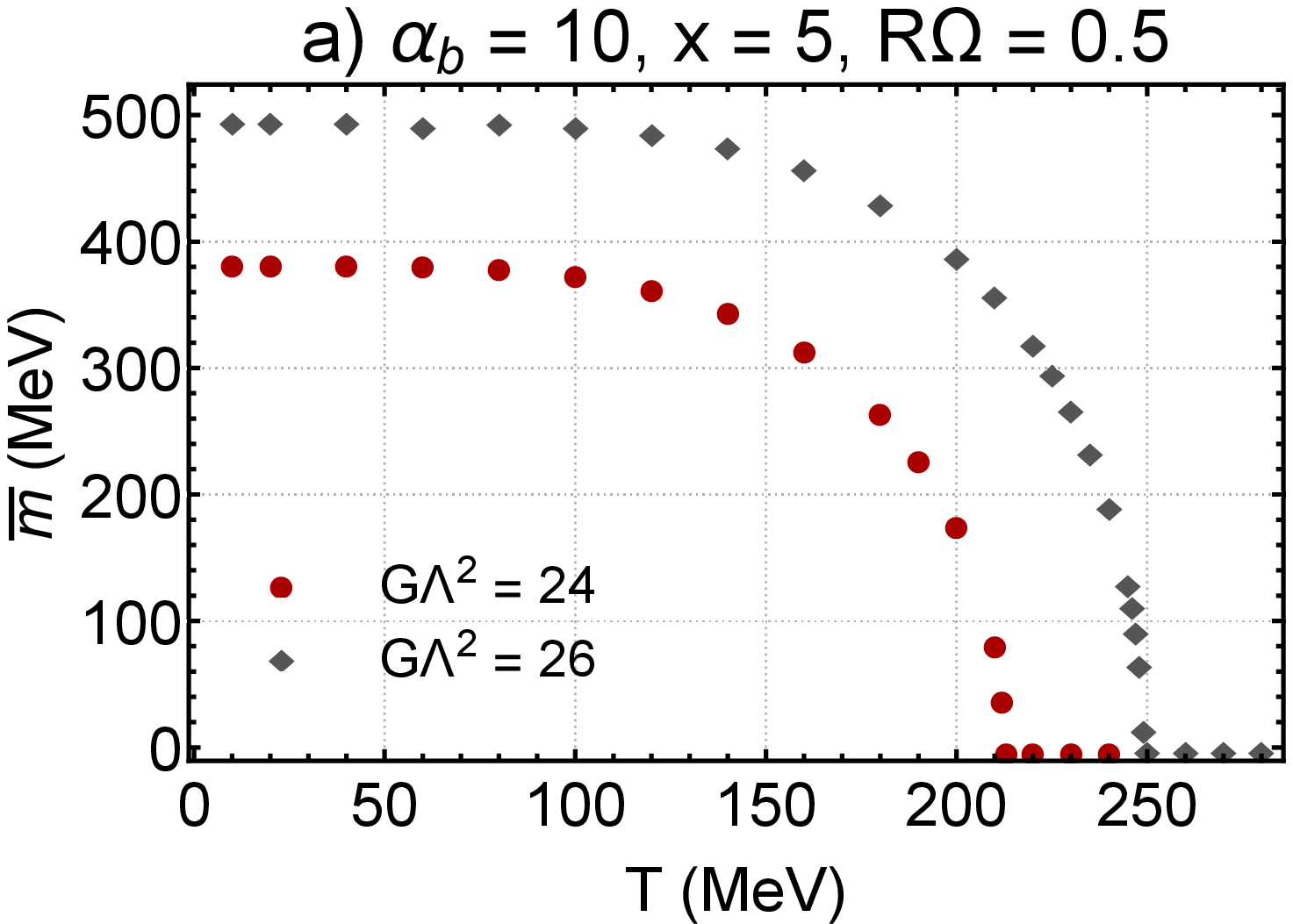}
	\includegraphics[width=8cm,height=6cm]{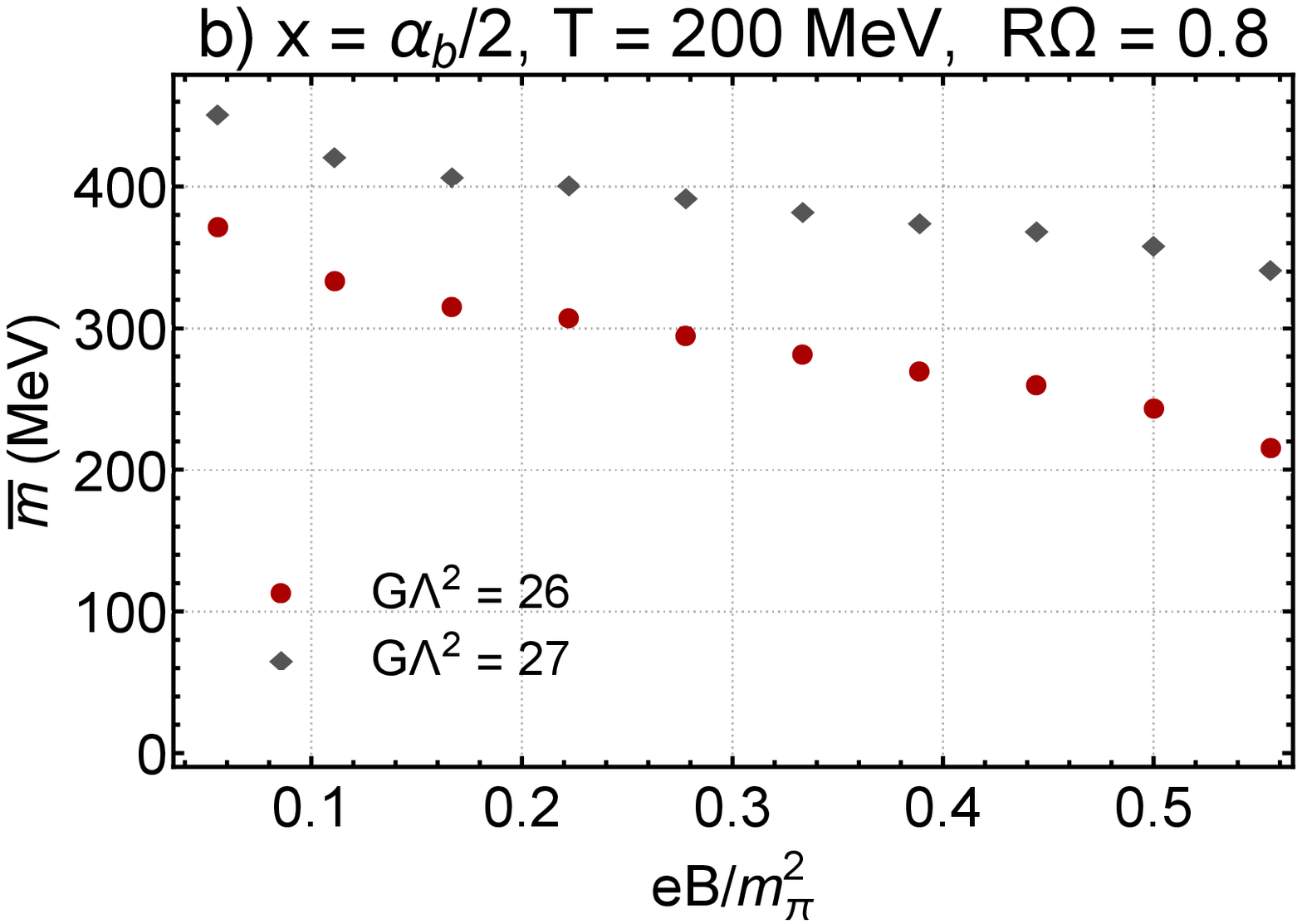}\vspace{.5cm}
	\includegraphics[width=8cm,height=6cm]{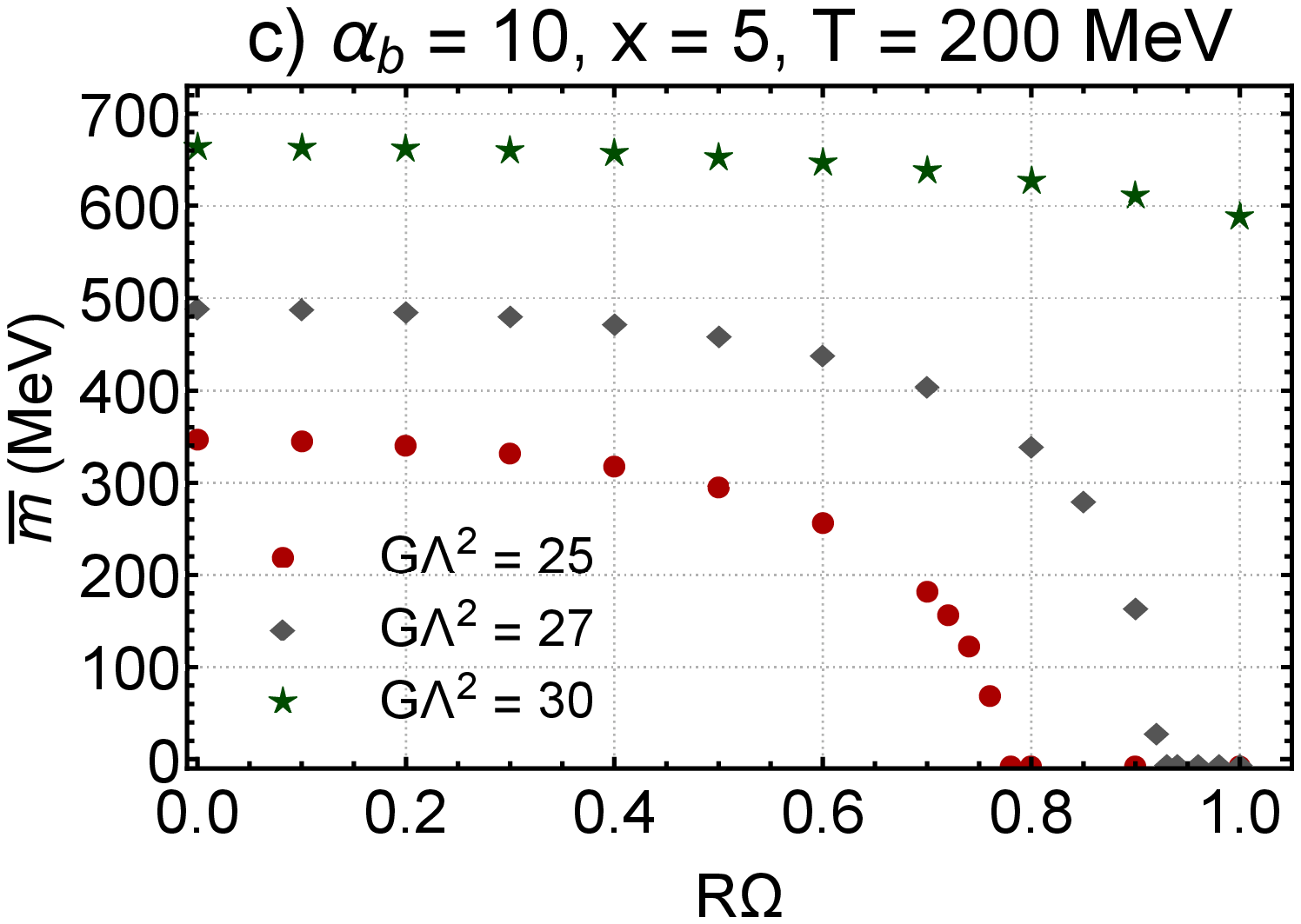}
	\includegraphics[width=8cm,height=6cm]{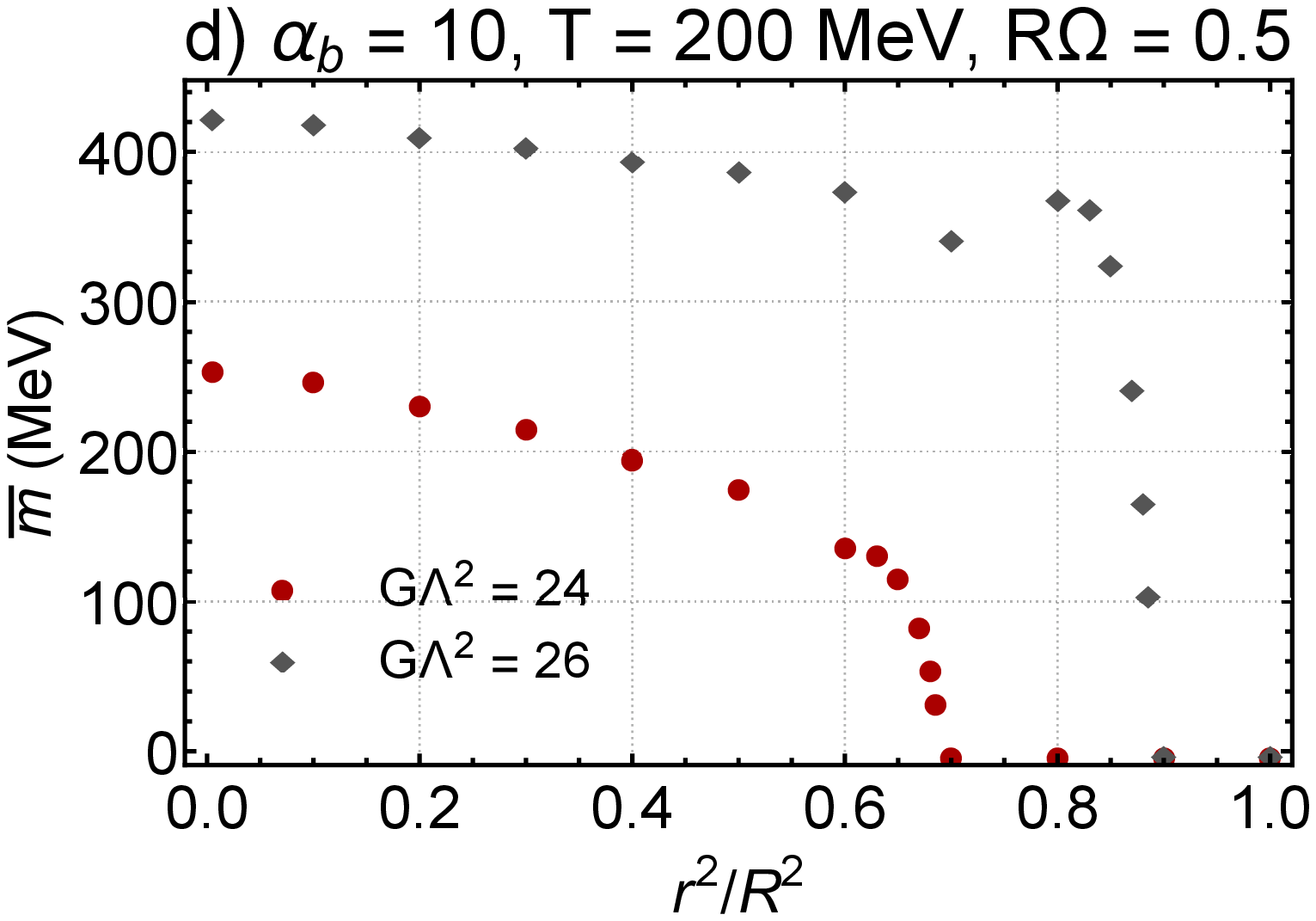}
	\caption{color online. a) The $T$ dependence of $\bar{m}$ is plotted for $G\Lambda^2=24,26$, $\alpha_b=10, x=5$ and $R\Omega=0.5$. The results indicate a second-order phase transition at certain critical temperature $T_{c}$.
		b) The $eB/m_{\pi}^{2}$ dependence of $\bar{m}$ is plotted for $G\Lambda^2=26,27$, $x=\alpha_{b}/2,T=200 $ MeV, and $R\Omega=0.8$. In the regime of weak magnetic field $eB\leq 0.5 m_{\pi}^{2}$, $\bar{m}$ decreases with increasing $eB$. This is an indication for the IMRC.
		c) The $R\Omega$ dependence of $\bar{m}$ is plotted for $G\Lambda^2=25,27,30$, $\alpha_b=10, x=5$, and $T=200$ MeV.  While for large couplings, $\bar{m}$ turns out to be almost constant, it decreases with increasing $R\Omega$ for small $G\Lambda^2=25,27$. This is another indication of the IMRC, in particular for small couplings. This effect can also be observed in the $R\Omega$ dependence of the critical temperature $T_c$ in Fig. \ref{fig9}. The critical $R\Omega$, for which the dynamical mass vanishes, increases with increasing coupling $G\Lambda^2$ (see also Fig. \ref{fig11}).
		d) The $r^2/R^2$ dependence of $\bar{m}$ is plotted for $G\Lambda^2=24,26$, $\alpha_b=10, T=200$ MeV, and $R\Omega=0.5$. As it turns out, for small couplings, $\bar{m}$ decreases with increasing $r$.
		The larger the velocity, and consequently the kinetic energy as well as the centrifugal force of a rotating system is, it is most probably in the chirally restored phase, where the dynamical mass vanishes.}\label{fig8}
\end{figure*}
In Fig. \ref{fig8}(a), the $T$ dependence of the constituent mass $\bar{m}$ is plotted for fixed $\alpha_{b}=10,x=5$ and $R\Omega=0.5$, and two different choices of $G\Lambda^{2}=24$ and $G\Lambda^{2}=26$.\footnote{By combining the definitions of $x$ and $\alpha_b$, we arrive at $x=\alpha_{b}r^2/R^2$. Hence, $\{\alpha_b=10,x=5\}$  corresponds to $r^2=0.5 R^2$. 
Moreover, the choice $x=\alpha_b/2$ in Figs. \ref{fig8}, \ref{fig9}, and \ref{fig11} corresponds to $r^{2}=0.5R^{2}$ in the whole range of $eB$.} 
As expected, for fixed $\alpha_b,R\Omega$ and $T$, $\bar{m}$ increases with increasing $G\Lambda^2$.  The same is also true for the critical temperature $T_c$. As it is demonstrated in Fig. \ref{fig8}(a), the corresponding critical temperatures for $G\Lambda^{2}=24$ and $26$ are $T_c\sim 220$ MeV and $\sim 250$ MeV, respectively. As expected, $T_c$ increases with increasing coupling.
The results presented in Fig. \ref{fig8}(a) indicate also a second-order chiral phase transition.
This is in contrast to the results presented in \cite{fayazbakhsh2011}, where it is shown that the presence of external magnetic fields leads principally to a first-order chiral phase transition. 
\par
In Fig. \ref{fig8}(b), the $eB$ dependence of $\bar{m}$ is plotted for $G\Lambda^2=26,27$, $x=x_{\text{max}}/2=\alpha_{b}/2,T=200 $ MeV, and $R\Omega=0.8$. As it turns out, in the regime of weak magnetic fields $eB\leq 0.5 m_{\pi}^{2}$, the dynamical mass $\bar{m}$ decreases with increasing $eB$. Let us remind that in a nonrotating system, because of the magnetic catalysis effect, the dynamical mass increases with increasing $eB$ \cite{fayazbakhsh2011}. In contrast, the results presented in Fig. \ref{fig8}(b) show that the rotation of a bounded system neutralizes this effect, and leads to IMRC.  A similar effect is introduced in \cite{fukushima2015} for an unbounded system. It has been dubbed "the rotational magnetic inhibition". Let us notice that the IMRC is best demonstrated for large $R\Omega$ and small $G\Lambda^2$. This is because it is mainly an effect of rotation in combination with the magnetic field.  
\par
The results presented in Fig. \ref{fig8}(c) and, in particular, Fig. \ref{fig9}(c) are another demonstration for this effect. In Fig. \ref{fig8}(c), the $R\Omega$ dependence of $\bar{m}$ is plotted for $G\Lambda^2=25,27,30$, $\alpha_b=10, x=5$, and $T=200$ MeV. For small couplings $G\Lambda^2=25,27$, $\bar{m}$ decreases significantly with $R\Omega$. It even vanishes at some critical $R\Omega_{c}$. The value of $R\Omega_{c}$ increases with increasing $G\Lambda^2$. This is because larger coupling enhances the production of the condensate, whereas rotation has a
counter-effect. There is thus a competition between rotation/coupling to destroy/produce chiral condensates. For a larger value of $G\Lambda^2\geq 30$, $\bar{m}$ decreases with increasing $R\Omega$, but it does not vanish, at least in the allowed regime of $ 0\leq R\Omega\leq 1$. According to the above results, for a constant magnetic field, the IMRC occurs in a bounded system of quark matter.
\par
\begin{figure}
\includegraphics[width=6.5cm,height=6.5cm]{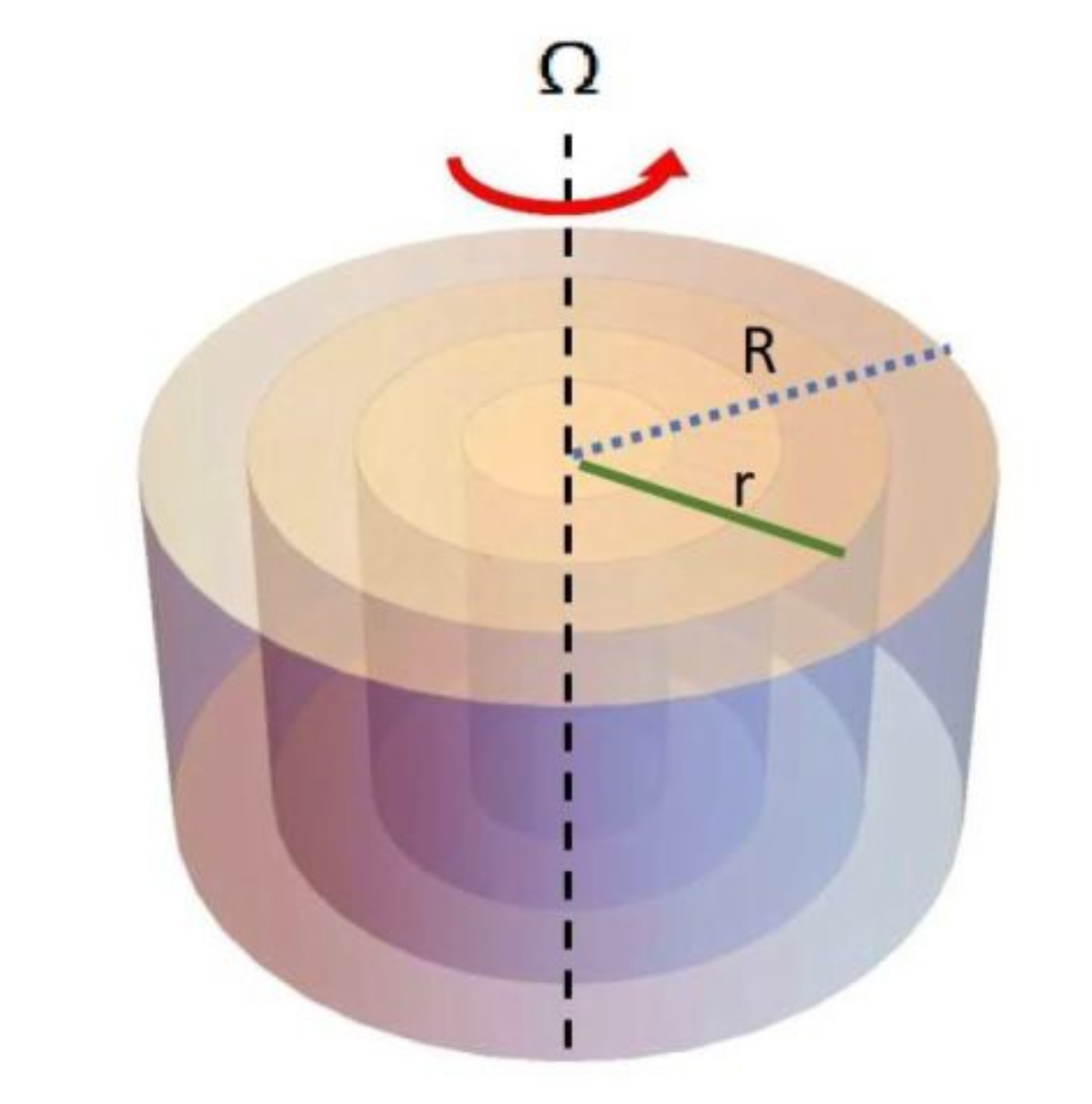}
\caption{color online. A sketch of a system of quark matter in a rotating cylinder. It can be imagined to consist of infinite number of rotating cylinders with radii $r_{i}\leq R, i=1,2,3,\cdots$, made of chiral condensate. According to the results from Fig. \ref{fig8}(c), for a given angular velocity $\Omega$, the larger $r$ is, the smaller is $\bar{m}$, thus the most probable is the chiral symmetry restoration.	
Moreover, the larger $r$ is, the larger are the kinetic energy and the centrifugal force applied on each layer. The fact that $\bar{m}$ decreases with increasing $r$ indicates that larger kinetic energy and centrifugal force have a positive impact on destroying the chiral condensate, and thus restoring the chiral symmetry in a rotating system.}\label{fig11-x}
\end{figure}

In Fig. \ref{fig8}(d) the $r^2/R^2$ dependence of $\bar{m}$ is plotted for $G\Lambda^2=24,26$, $\alpha_b=10, T=200$ MeV, and $R\Omega=0.5$. As it turns out, for intermediate value of $R\Omega$, and fixed $\alpha_{b}$ and $T$, the dynamical mass decreases with increasing $r^2/R^2$. Moreover, the results show that in order to keep $\bar{m}$ almost constant in the whole range of $0<r\leq R$, the coupling $G\Lambda^2$ has to be large enough.
The $r$ dependence of $\bar{m}$, demonstrated in Fig. \ref{fig8}(d) indicates that the IMRC is in fact induced by the linear velocity $v=r\Omega$. For a given $\Omega$, the farther the condensate is from the axis of rotation, i.e. the larger $r$ is, the larger is the rotational kinetic energy of the condensate as well as the centrifugal force it feels, and the smaller is the value of the dynamical mass $\bar{m}$. This has, by itself, positive consequences for the chiral symmetry restoration in rotating systems (see Fig. \ref{fig11-x} for a sketch of a system of quark matter in a rotating cylinder). 
\par
Let us notice at this stage that the above results are only valid for intermediate values of the magnetic fields. For a larger value of $\alpha_b$, the dynamical mass increases near the boundary. This is known as "the surface magnetic catalysis", and is elaborated, \textit{inter alia}, in \cite{fukushima2017}. 
\subsubsection{Critical temperature as a function of $G\Lambda^2, eB, R\Omega$, and $r^2/R^2$}\label{sec3b2} 
\begin{figure*}
\includegraphics[width=8cm,height=6cm]{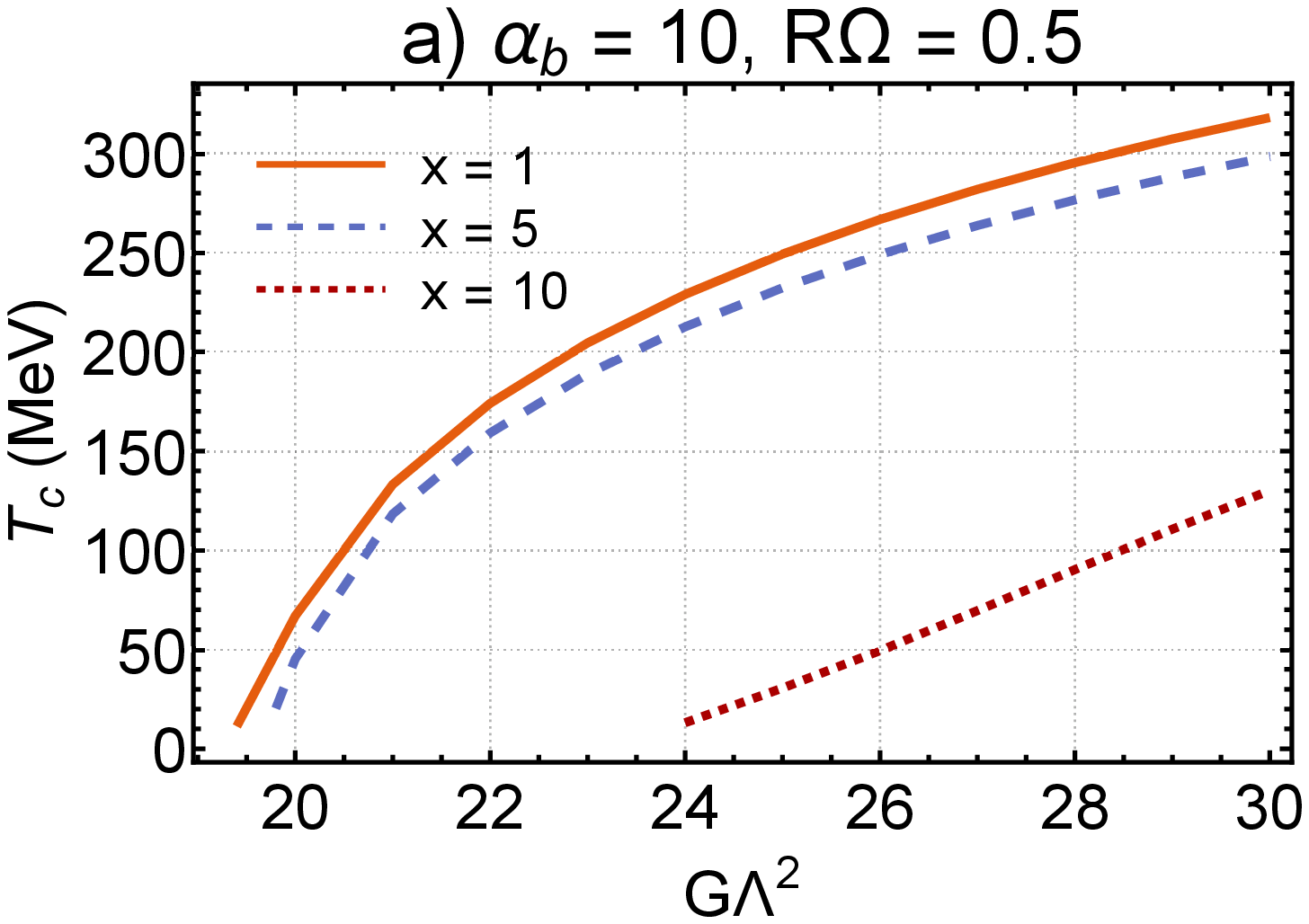}
\includegraphics[width=8cm,height=6cm]{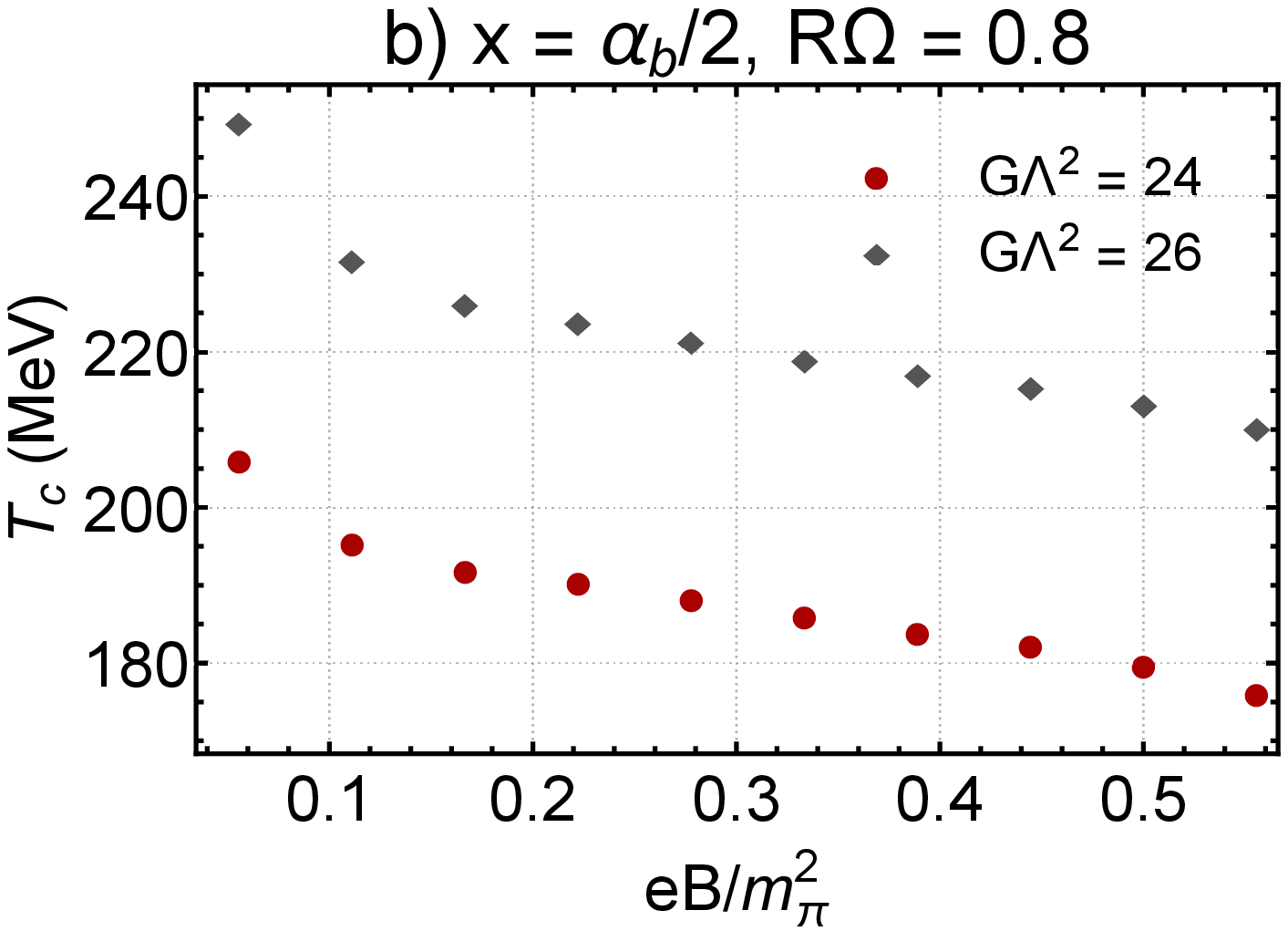}\vspace{.3cm}
\includegraphics[width=8cm,height=6cm]{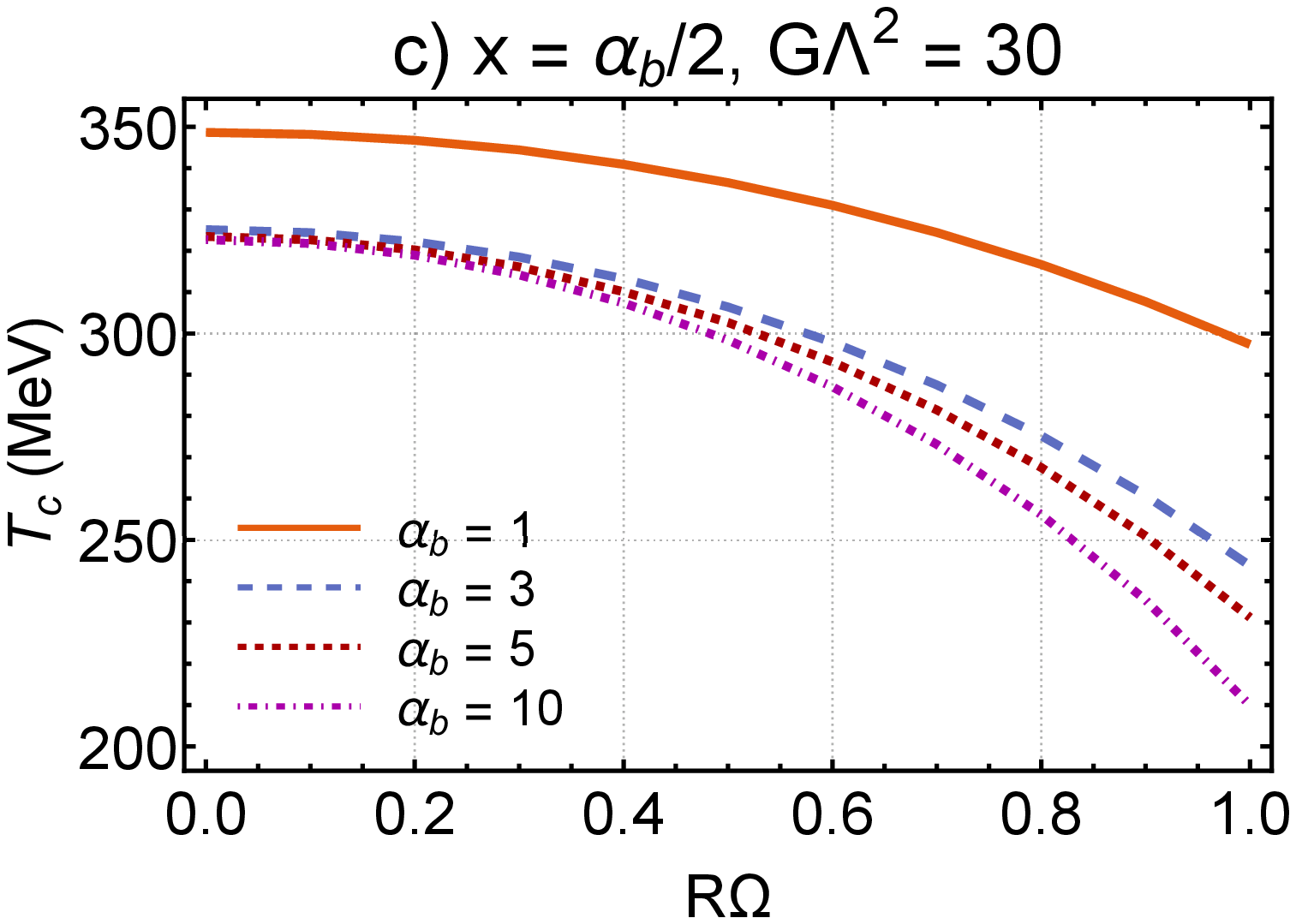}
\includegraphics[width=8cm,height=6cm]{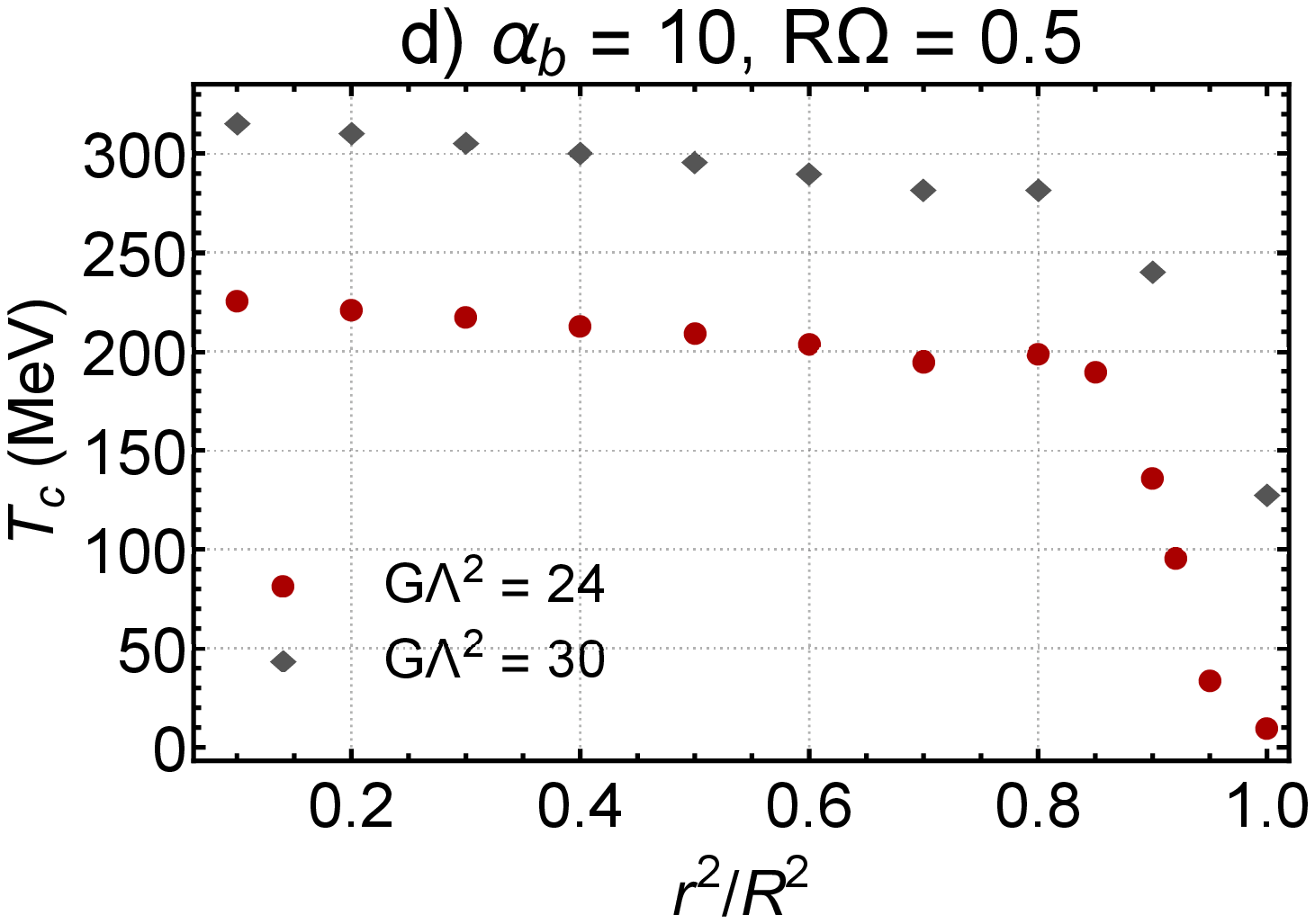}
\caption{color online. a) The $G\Lambda^2$ dependence of $T_c$ is plotted for $\alpha_{b}=10,R\Omega=0.5$ and $x=1,5,10$. As expected, $T_c$ increases with increasing $G\Lambda^2$. This confirms the fact that larger couplings enhance the formation of the dynamical mass.
b) The $eB/m_{\pi}^{2}$ dependence of $T_{c}$ is plotted for $x=\alpha_b/2, R\Omega=0.8$, and $G\Lambda_{2}=24,26$. Because of the IMRC, $T_{c}$ decreases with increasing $eB$. The effect is mainly induced by the rotation so that for a smaller value of $R\Omega$, the slope of the curves becomes smaller. c)  The  $R\Omega$ dependence of $T_c$ is plotted for $x=\alpha_{b}/2,G\Lambda^2=30$ and
$\alpha_{b}=1,3,5,10$. As it turns out, $T_{c}$ decreases with increasing $R\Omega$. This indicates an IMRC induced apparently by the interplay between rotation and the presence of a magnetic field in a bounded system of quark matter.
d) The $r^{2}/R^{2}$ dependence of $T_{c}$ is plotted for $\alpha_b=10, R\Omega=0.5$, and $G\Lambda^2=24,30$. Only at the boundary $0.8 R^2<r^2<R^2$, the critical temperature $T_c$ decreases with $r/R$, otherwise it remains almost constant.}\label{fig9}
\end{figure*}
\begin{figure*}
\includegraphics[width=5.5cm,height=5cm]{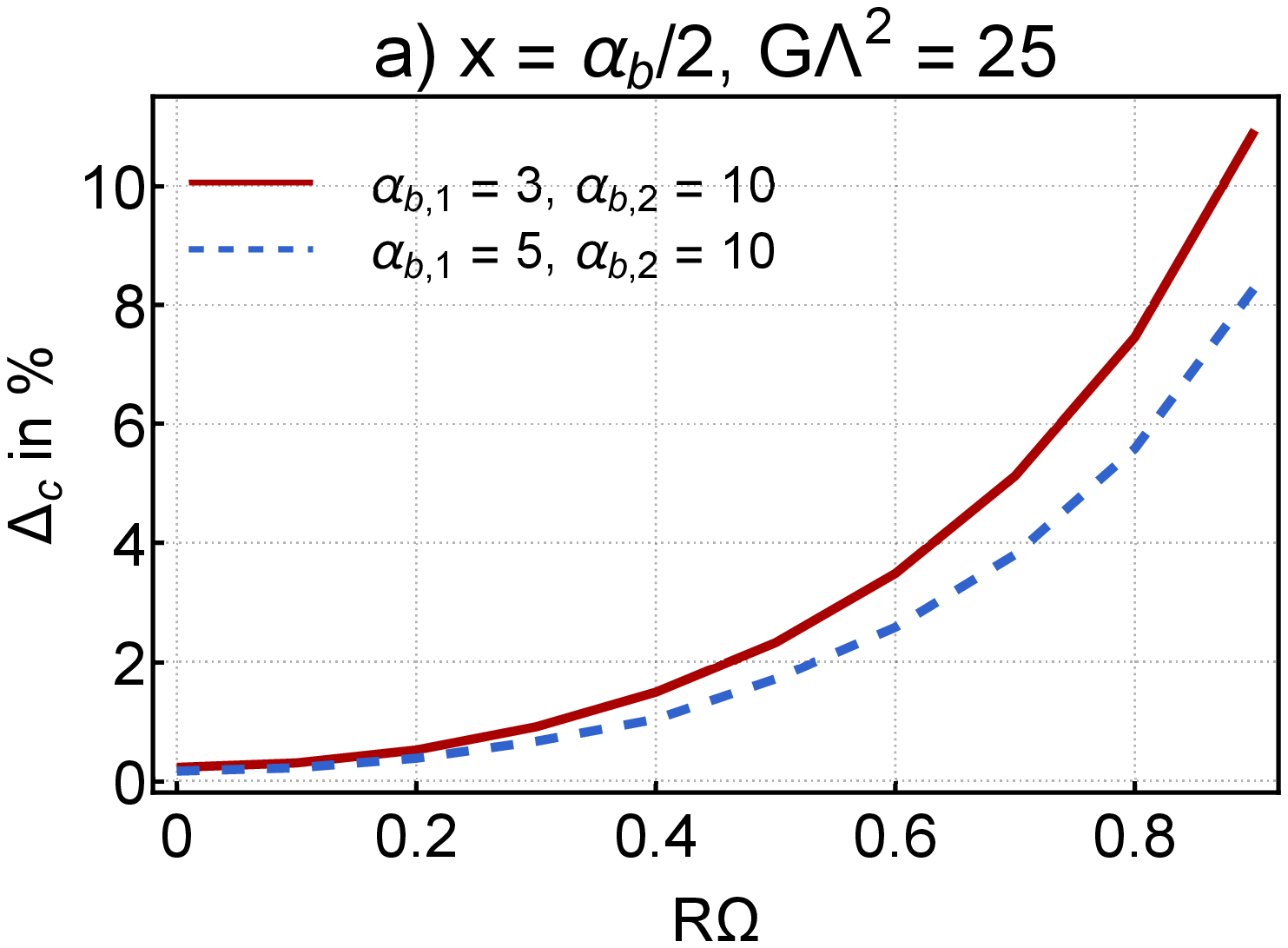}
\includegraphics[width=5.5cm,height=5cm]{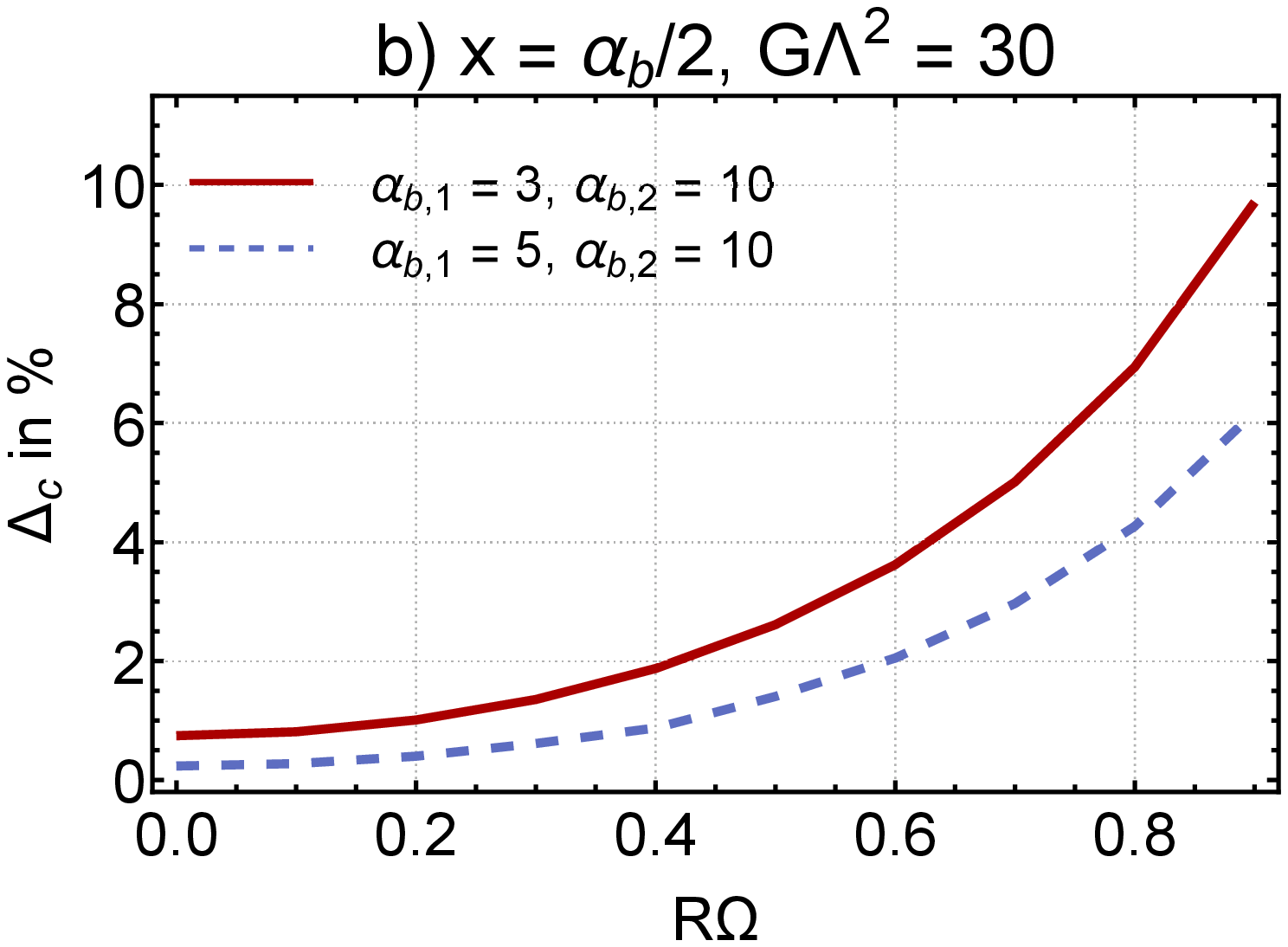}
\includegraphics[width=5.5cm,height=5cm]{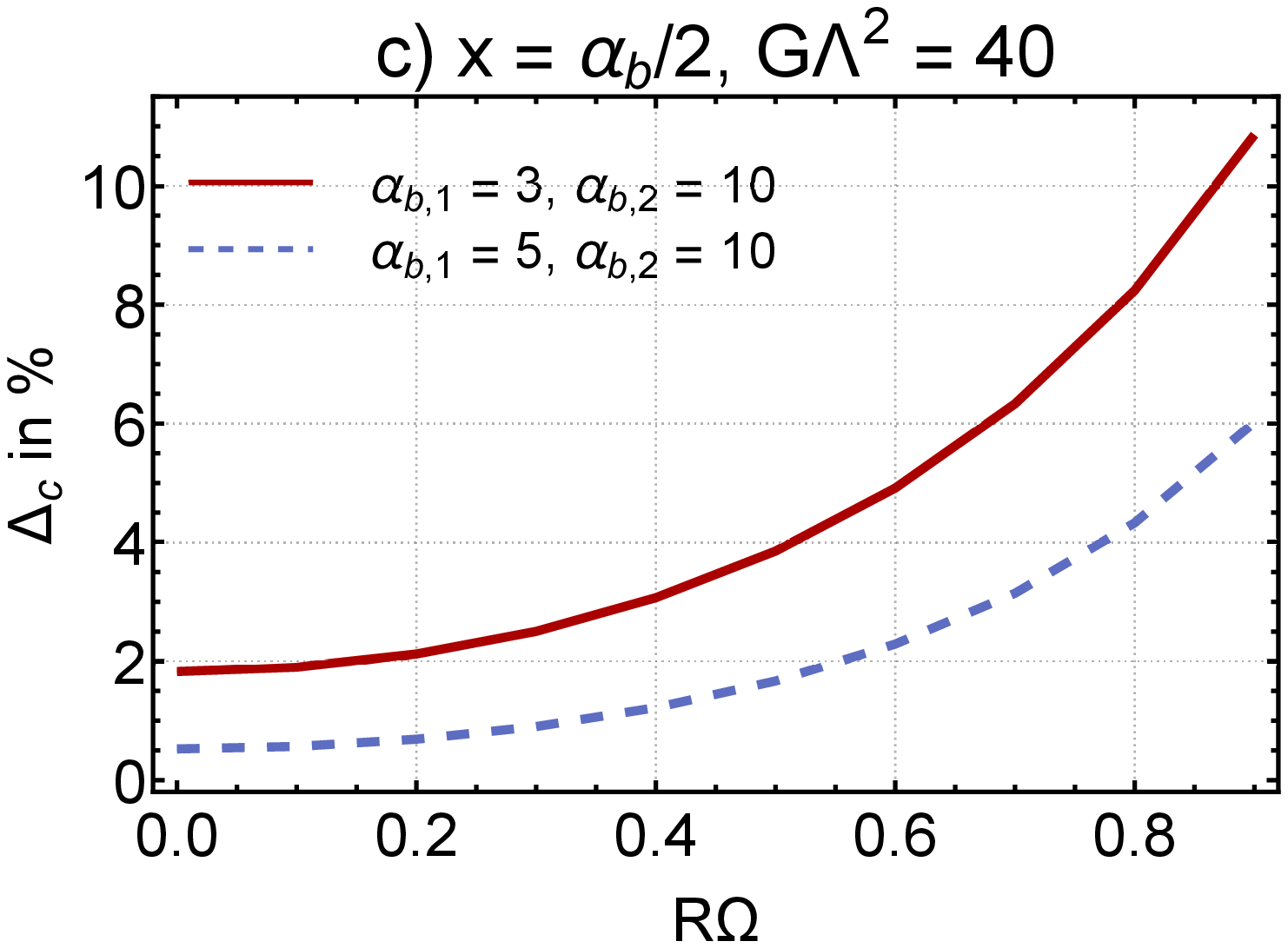}
\caption{color online. The $R\Omega$ dependence of $\Delta_{c}$ is plotted for $x=\alpha_{b}/2$ as well as $G\Lambda^2=25$ (panel a), $G\Lambda^2=30$ (panel b), and $G\Lambda^2=40$ (panel c). As it turns out $\Delta_{c}$ increases with increasing $R\Omega$. A comparison between the results in the panels a, b, and c shows that the maximum value of $\Delta_{c}$ decreases with increasing $G\Lambda^2$.}\label{fig10}
\end{figure*}
To explore the interplay between the background magnetic field and the rotation of a bounded system of quark matter, and in particular, to scrutinize their possible effects on the critical temperature $T_c$ of a chiral phase transition, the $G\Lambda^2, eB, R\Omega$, and $r$ dependence of the critical temperature $T_{c}$ are plotted for various fixed parameters (see Fig. \ref{fig9}). In Fig. \ref{fig9}(a), the $G\Lambda^2$ dependence of $T_c$ is plotted for $\alpha_{b}=10,R\Omega=0.5$ and $x=1,5,10$. The latter correspond to $r^2/R^2=0.1, 0.5, 1$, respectively. As expected, $T_c$ increases with increasing coupling. This is because of large values of $G\Lambda^2$ the formation of a chiral condensate is enhanced, and a phase transition to a chiral symmetry restored phase becomes only possible at higher temperatures. Moreover, the transition temperature depends on the distance (velocity/angular kinetic energy/centrifugal force) of the system with respect to the rotation axis. The farther bound states are from the origin (rotation axis), the lower is the critical temperature of a chiral phase transition.
\par
In Fig. \ref{fig9}(b), the $eB$ dependence of $T_{c}$ is plotted for $x=\alpha_{b}/2, R\Omega=0.8$, and $G\Lambda^{2}=24,26$. We choose an appropriate large value of $R\Omega$, and small values of $G\Lambda^2$, to best demonstrate the IMRC. As it turns out, $T_{c}$ decreases with increasing $eB$. This is definitely a sign for an IMRC, induced solely by the rotation of a bounded system of quark matter. Let us remind that the results presented in \cite{fayazbakhsh2011} for a nonrotating two-flavor NJL model at zero chemical potential $\mu$ show an increase of $T_c$ as a function of $eB$. This is because of the catalytic effect of the magnetic field. In contrast, it is also shown that for nonvanishing chemical potential, $T_{c}$ decreases with increasing $eB$, and thus inverse magnetic catalysis occurs for $\mu\neq 0$ \cite{fayazbakhsh2011}. There are, however, pieces of evidence for inverse magnetic catalysis arising from an \textit{ab initio} lattice QCD computations \cite{bali2013}.  Here, it is shown that $T_{c}$ decreases with $eB$ even in a system with zero chemical potential.
\par
The IMRC is best demonstrated in the $T_{c}$ versus $R\Omega$ phase portrait in Fig. \ref{fig9}(c). Here, the  $R\Omega$ dependence of $T_c$ is plotted for $x=\alpha_{b}/2,G\Lambda^2=30$ and $\alpha_{b}=1,3,5,10$. The critical temperature decreases with increasing $R\Omega$, but the slope of the corresponding curves increases with increasing $\alpha_{b}$. Moreover, for a constant $R\Omega$, $T_{c}$ decreases with increasing $\alpha_{b}$.  These effects become more significant in the regime of large $R\Omega$, e.g. $0.4<R\Omega\leq 1$  (see also Fig. \ref{fig10}).
\par
In Fig. \ref{fig9}(d), the $r^{2}/R^{2}$ dependence of $T_{c}$ is plotted for $\alpha_b=10, R\Omega=0.5$, and $G\Lambda^2=24,30$. In the regime $0<r^2<0.8 R^2$, $T_c$ decreases
slightly with $r/R$. Only near the boundary, for $0.8 R^2<r^2<R^2$ the critical temperature $T_c$ decreases drastically with $r/R$.
These results indicate that as long as the ratio $r^2/R^2$ is not larger than $r^2/R^2=0.8$, the above conclusions concerning the appearance of the IMRC do not depend on $r$. The reason for the slight decrease of $T_c$ as a function of $r$ is the fact that for small values of coupling, the dynamical mass decreases with increasing $r$ [see Fig. \ref{fig8}(d)], and thus smaller temperatures are necessary to destroy the bound states.
\par
To have a measure for the effect of large $R\Omega$ on IMRC, let us define a quantity $\Delta_c$ as
\begin{eqnarray}
\Delta_{c}\equiv\frac{T_{c}(\alpha_{b,1})-T_{c}(\alpha_{b,2})}{T_{c}(\alpha_{b,1})}\time 100\qquad\mbox{in \%},
\end{eqnarray}
for fixed $x$ and $G\Lambda^2$. In Fig. \ref{fig10}, the $R\Omega$ dependence of $\Delta_{c}$ is plotted for $\alpha_{b,1}=1,5$ and $\alpha_{b,2}=10$, fixed $x=\alpha_{b}/2$, and $G\Lambda^2=25,30,40$ [See Figs. \ref{fig10}(a)-\ref{fig10}(c)].\footnote{Note that $T_{c}(\alpha_{b}=10)\leq T_{c}(\alpha_{b}=1),\mbox{and},~T_{c}(\alpha_{b}=3)$. Hence, $\Delta_{c}\geq 0$ in these two cases.} The results reveal that independent of the value of the coupling $G\Lambda^2$, $\Delta_{c}$ increases with increasing $R\Omega$.
\subsubsection{Critical $R\Omega_{c}$ as a function of $G\Lambda^2, eB,T$, and $r^2/R^2$}\label{sec3b3} 
Focusing on the behavior of the dynamical mass and the critical temperature of the chiral phase transition with respect to $eB$ and $R\Omega$, the IMRC is demonstrated in Figs. \ref{fig8}(b), \ref{fig8}(c), as well as \ref{fig9}(b) and \ref{fig9}(c). As it is shown in Fig. \ref{fig8}(c), for fixed values of $\alpha_b,x,T$, and, in particular, for small values of $G\Lambda^{2}$, the dynamical mass $\bar{m}$ vanishes for certain critical velocity $R\Omega_c\leq 1$. For $G\Lambda^2=25$ and $G\Lambda^{2}=27$, $R\Omega_{c}\sim 0.78$ and $R\Omega_c=0.93$, respectively. In what follows, the dependence of $R\Omega_{c}$ on $G\Lambda^2, eB,T$, and $r^2/R^2$ is explored.
\par
\begin{figure*}
\includegraphics[width=8cm,height=6cm]{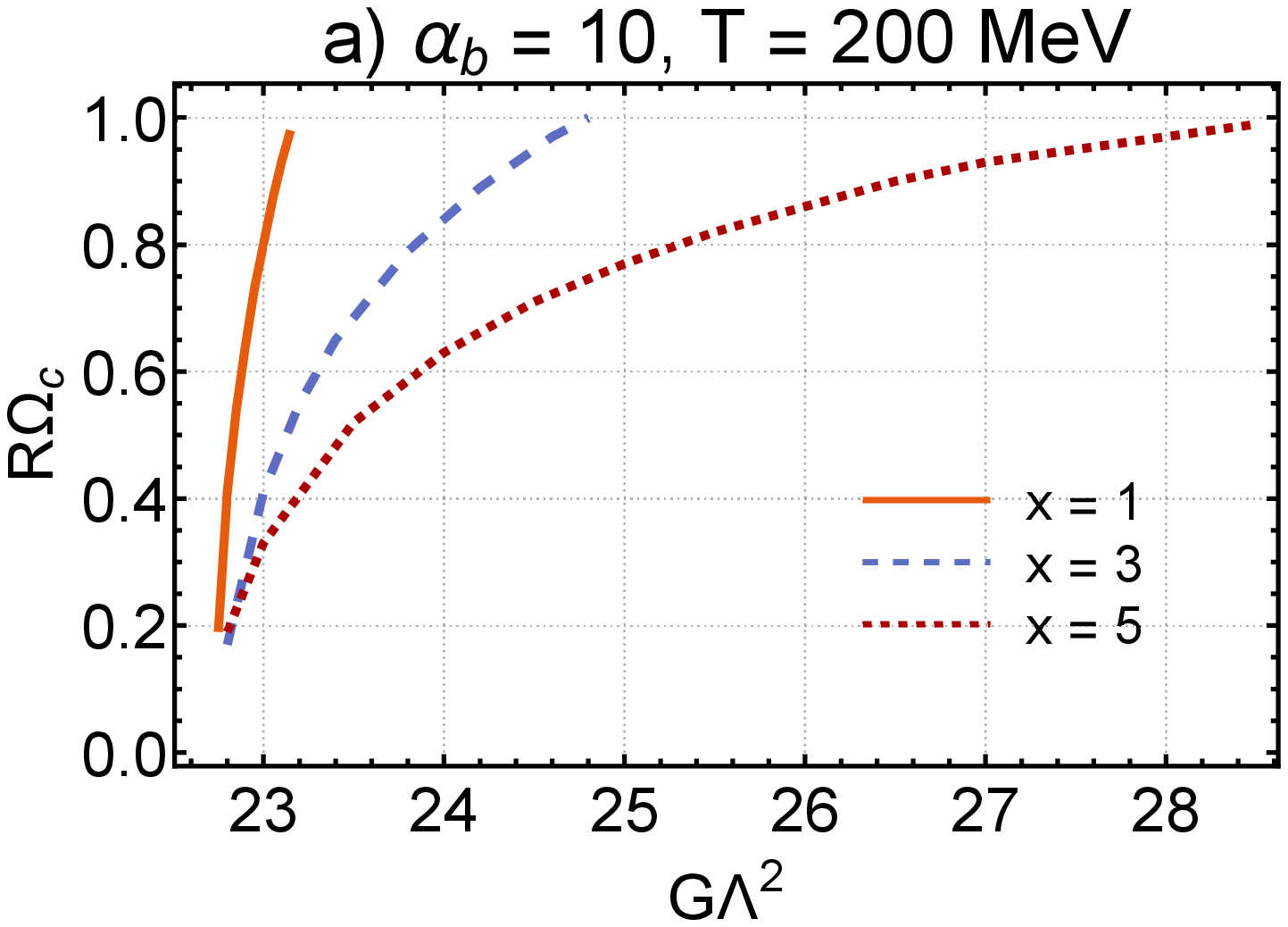}
\includegraphics[width=8cm,height=6cm]{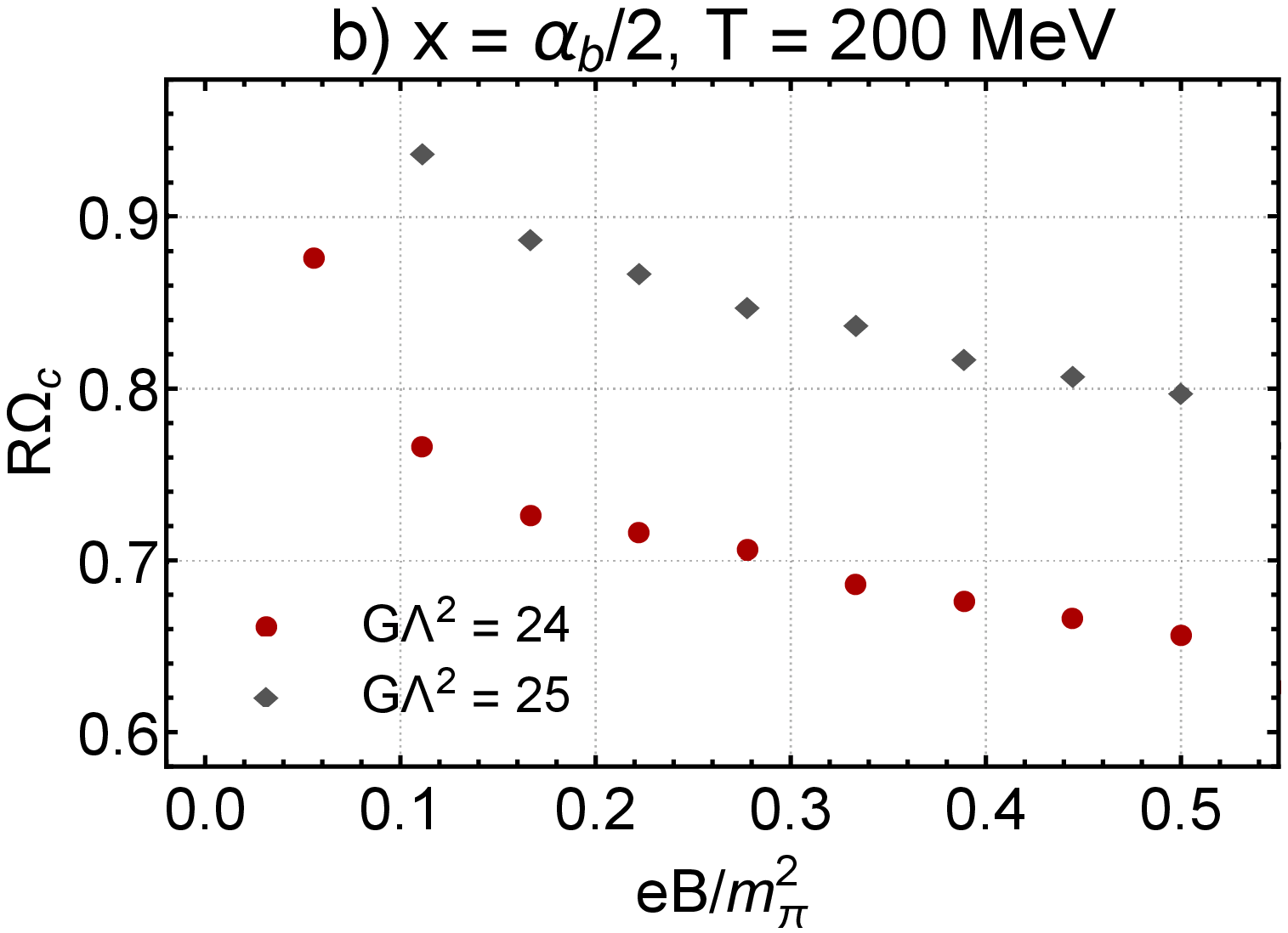}\vspace{0.3cm}
\includegraphics[width=8cm,height=6cm]{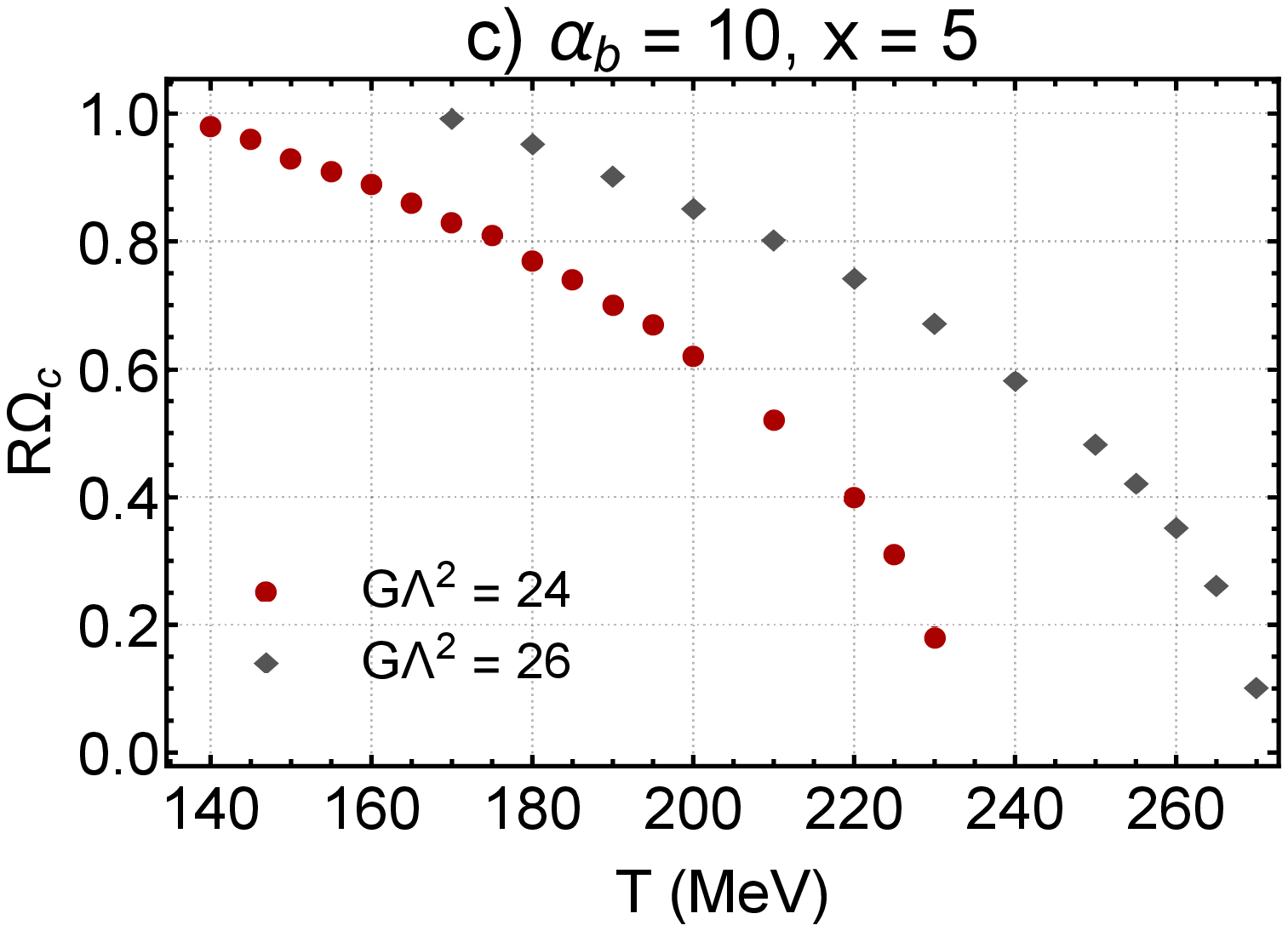}
\includegraphics[width=8cm,height=6cm]{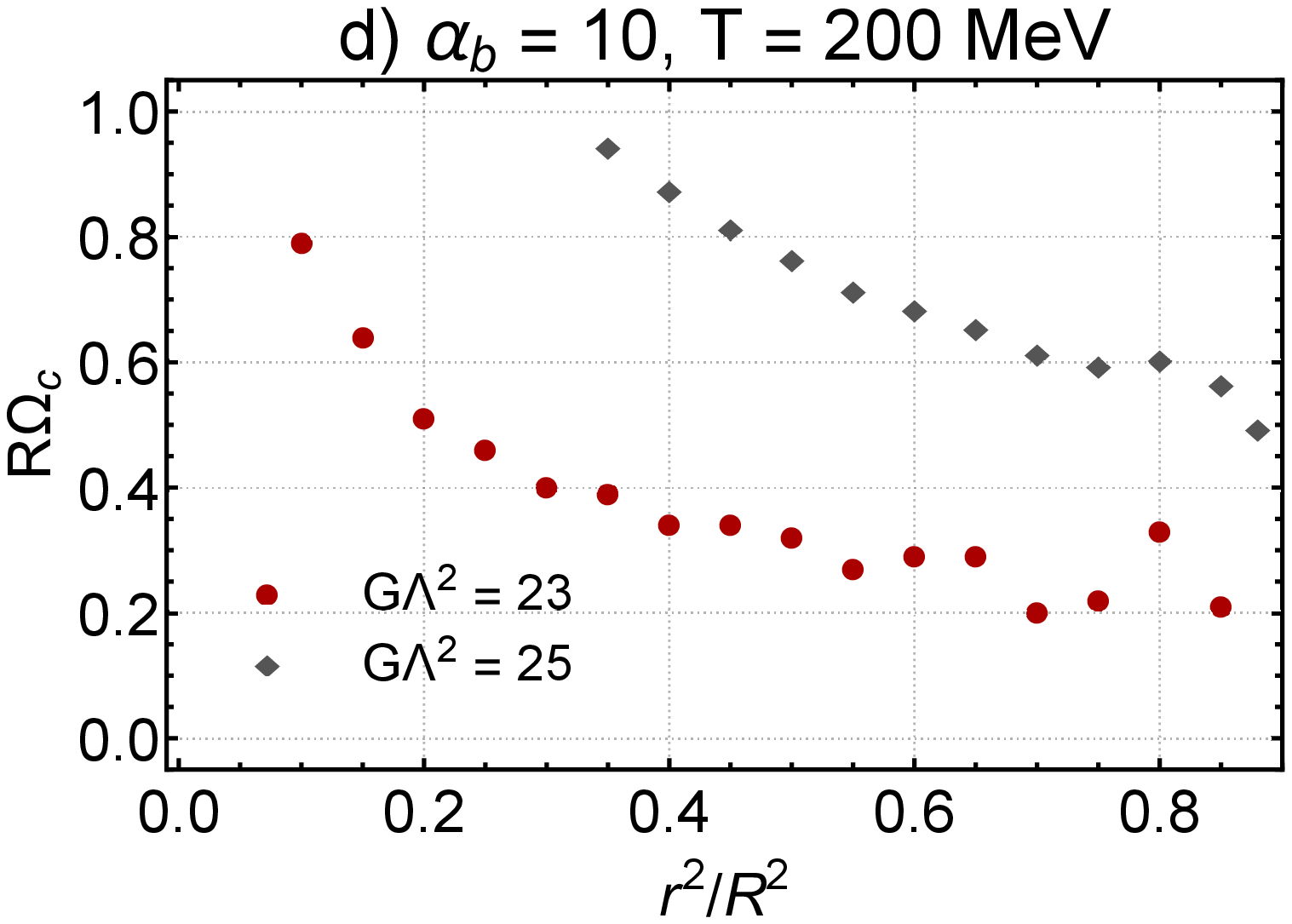}
\caption{
color online. a) The $G\Lambda^2$ dependence of $R\Omega_c$ is plotted for $\alpha_{b}=10,T=200$ MeV, and $x=1,3,5$. As expected, $R\Omega_c$ increases with increasing $G\Lambda^2$. For a given value of $G\Lambda^2$, the smaller $x$ is, the larger values of $\Omega$ is necessary to restore the chiral symmetry.
b) The $eB/m_{\pi}^{2}$ dependence of $R\Omega_{c}$ is plotted for $x=\alpha_{b}/2, T=200$ MeV, and $G\Lambda_{2}=24,25$.  Because of the IMRC,  $R\Omega_{c}$ decreases with increasing $eB$.
c)  The  $T$ dependence of $R\Omega_c$ is plotted for $\alpha_b=10, x=5$, and $G\Lambda^2=24,26$.  As expected, $R\Omega_{c}$ decreases with increasing temperature, which, by itself, trigger the chiral symmetry restoration.
d) The $r^{2}/R^{2}$ dependence of $R\Omega_{c}$ is plotted for $\alpha_b=10, T=200$ MeV, and $G\Lambda^2=23,25$. For a given value of $eB,\Omega,T$ and $G$, the farther the distance from the rotational axis is, the larger is the kinetic energy of a bound state as well as the centrifugal force applied on it, thus the smaller values of $\Omega$ are necessary to destroy the bound state, and restore the chiral symmetry.
}\label{fig11}
\end{figure*}

In Fig. \ref{fig11}(a), the $G\Lambda^2$ dependence of $R\Omega_{c}$ is plotted for $\alpha_b=10, T=200$, and $x=1,3,5$. Although for all values of $x$, $R\Omega_c$ increases with $G\Lambda^2$, but the slope of the curves decreases with increasing $x$. For a given value of $G\Lambda^2$, the smaller the value of $x$ is, the larger $\Omega$ is necessary to break the condensate, and to restore the chiral symmetry. This is because, for a given $\Omega$, the larger $x$ is, the larger is the kinetic energy of a bound state in the rotating system of quark matter and the centrifugal force applied on it. Apparently, larger kinetic energy/centrifugal force applied on constituents helps to destroy the corresponding bound state.
The same effect is also observed in Fig. \ref{fig11}(d), where the $r^2/R^2$ dependence of $R\Omega_c$ is plotted for $\alpha_b=10, T=200$ MeV and $G\Lambda^2=23, 25$. Here, for a fixed $\Omega$, the larger $r$ is, the larger is the rotational kinetic energy as well as the centrifugal force, thus the smaller values of $\Omega$ are necessary to destroy the chiral condensate and restore the chiral symmetry.
\par
In Fig. \ref{fig11}(b), the $eB$ dependence of $R\Omega_c$ is plotted for fixed $x=\alpha_{b}/2, T=200$ MeV, and $G\Lambda^2=24,25$. Because of the IMRC, in which the dynamical mass decreases with increasing $eB$ [see Fig. \ref{fig8}(b)], once the larger $eB$ is, the smaller values of $R\Omega$ are necessary to restore the chiral symmetry. The same is also true for the temperature dependence of $\Omega_{c}$. Since higher temperatures help to destroy the chiral condensate, the larger $T$ is, the smaller values of $\Omega$ are necessary to restore the chiral symmetry.

\section{Concluding remarks}\label{sec4}
\setcounter{equation}{0}
One of the key ingredients in studying the possible effects of the simultaneous presence of the rotation and constant magnetic field on fermionic systems is the solution of the corresponding Dirac equation. In the first part of the present work, we presented a systematic derivation of the solutions of the Dirac equation in a rotating and magnetized system, using the Ritus eigenfunction method. Using the Ritus eigenfunctions and energy spectrum, we derived the corresponding quantization relations for fermions in an infinitely extended system with no boundary condition. We then imposed a global boundary condition on this fermionic system to avoid causality-violating effects, and scrutinized the effect of boundary conditions on the energy spectrum of fermions once they are bounded in a cylinder with radius $R$. This derivation, which did not appear in the literature before, gives us new insight, in particular, into the way how fermions occupy the lowest energy levels. We showed, in particular, that for a bounded system the lowest energy level is to be determined numerically. Using the aforementioned quantization relation of a bounded magnetized and rotating system, we derived the corresponding fermion propagator, from which appropriate expressions for the gap equations of a QCD-like model at zero and nonzero temperatures were found. 
\par
We showed that at zero temperature, the dynamical mass $\bar{m}$ has no dependence on the angular frequency $\Omega$. This confirms the statement in \cite{chernodub2016-1} that "cold vacuum does not rotate". Its dependence on the distance $r$ from the rotation axis could be regulated by choosing appropriate coupling $G_m$ for each fixed $eB$. The values of $G_{m}$ are determined so that the dynamical mass remains almost constant in a relatively large interval of $r$ (see Table \ref{tab5}). We then studied the $r$ and $eB$ dependence $\bar{m}$ for different values of $G_m$ and fixed values of $eB$ and $r$, respectively, and showed how the $eB$ dependence of $G_m$ is reflected in the $eB$ dependence of $\bar{m}$ [compare the results demonstrated in Figs. \ref{fig3x} and \ref{fig6}(b)].    
\par
In the finite temperature case, we numerically determined the $T, eB,\Omega$, and $r$ dependence of $\bar{m}$, and plotted the complete phase portraits of $T_c$ versus $G\Lambda^2$, $eB/m_{\pi}^{2}$, $R\Omega$, and $r^2/R^2$ as well as $R\Omega_c$ versus $G\Lambda^2$, $eB/m_{\pi}^{2}$, $R\Omega$, and $r^2/R^2$. Our results show that there are, at least, three signatures for the fact that rigid rotation initiates the IMRC. The first one is the decrease of $\bar{m}$ with $eB$, once $G\Lambda^2, T, r$, and $\Omega$ are kept fixed [see Fig. \ref{fig8}(b)]. As it turns out, the slope is larger for smaller values of $G\Lambda^2$ and $R\Omega$. Hence, an appropriate choice of these two parameters negatively affects the production of the dynamical mass and leads to IMRC. Apart from this specific behavior of $\bar{m}$, the fact that $T_{c}$ decreases with increasing $eB$ is a result of the IMRC [see Fig. \ref{fig9}(b)]. This is similar to ordinary inverse magnetic catalysis in a nonrotating quark matter, whose evidence from lattice QCD is exactly the same phenomenon. But, there is also a third and novel evidence of IMRC. This is the decrease of $\Omega_c$ with $eB$ [see Fig. \ref{fig11}(b)]. Recent studies show similar results for a two-flavor quark matter \cite{tabatabaee2021-2}.          
\par
Apart from the IMRC, the $r$ dependence of $\bar{m}$ is striking. It turns out that, for small enough coupling $G\Lambda^2$ and even for large $eB$, the dynamical mass decreases with increasing $r/R$ [see Fig. \ref{fig8}(d)]. A similar conclusion is also made in \cite{chernodub2016-1}. Here, we must be cautious. Actually, we solved the gap equation with the assumption of $\partial_r\bar{m}\ll \bar{m}(r)^2$ \cite{fukushima2017}. Hence,  $\bar{m}(r)$  curves with very large slopes in Fig. \ref{fig8}(d) are unacceptable. As it turns out, for fixed values of $eB,T$, and $R\Omega$,  an appropriate choice of $G\Lambda^2$ affects the slope of the $\bar{m}(r)$ curves. Without this assumption, we are facing an integral equation, whose solution leads to an inhomogeneous mass gap. In \cite{wang2019-1, wang2019-2}, the nonlocal chiral condensate in $2+1$ dimensions is treated using a Bogoliubov-de Gennes like method \cite{buballa2015}. It is shown that for sufficiently large angular frequency, chiral vortices are built in the ground state. These are topological defects in analogy to the ones appearing in superfluids and superconductors \cite{wang2019-2}. It would be interesting to extend this work to the case of magnetized quark matter, which is relevant in the physics of neutron stars. These compact stars are subject to extremely large angular frequencies and very large magnetic fields. Other methods to treat inhomogeneous chiral condensates are presented in \cite{ferrer2015}, where the Magnetic Dual Chiral Density Wave phase is studied. A possible extension of this work is to introduce the rotation, and to study its effect on the phase diagram of this system \cite{sadooghi2021-3}. Implications of rotation on the phenomenological observables of HIC, like dilepton and photon production rates, are also most relevant in HIC's physics. We postpone these interesting subjects to our future publications.  
\section{Acknowledgments}
N. S. thanks M. H. Gholami for useful discussions. This work is supported by Sharif University of Technology's Office of Vice President for Research under Grant No: G960212/Sadooghi.



\end{document}